\documentclass[prd,preprint,superscriptaddress,preprintnumbers,eqsecnum,showpacs,nofootinbib,nobibnotes]{revtex4}
\usepackage{amsfonts,bm}
\usepackage{amsfonts,amssymb,amsmath}
\usepackage{graphicx}
\usepackage{paralist}
\usepackage[english]{babel}
\usepackage{slashed}
\usepackage[usenames]{xcolor}
\usepackage{float}
\usepackage{pstricks}
\usepackage{tabularx}
\newcolumntype{C}[1]{>{\centering\arraybackslash}p{#1}}

\newcommand{\be}{\begin{equation}}
\newcommand{\bea}{\begin{eqnarray}}
\newcommand{\ee}{\end{equation}}
\newcommand{\eea}{\end{eqnarray}}

\def\s#1{{\scriptscriptstyle #1}}
\def\ie{{\it i.e.}, }
\def\eg{{\it e.g.}, }

\def\1eq#1{Eq.~(\ref{#1})}

\def\2eqs#1#2{Eqs.~(\ref{#1}) and~(\ref{#2})}
\def\3eqs#1#2#3{Eqs.~(\ref{#1}),~(\ref{#2}), and~(\ref{#3})}

\def\cd{\!\cdot\!}

\def\K{{\cal K}}

\newcommand{\Mass}{\s{\rm{1M}}}

\newcommand{\inpt}{\s{\rm{in}}}

\newcommand{\fatg}{{\rm{I}}\!\Gamma}   
\newcommand{\Gnp}{\Gamma}        
\newcommand{\Gp}{V}          

\begin{document}
\title{Nonperturbative structure of the ghost-gluon kernel}

\author{A.~C. Aguilar}
\affiliation{University of Campinas - UNICAMP, 
Institute of Physics ``Gleb Wataghin'',
13083-859 Campinas, SP, Brazil}

\author{M.~N. Ferreira}
\affiliation{University of Campinas - UNICAMP, 
Institute of Physics ``Gleb Wataghin'',
13083-859 Campinas, SP, Brazil}
\affiliation{\mbox{Department of Theoretical Physics and IFIC, 
University of Valencia and CSIC},
E-46100, Valencia, Spain}

\author{C.~T. Figueiredo}
\affiliation{University of Campinas - UNICAMP, 
Institute of Physics ``Gleb Wataghin'',
13083-859 Campinas, SP, Brazil}
\affiliation{\mbox{Department of Theoretical Physics and IFIC, 
University of Valencia and CSIC},
E-46100, Valencia, Spain}

\author{J. Papavassiliou}
\affiliation{\mbox{Department of Theoretical Physics and IFIC, 
University of Valencia and CSIC},
E-46100, Valencia, Spain}

\begin{abstract}

The ghost-gluon scattering kernel is a special correlation function 
 that is intimately connected with two fundamental
 vertices of the gauge sector of QCD: the ghost-gluon vertex, which
 may be obtained from it through suitable contraction,
 and the three-gluon vertex, whose Slavnov-Taylor identity
 contains that kernel as one of its main ingredients.  
In this work we 
present a detailed nonperturbative study of the five form factors
comprising it, using as starting point the ``one-loop dressed''
approximation of the dynamical equations governing their evolution. 
The analysis is carried out for arbitrary Euclidean momenta, 
and makes extensive use of the gluon propagator and the ghost dressing function,
whose infrared behavior has been firmly established from 
a multitude of  continuum studies and large-volume lattice simulations.  
In addition, special Ans\"atze are employed for the vertices entering in the
relevant equations, and their impact on the results is scrutinized in detail.
Quite interestingly, the veracity of the approximations employed
may be quantitatively tested by appealing to an exact relation,  
which fixes the value of a special combination of the form factors under construction. 
The results obtained furnish the two form factors of the ghost-gluon vertex
for arbitrary momenta, and, more importantly,  
pave the way towards the nonperturbative generalization
of the Ball-Chiu construction for the longitudinal part of the three-gluon vertex.

\end{abstract}


\maketitle

\section{\label{sec:intro} Introduction}

The nonperturbative behaviour of the  
fundamental Green's functions of QCD, such as propagators and vertices, has received considerable attention in recent years~\cite{Roberts:1994dr,Binosi:2014aea,Binosi:2016nme,Binosi:2016rxz,Fischer:2006ub,Fischer:2008uz,Aguilar:2008xm, RodriguezQuintero:2010wy,Boucaud:2008ky,Pennington:2011xs,Campagnari:2010wc,Alkofer:2000wg,Schleifenbaum:2004id,Maris:1997hd,Maris:2003vk,Aguilar:2004sw,Kondo:2006ih,Braun:2007bx,Binosi:2007pi,Binosi:2008qk,Binosi:2009qm,Kondo:2011ab,Watson:2010cn,Watson:2011kv,Aguilar:2010cn,Cloet:2013jya,Mitter:2014wpa,Braun:2014ata,Heupel:2014ina,Binosi:2016wcx,Aguilar:2018epe,Blum:2014gna, Eichmann:2014xya,Aguilar:2013vaa,Pelaez:2013cpa,Siringo:2016jrc, Siringo:2018uho, Bermudez:2017bpx,  
Aguilar:2014lha,Aguilar:2015nqa,Aguilar:2016lbe,
Cornwall:1989gv,Cornwall:1981zr,Aguilar:2001zy,Aguilar:2002tc,
Vandersickel:2012tz,
Dudal:2012zx,Boucaud:2011eh,Huber:2012kd,Mintz:2017qri,Cucchieri:2007md,Cucchieri:2010xr,Cucchieri:2007rg,Bogolubsky:2007ud,Bogolubsky:2009dc,Oliveira:2009eh,Oliveira:2018lln,Oliveira:2010xc, Oliveira:2012eh, Bowman:2005vx, Ayala:2012pb,Sternbeck:2017ntv,Ilgenfritz:2006he,Sternbeck:2006rd,Kizilersu:2006et,Skullerud:2003qu,Boucaud:2017obn,Athenodorou:2016oyh,Cucchieri:2008qm,Cucchieri:2006tf,Aguilar:2013xqa,Huber:2012zj,Maas:2011se,Williams:2015cvx,Eichmann:2015nra,Eichmann:2008ef,Corell:2018yil,Cyrol:2017ewj,Cyrol:2016tym}, and is believed to
be essential for acquiring a deeper understanding
of the strong interactions. In this particular quest, the combined efforts 
between various continuum 
approaches\mbox{~\cite{Roberts:1994dr,Boucaud:2008ky,Aguilar:2008xm,Fischer:2008uz,Mitter:2014wpa,Vandersickel:2012tz}}, and large-volume lattice 
simulations~\cite{Cucchieri:2007md,Cucchieri:2010xr,Cucchieri:2007rg,Bogolubsky:2007ud,Bogolubsky:2009dc,Oliveira:2009eh,Oliveira:2018lln,Oliveira:2010xc, Oliveira:2012eh, Bowman:2005vx, Ayala:2012pb} 
have furnished a firm control on the infrared structure of 
the two-point sector of the theory (gluon, ghost, and quark propagators).

The case of the three-point functions (vertices) represents currently a
major challenge, because, while their knowledge is considered to be crucial 
for both theory and phenomenology, their
first-principle determination 
by means of conventional approaches is technically
rather involved.
In particular, such vertices possess, in general, rich tensorial structures, 
and their form factors contain three independent momenta.
In order to determine the momentum dependence of vertex form factors,    
one may perform lattice simulations~\cite{Sternbeck:2017ntv,Ilgenfritz:2006he,Sternbeck:2006rd,Kizilersu:2006et,Skullerud:2003qu,Boucaud:2017obn,Athenodorou:2016oyh,Cucchieri:2008qm,Cucchieri:2006tf} 
or resort to continuum methods such as
Schwinger-Dyson equations (SDEs)~\cite{Schleifenbaum:2004id,Huber:2012kd,Aguilar:2013xqa,Huber:2012zj,Blum:2014gna,Eichmann:2014xya,Williams:2015cvx, Binosi:2016wcx,Eichmann:2015nra,Eichmann:2008ef} or the functional renormalization group~\cite{Corell:2018yil,Cyrol:2017ewj,Cyrol:2016tym}.
Within these latter formalisms, the dynamical equations governing the momentum evolution of the vertices
are derived and solved, under a variety of simplifying assumptions 
that reduce the inherent complexity of these calculations.  

In a series of recent works~\cite{Aguilar:2010cn,Aguilar:2014lha,Aguilar:2016lbe, Aguilar:2018epe,Rojas:2013tza,Oliveira:2018fkj}, the aforementioned approaches have been complemented by 
an alternative procedure, which exploits the Slavnov-Taylor identities (STIs) 
satisfied by a given vertex, and constitutes a
modern version of the so-called ``gauge technique''~\cite{Salam:1963sa,Salam:1964zk,Delbourgo:1977jc,Delbourgo:1977hq}.
The main upshot of this method is to determine 
the non-transverse part of the vertex\footnote{This part is  
usually referred to as ``longitudinal'', or ``gauge'', or ``STI-saturating''.}, in terms
of the quantities that enter in the STIs, such as two-point functions and the
so-called ``ghost scattering kernels''.
These kernels correspond to the
Fourier transforms of composite operators, where a ghost field and a quark or a gluon
are defined at the same space-time point.
In the case of the quark-gluon vertex considered in the recent literature,
the quantity in question is the ``ghost-quark'' kernel; its 
form factors have been reconstructed from the corresponding SDE in~\cite{Aguilar:2016lbe,Aguilar:2018epe}, 
and certain special kinematic configurations have been computed in~\cite{Aguilar:2010cn,Rojas:2013tza}. 

In the present work we turn our attention to the ghost-gluon kernel, to be
denoted by $H_{\nu\mu}^{abc}(q,p,r)= -gf^{abc}H_{\nu\mu}(q,p,r)$. 
The main objective is to compute from an appropriate SDE [see Fig.~\ref{fig:Hfull}]
the five form factors comprising this quantity,
to be denoted by $A_i(q,p,r)$ $(i=1,...,5)$, for arbitrary Euclidean values of
the momenta.

The interest in $H_{\nu\mu}$ and its form factors is mainly related with the
two fundamental Yang-Mills vertices shown in Fig.~\ref{fig:ggvertex}~\cite{Marciano:1977su}.
First, as was shown in the classic work of Ball and Chiu (BC)~\cite{Ball:1980ax},
the ``longitudinal'' part of the three-gluon vertex, $\fatg_{\alpha\mu\nu}(q,r,p)$,
may be fully reconstructed from the set of  STIs that it satisfies [see \1eq{eq:sti_delta}].
The ingredients entering in the BC ``solution'' are the gluon propagator, the ghost dressing function, and three of the  
form factors of $H_{\nu\mu}(q,p,r)$. Thus, in order to obtain reliable 
information on the infrared behaviour of $\fatg_{\alpha\mu\nu}(q,r,p)$ by means of this method,
the nonperturbative structure of the ghost-gluon kernel must be firmly established.
Second, by virtue of an exact relation [see \1eq{eq:H_and_Gamma}],
the ghost-gluon vertex, $\Gamma_{\mu}(q,p,r)$, which constitutes an important ingredient
for a variety of SDE studies, is completely determined from the contraction of $H_{\nu\mu}(q,p,r)$
by $q^{\nu}$. Thus, knowledge of the $A_i(q,p,r)$ furnishes {\it both} form factors of $\Gamma_{\mu}(q,p,r)$~\cite{Aguilar:2009nf}.

The methodology used for the computation of the $A_i(q,p,r)$ may be described as follows.
The diagrammatic definition of $H_{\nu\mu}(q,p,r)$ shown in Fig.~\ref{fig:Hfull} involves the connected kernel 
$A^{\mu} A^{\rho} {\bar c} c$ (grey ellipse), whose skeleton expansion will be approximated by the
``one-loop dressed'' diagrams, depicted in Fig.~\ref{fig:H_truncated}; the basic quantities entering at this level 
are the gluon and ghost propagators, and the fully dressed vertices $\fatg_{\alpha\mu\nu}(q,r,p)$ and $\Gamma_{\mu}$.
The individual form factors of $H_{\nu\mu}$ may then be isolated from the resulting
equations by means of an appropriate set of projection operators.
In the final numerical treatment we use the results of large-volume lattice simulations as input for the  propagators, 
while for the vertices we resort to certain simplified Ans\"atze.

We next list the main highlights of our analysis: 
({\it i}) we determine the form factors $A_i$ for 
general values of the Euclidean momenta, presenting the results 
in 3-D plots, where $q^2$ and $p^2$ will be varied, for fixed values of the angle $\theta$ between them;
({\it ii})
the nonperturbative results obtained for $A_i$ are compared 
with their one-loop counterparts  in three special kinematic limits;
({\it iii})
with the help of a constraint imposed by the STI [see Eqs.~\eqref{eq:constraint_ratio}
and~\eqref{eq:constraint1}], we quantify  the  accuracy and veracity of our truncation scheme;
({\it iv}) 
as a direct application, the various $A_i$  are fed into the Euclidean version of Eq.~\eqref{eq:Bi_from_Ai},
giving rise to  both form factors of  the ghost-gluon vertex, for arbitrary  momenta.

The article is organized as follows. In  section~\ref{sec:theory}
we introduce the notation and set up the relevant theoretical framework.
In section~\ref{sec:H_sde}, we 
discuss the truncation scheme employed and we define the set of projectors necessary
for the derivation of the dynamical equations governing the form factors  $A_i$.
In section~\ref{inputs} we present the inputs 
and the additional approximations necessary for the numerical calculation of the $A_i$. 
Then, in section~\ref{Ai_res} we present the numerical solution of the $A_i$ for general Euclidean momenta,
and compare them with the one-loop results for some special kinematic limits.
Next, in section~\ref{nun_const} we discuss how the constraint imposed by the STI may help us
optimize the quality of the inputs used for the computation of the $A_i$.
In section~\ref{sec:ggvertex} we 
construct the two form factors of the ghost-gluon vertex, carry out a comparison 
with the results of various approaches in the literature,
and study their impact on the SDE of the ghost propagator.
In section~\ref{sec:conc} we present our discussion and conclusions.
Finally, in two Appendices we present the  
one-loop results for the $A_i$ in some special kinematic limits, for both ``massive'' and massless gluons,
and certain lengthy expressions appearing in the derivation of the $A_i$.

\section{\label{sec:theory} Theoretical background}

In this section we introduce the basic concepts and ingredients necessary for
the study of $H_{\nu\mu}$, and elucidate on its connection with the
ghost-gluon and three-gluon vertices.
In addition, we introduce  a particular relation,
 which is a direct consequence of the STI that $H_{\nu\mu}$ satisfies~\cite{Ball:1980ax,Binosi:2011wi}, and 
provides a nontrivial constraint on a combination of its form factors.
We emphasize that throughout this article we work in the {\it Landau gauge}, where the 
gluon propagator $\Delta^{ab}_{\mu\nu}(q)=\delta^{ab}\Delta_{\mu\nu}(q)$ assumes the completely transverse form, 
\begin{align}
\Delta_{\mu\nu}(q) = -i\Delta(q)P_{\mu\nu}(q)\,, \qquad P_{\mu\nu}(q) = g^{\mu\nu} - \frac{q^\mu q^\nu}{q^2}\,.
\end{align}
In the case of a gluon propagator that saturates to a nonvanishing value in the deep infrared,
it is natural to set (Euclidean space)~\cite{Aguilar:2011ux,Binosi:2012sj}~\footnote{
Contrary to the case of the quark propagator, this decomposition is not
  mathematically unique; however, the relevant dynamical equations severely restrict the possible structures~\cite{Aguilar:2014tka}.} 
\be
\label{eq:gluonp}
\Delta^{-1}(q) = q^2J(q) + m^2(q) \,,
\ee
where $q^2 J(q)$ denotes the ``kinetic term'' of the gluon propagator, and
$m^2(q)$ is the dynamical gluon mass~\cite{Cornwall:1981zr,Aguilar:2008xm,Binosi:2012sj,Aguilar:2017dco}. 

\begin{figure}[t]
\begin{center}
\includegraphics[scale=0.6]{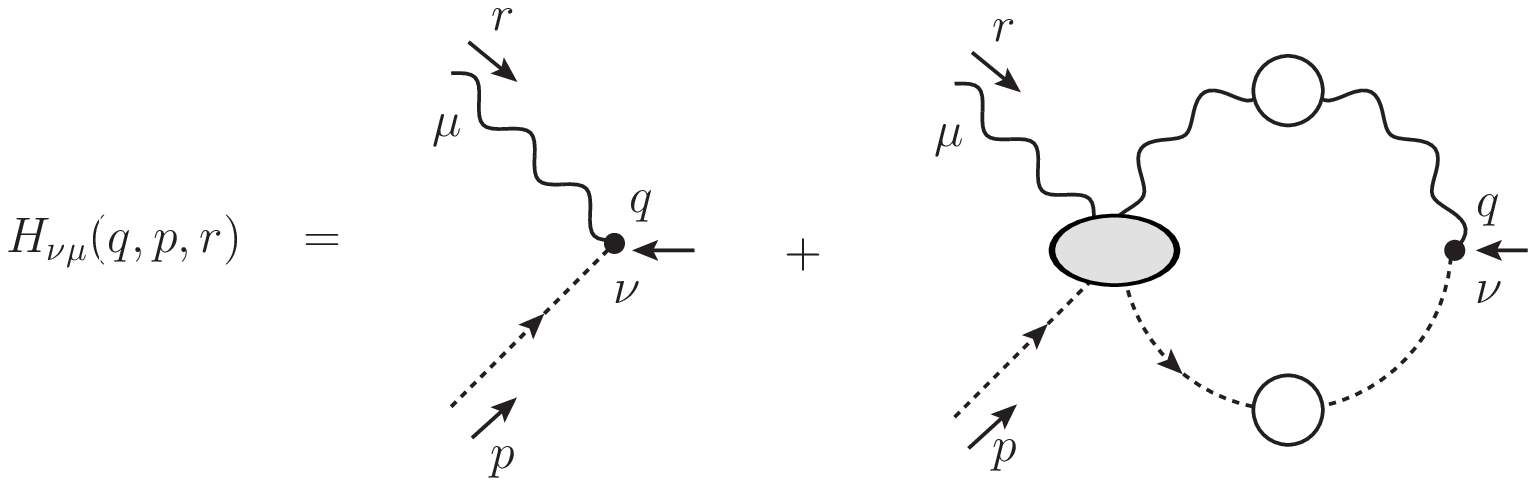}
\end{center}
\caption{The diagrammatic representation of the ghost-gluon scattering kernel. The tree-level contribution is given by $g_{\mu\nu}$. 
}\label{fig:Hfull}
\end{figure}

The ghost-gluon scattering kernel $H^{abc}_{\nu\mu}(q,p,r) = -gf^{abc}H_{\nu\mu}(q,p,r)$ is
diagrammatically depicted in Fig.~\ref{fig:Hfull}.
The most general tensorial decomposition of $H_{\nu\mu}(q,p,r)$ is given by
~\cite{Ball:1980ax,Davydychev:1996pb}
\be\label{eq:H}
H_{\nu\mu}(q,p,r) = A_1 g_{\mu\nu} + A_2 q_\mu q_\nu +  A_3 r_\mu r_\nu  + A_4 q_\mu r_\nu  + A_5 r_\mu q_\nu \,,
\ee
where the momentum dependence, $A_i \equiv A_i(q,p,r)$, has been suppressed for compactness.  Notice that, at tree-level,
$H^{(0)}_{\nu\mu}(q,p,r) = g_{\nu\mu}$, so that the form factors assume the values $A^{(0)}_1 = 1$ and $A^{(0)}_i = 0$, for $i=2,\ldots,5$.

\begin{figure}[!ht]
\begin{center}
\includegraphics[scale=0.6]{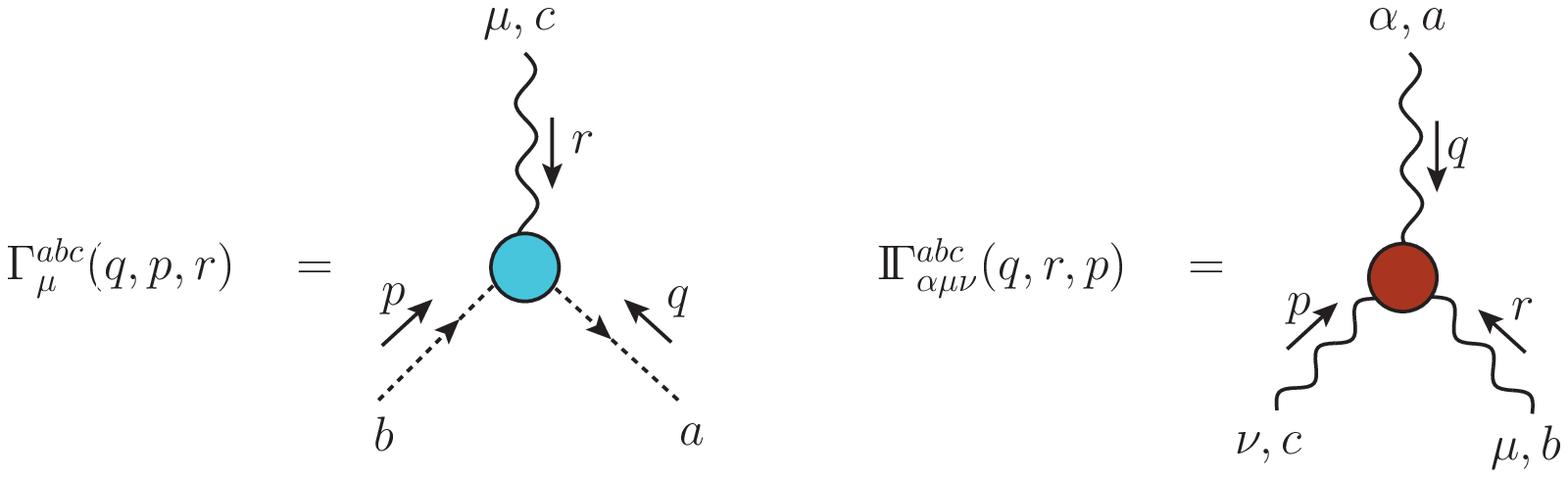}
\end{center}
\caption{Diagrammatic representations of the ghost-gluon (left) and three-gluon (right) vertices, 
and the adopted convention for their momenta dependence.}\label{fig:ggvertex}
\end{figure}

As mentioned in the Introduction, 
our interest in the dynamics of $H_{\nu\mu}$ stems mainly from its connection to two of
the most fundamental Yang-Mills vertices~\cite{Marciano:1977su}, namely
the ghost-gluon vertex, \mbox{$\Gamma^{abc}_{\mu}(q,p,r) =-g f^{abc} \Gamma_{\mu}(q,r,p)$}, 
and the three-gluon vertex, \mbox{$\fatg^{abc}_{\alpha\mu\nu}(q,r,p)=g f^{abc} \fatg_{\alpha\mu\nu}(q,r,p)$},
where $g$  denotes the gauge coupling, and \mbox{$q+r+p=0$}; both vertices are shown diagrammatically in
Fig.~\ref{fig:ggvertex}.

In particular, $H_{\nu\mu}$ and the aforementioned vertices are related by the followings STIs,  
\be
\label{eq:H_and_Gamma}
q^\nu H_{\nu\mu}(q,p,r) = \Gamma_\mu(q,p,r)\,,
\ee
and
\begin{align}
r^\mu\fatg_{\alpha\mu\nu}(q,r,p) =& F(r)[\Delta^{-1}(q) P^{\mu}_\alpha(q)H_{\mu\nu}(q,r,p) - \Delta^{-1}(p)P^{\mu}_\nu(p)H_{\mu\alpha}(p,r,q)] \,, 
\label{eq:sti_delta}
\end{align}
where $F(q)$ stands for the ghost dressing function, which is obtained from the 
ghost propagator, \mbox{$D^{ab}(q) = \delta^{ab}D(q)$}, through 
\be\label{eq:ghost_dressing}
D(q) = \frac{iF(q)}{q^2}\,.
\ee
Evidently, the contraction of $\fatg_{\alpha\mu\nu}(q,r,p)$ with respect to $q^\alpha$ or $p^\nu$
leads to cyclic permutations of the STI in Eq.~\eqref{eq:sti_delta}.

Employing the standard tensorial decomposition of $\Gamma_\mu (q,p,r)$, 
\be\label{Gammamu}
\Gamma_\mu (q,p,r) = q_\mu B_1(q,p,r) + r_\mu B_2(q,p,r)\,,
\ee
where, at tree-level, $B_1^{(0)}=1$ and $B_2^{(0)}=0$, it is straightforward to establish 
from the STI of Eq.~\eqref{eq:H_and_Gamma} that~\cite{Aguilar:2009nf} 
\begin{align}
B_1 &= A_1 + q^2A_2 + (q\cdot r)A_4\,; \nonumber \\
B_2 &= (q\cdot r)A_3 + q^2A_5 \,. 
\label{eq:Bi_from_Ai}
\end{align}
Thus, knowledge of the form factors of $H_{\nu\mu}$ determines fully the corresponding form factors of the ghost-gluon vertex $\Gamma_\mu (q,p,r)$.

On the other hand, the extraction of information
on the structure of $\fatg_{\alpha\mu\nu}(q,r,p)$ from \1eq{eq:sti_delta} (and its permutations)
is significantly more involved, both conceptually and operationally.
Note in particular, that, in the framework composed by the union between   
the Pinch Technique and the Background Field Method (PT-BFM)~\cite{Binosi:2009qm}, the form of 
$\fatg_{\alpha\mu\nu}(q,r,p)$ is intimately connected with the mechanism that is responsible for the
infrared finiteness of the gluon propagator, and especially the form employed in \1eq{eq:gluonp}.
Specifically, the full vertex is composed by two characteristic pieces~\cite{Aguilar:2011ux,Binosi:2012sj} 
\be
\label{eq:Gnp}
\fatg_{\alpha\mu\nu}(q,r,p) =\Gnp_{\alpha\mu\nu}(q,r,p) +\Gp_{\alpha\mu\nu}(q,r,p)\,,
\ee
where the term $\Gp_{\alpha\mu\nu}(q,r,p)$ is very special, in the sense that it contains
``longitudinally coupled'' massless poles, \ie has the general form 
\be
\Gp_{\alpha\mu\nu}(q,r,p) = \left(\frac{q_{\alpha}}{q^2}\right)A_{\mu\nu}(q,r,p) +
\left(\frac{r_{\mu}}{r^2}\right)B_{\alpha\nu}(q,r,p) + \left(\frac{p_{\nu}}{p^2}\right)C_{\alpha\mu}(q,r,p)\,,
\label{longcoupl}
\ee
which trigger the Schwinger mechanism and the subsequent
emergence of a gluonic mass scale~\cite{Aguilar:2015bud}.
Note that, by virtue of \1eq{longcoupl}, $\Gp^{\alpha\mu\nu}(q,r,p)$ satisfies the important projection property
\mbox{$P_{\alpha\alpha^{\prime}}(q)P_{\mu\mu^{\prime}}(r)P_{\nu\nu^{\prime}}(p)\Gp^{\alpha\mu\nu}(q,r,p) = 0$}.

As has been explained in detail in the literature mentioned above, the decompositions of $\Delta^{-1}$ and $\fatg$ 
put forth in \2eqs{eq:gluonp}{eq:Gnp}, respectively, prompt a particular realization
of \1eq{eq:sti_delta}. Specifically, the initial STI is decomposed into two partial STIs, one for $\Gnp_{\alpha\mu\nu}(q,r,p)$
and one for $\Gp_{\alpha\mu\nu}(q,r,p)$, namely (Minkowski space)\footnote{In Minkowski space, $\Delta^{-1}(q) = q^2J(q) - m^2(q)$;
to recover \1eq{eq:gluonp}, set $q^2 \to -q^2_{\s{\rm E}}$ and use 
the transformation conventions of Eq.~\eqref{eq:d_euc}; finally, drop the subscript ``E''.} 
\bea
r^\mu\Gnp_{\alpha\mu\nu}(q,r,p) &=& F(r)[q^2J(q) P^{\mu}_\alpha(q)H_{\mu\nu}(q,r,p) - p^2J(p)P^{\mu}_\nu(p)H_{\mu\alpha}(p,r,q)] \,,
\label{stig}
\\
r^\mu\Gp_{\alpha\mu\nu}(q,r,p) &=& F(r)[m^2(p)P^{\mu}_\nu(p)H_{\mu\alpha}(p,r,q)-m^2(q) P^{\mu}_\alpha(q)H_{\mu\nu}(q,r,p) ] \,.
\label{stiv}
\eea
The correspondence $\Gnp \leftrightarrow q^2 J(q)$ and $\Gp \leftrightarrow m^2(q)$ 
leading to \2eqs{stig}{stiv} is natural, in the sense that, the term $\Gp$
that triggers the generation of the mass saturates, at the same time,  the ``mass-dependent'' part of the STI in \1eq{eq:sti_delta};
however, a comment on its uniqueness is in order (see also footnote 1).
In particular, one may envisage the possibility of defining \mbox{$J^{\prime}(q) = J(q) + f(q)$} and
\mbox{${m^{\prime}}^2(q) = m^2(q) + h(q)$}, such that \mbox{$\Delta^{-1}(q) =q^2J(q) + m^2(q) = q^2 J^{\prime}(q) + {m^{\prime}}^2(q)$}, 
which forces the constraint $h(q)= -q^2 f (q)$. 
The form of $f (q)$, in turn, will be severely constrained 
by the nonlinear SDEs satisfied by $J(q)$ and  $m^2(q)$, in conjunction with additional requirements
such as the positive-definiteness and monotonicity of the final $m^2(q)$. However, in the
absence of a concrete proof stating that $f(q)=0$, the correspondence employed above should be
understood as a physically motivated {\it Ansatz}. 

It turns out that the STI of \1eq{stiv} and its permutations, together with the aforementioned projection property,  
determine completely the form of $\Gp_{\alpha\mu\nu}(q,r,p)$, which has been worked out in~\cite{Ibanez:2012zk}.

$\Gamma_{\alpha\mu\nu}(q,r,p)$ contains the bulk of the nonperturbative corrections not related to the
poles, and is precisely the part that survives when the ``transversely projected'' vertex 
$P_{\alpha\alpha^{\prime}}(q)P_{\mu\mu^{\prime}}(r)P_{\nu\nu^{\prime}}(p)\fatg^{\alpha\mu\nu}(q,r,p)$ is considered. 
The  STI in \1eq{stig}, together with its two cyclic permutations, permits the reconstruction 
of its ``longitudinal'' form factors, through the application of the procedure described in~\cite{Ball:1980ax}.
In practice, the complete construction of the BC ``solution''  
depends not only on the infrared behaviour of $J$ and $F$, discussed in section~\ref{inputs},
but also on the details of $A_1$, $A_3$, and $A_4$, which are largely unexplored,
and are the focal point of the present study.


Quite interestingly, the BC construction for the longitudinal part of $\Gamma_{\alpha\mu\nu}$  hinges on the validity of 
a special relation between $A_1$, $A_3$, and $A_4$,  
which in the original work of~\cite{Ball:1980ax} was shown to
hold at the one-loop level (in the Feynman gauge).
Subsequently, this relation was derived from the fundamental STI that $H_{\nu\mu}$ satisfies when contracted by the
momentum of the incoming gluon~\cite{Binosi:2011wi}, 
and is therefore exact both perturbatively to all orders as well as nonperturbatively.
The relation in question may be expressed in terms of the ratio 
\be
\mathcal{R}(q^2,p^2,r^2) := \frac{F(r)[A_1(q,r,p) + p^2A_3(q,r,p) + (q\cdot p)A_4(q,r,p)]}{F(p)[A_1(q,p,r) + r^2A_3(q,p,r) + (q\cdot r)A_4(q,p,r)]} \,,
\label{eq:constraint_ratio}
\ee
and states simply that, by virtue of the aforementioned STI,  
one must have\footnote{An approximate version of this identity was first derived in~\cite{vonSmekal:1997ern} and further analyzed in~\cite{Watson:1999ha}.}  
\be
\mathcal{R}(q^2,p^2,r^2) = 1 \,, 
\label{eq:constraint1}
\ee
for any value of $q$, $r$, and $p$.

As we will see in sections~\ref{inputs} and~\ref{nun_const}, the constraint of 
Eq.~\eqref{eq:constraint1} is particularly useful for optimizing the form of the
ingredients entering into the computation of the $A_i$, and 
for quantifying the veracity of the truncations and approximations employed.

\section{\label{sec:H_sde} Ghost-gluon kernel at the one-loop dressed level}

\begin{figure}[!ht]
 	\includegraphics[scale=0.5]{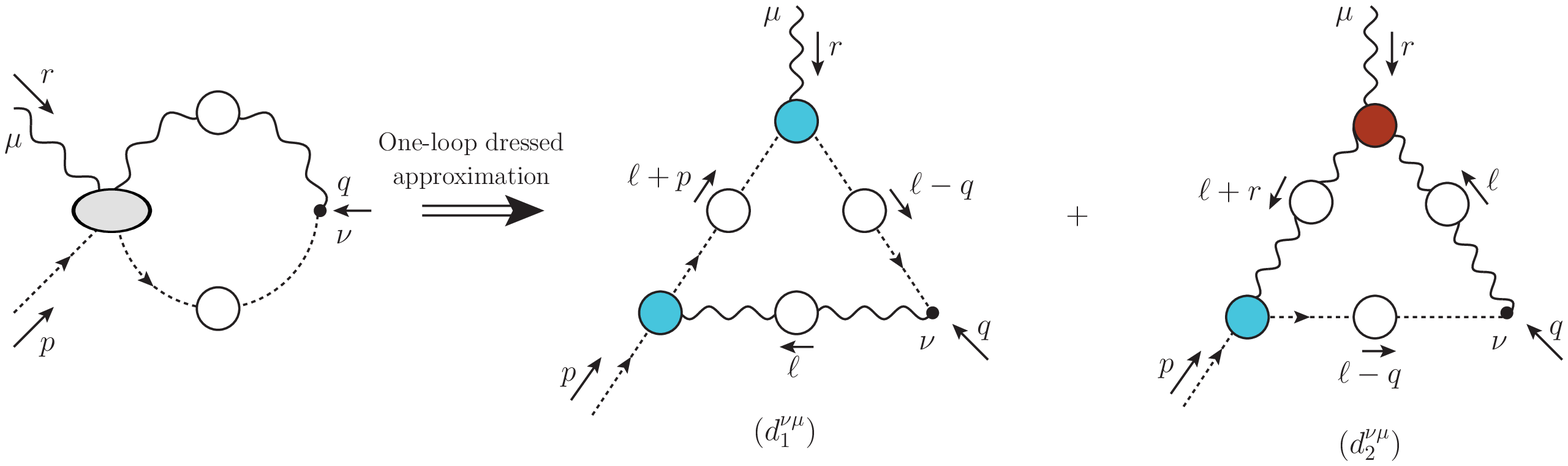}
 	\caption{One-loop dressed approximation of the SDE governing the ghost-gluon scattering kernel.}
        \label{fig:H_truncated}
\end{figure}

In this section we derive the expressions for the form factors $A_i$
within the one-loop dressed approximation. In particular, the four point ghost-gluon scattering amplitude, entering in the diagrammatic definition
of $H_{\nu\mu}(q,p,r)$ in 
Fig.~\ref{fig:Hfull}, is approximated by its lowest
order contributions, including the one-gluon and one-ghost exchange terms, which are subsequently  ``dressed'' as shown in Fig.~\ref{fig:H_truncated}. 
%

Note that the terms kept in the above truncation correspond to the {\it  one-particle reducible} part of the connected kernel $A^{\mu} A^{\rho} {\bar c} c$ (grey ellipse),
while the omitted terms comprise the {\it one-particle irreducible} two-ghost-two-gluon Green's function, whose lowest-order diagrammatic representation may be
found in Fig.~6 of~\cite{Huber:2018ned}\footnote{ We emphasize that all graphs of that figure are one-loop dressed; therefore, 
if inserted into the fundamental diagram [first one in Fig.~\ref{fig:H_truncated}], they would give rise 
to two-loop dressed contributions for $H_{\nu\mu}(q,p,r)$.}.
As was shown in a recent study~\cite{Huber:2017txg}, the inclusion of this subset of corrections    
into the SDE analysis for $\Gamma_\mu$ provides a small contribution of the order of $2\%$. Therefore, given that $\Gamma_\mu$ and $H_{\nu\mu}$ 
are intimately connected by \1eq{eq:H_and_Gamma}, it is reasonable to expect that the truncation implemented in this work will capture faithfully the main bulk of the result. 

Thus, the approximate  version of the SDE that we employ reads
\begin{equation}\label{eq:H_truncated}
H_{\nu\mu}(q,p,r) = g_{\nu\mu} + (d_1)_{\nu\mu} + (d_2)_{\nu\mu} \,,
\end{equation}
with 
\begin{eqnarray}
\label{eq:d1d2}
  (d_1)_{\nu\mu} &=& \frac{1}{2}C_\mathrm{A}g^2\, p_\rho \int_{\ell} \Delta^{\rho}_\nu(\ell)D(\ell + p) D(\ell - q)
  \Gamma_\mu(q - \ell, \ell + p, r)B_1( - \ell - p, p, \ell)\,,  \\
  (d_2)_{\nu\mu} &=& \frac{1}{2}C_\mathrm{A}g^2\,  p_\rho \int_{\ell}\Delta^{\beta}_\nu(\ell)\Delta^{\alpha\rho}(\ell + r)D(\ell - q)
 \Gamma_{\mu\alpha\beta}(r, - \ell - r, \ell)B_1(q - \ell, p, \ell + r)\nonumber  \,, 
\end{eqnarray}
where $C_\mathrm{A}$ is the eigenvalue of the Casimir operator in the adjoint representation, 
and  we have defined the
integration measure
\be
\int_{\ell}\equiv\!\int\!\frac{\mathrm{d}^4 \ell}{(2\pi)^{4}}\,. 
\label{dqd}
\ee
Note that in arriving at Eq.~\eqref{eq:d1d2} we have exploited 
the full transversality of the gluon propagator in the Landau gauge in order to eliminate the $B_2$ form factors of two of the ghost-gluon vertices. 

It is obvious from Eq.~\eqref{eq:d1d2}
that in the soft ghost limit, {\it i.e.} $p\to 0$, the 
one-loop dressed corrections vanish, {\it i.e.} $H_{\nu\mu}(q,p,r)=g_{\nu\mu}$. 
This result is valid to all orders, independently of  the truncation scheme adopted
(see, {\it e.g.},  Eqs.~(6.24) and (6.25) of~\cite{Ibanez:2012zk}), being a plain manifestation  
of Taylor's theorem~\cite{Taylor:1971ff}.


The renormalization of \1eq{eq:H_truncated} proceeds through the replacements~\cite{Aguilar:2013xqa} 
\bea
\Delta_{\s R}(q^2)&=& Z^{-1}_{A} \Delta(q^2),\nonumber\\
F_{\s R}(q^2)&=& Z^{-1}_{c} F(q^2),\nonumber\\ 
\Gamma^{\mu}_{\s R}(q,p,r) &=& Z_1 \Gamma^{\mu}(q,p,r),\nonumber\\ 
\Gamma^{\mu\alpha\beta}_{\s R}(q,r,p) &=&  Z_3 \Gamma^{\mu\alpha\beta}(q,r,p),\nonumber\\ 
g_{\s R} &=& Z_g^{-1} g = Z_1^{-1} Z_A^{1/2} Z_c\, g  = Z_3^{-1} Z_A^{3/2} \, g\,,
\label{renconst}
\eea
where  $Z_{A}$, $Z_{c}$, $Z_{1}$, $Z_{3}$, and $Z_g$ are the corresponding renormalization constants.
Within the momentum subtraction (MOM) scheme that we employ,
propagators assume their tree-level values at the subtraction point  $\mu$, while 
an analogous condition is imposed on the vertices, usually implemented at a common value
of all their momenta (``symmetric'' point).

A well-known consequence of Eq.~\eqref{eq:H_and_Gamma} is that $H^{\nu\mu}$ 
renormalizes as $\Gamma^{\mu}$, namely \mbox{$H^{\nu\mu}_{\s R} = Z_1 H^{\nu\mu}$}. The (multiplicative) renormalization of Eq.~\eqref{eq:H_truncated} proceeds in the standard way, by
replacing the unrenormalized quantities by renormalized ones, using 
 the relations given in Eq.~\eqref{renconst}. Then, it is straightforward to show that the 
integrands of $(d_1)_{\nu\mu}$ and $(d_2)_{\nu\mu}$ can be
written exclusively in terms of the standard renormalization-group invariant quantities formed 
by $g \Gamma_{\mu} \Delta^{1/2} D$ and $g \Gamma^{\nu\sigma\alpha} \Delta^{3/2}$; therefore  
both terms maintain their original form after renormalization. Thus,  
the renormalized version of Eq.~\eqref{eq:H_truncated} reads 
\begin{equation}\label{eq:H_truncatedren}
H_{\s R}^{\nu\mu}(q,p,r) = Z_1 \left[ g^{\nu\mu} +  (d_1)_{\s R}^{\nu\mu} + (d_2)_{\s R}^{\nu\mu} \right] \,,
\end{equation}
where the $Z_1$ originates from the renormalization of the $H^{\nu\mu}(q,p,r)$ on the l.h.s.
The subscript ``R'' will be subsequently suppressed to avoid notation clutter. 

In what follows we will set $Z_1=1$. This particular choice is exact in the case of the soft ghost limit, 
being strictly enforced by the validity of Taylor's theorem~\cite{Taylor:1971ff}.
For any other MOM-related prescription,
$Z_1$ deviates only slightly (a few percent) from unity,
for the subtraction point $\mu=4.3$ GeV that we employ. For example,
as we have explicitly confirmed from our results, in the case where the MOM prescription
is imposed at the symmetric point ($p^2=r^2=q^2=\mu^2$), instead of the exact $A_1(\mu) = 1$ we have 
$A_1(\mu) = 1.03$.

The relation between $H_{\nu\mu}$ and $\Gamma_{\mu}$,
given by Eq.~\eqref{eq:H_and_Gamma}, prompts a final adjustment, which permits us to preserve  
the ghost--anti-ghost symmetry at the level of the approximate SDE that we consider\footnote{This
  special symmetry of the ghost-gluon vertex is valid {\it only}
  in the Landau gauge~\cite{Alkofer:2000wg}.}. Specifically, the form factor
$B_1(q,p,r)$ of the ghost-gluon vertex is symmetric under the exchange of the ghost
and anti-ghost momenta, $p$ and $q$, respectively.
However, the truncated SDE of Fig.~\ref{fig:H_truncated} does not respect 
this special symmetry, because the vertex where the ghost leg is entering is ``dressed'' while that of the anti-ghost is bare.
A simple expedient for restoring this 
property is to ``average'' the SDEs  dressed on either leg~\cite{Blum:2014gna,Cyrol:2014kca,Eichmann:2014xya}, which amounts to substituting into \1eq{eq:d1d2}
\begin{align}
B_1(-\ell-p,p,\ell) \to {\mathcal V}_1(\ell,q,p,r) =& \frac{1}{2}\left[ B_1(-\ell-p,p,\ell) + B_1(q,\ell-q,-\ell) \right]\,, \nonumber \\
B_1(q-\ell,p,\ell+r) \to {\mathcal V}_2(\ell,q,p,r) =& \frac{1}{2}\left[ B_1(q-\ell,p,\ell+r) + B_1(q,\ell-q,-\ell) \right]\,.
\label{eq:sym_sub}
\end{align}

In general, the individual $A_i$ may be projected  out from $H_{\nu\mu}(q,p,r)$ by means of
a set of suitable projectors, $\mathcal{T}^{\mu\nu}_{\!i}(q,r)$. In particular, 
\be
\label{eq:Ai-proj-gen}
A_i(q,p,r)=\frac{\mathcal{T}^{\mu\nu}_{\!i}(q,r) H_{\nu\mu}(q,p,r)}{2h^2(q,r)}\,,
\ee
where
\begin{align}
\mathcal{T}^{\mu\nu}_{\!1}(q,r) =& h(q,r)\left[ h(q,r)g^{\mu\nu}+ h^{\mu\nu}(q,r)\right] \,,\nonumber\\
\mathcal{T}^{\mu\nu}_{\!2}(q,r) =& -h(q,r)r^2g^{\mu\nu} -2 h(q,r)r^\mu r^\nu -3r^2 h^{\mu\nu}(q,r) \,,\nonumber\\
\mathcal{T}^{\mu\nu}_{\!3}(q,r) =& \mathcal{T}^{\mu\nu}_{\!2}(r,q) \,,\nonumber\\
\mathcal{T}^{\mu\nu}_{\!4}(q,r) =& h(q,r)(r\cdot q)g^{\mu\nu} +2h(q,r)q^{\mu}r^{\nu} +3(r\cdot q)
h^{\mu\nu}(q,r) \,,
\nonumber\\
\mathcal{T}^{\mu\nu}_{\!5}(q,r) =& \mathcal{T}^{\mu\nu}_{\!4}(r,q) \,,
\label{project}
\end{align}
and
\begin{align}
h(q,r) =& q^2r^2 - (q\cdot r)^2 \,,\nonumber\\
h^{\mu\nu}(q,r) =& (q\cdot r)\left[q^\mu r^\nu+q^\nu r^{\mu}\right] -r^2q^{\mu}q^{\nu} - q^2r^{\mu}r^{\nu} \,.
\end{align}
Clearly, since in the present work $H_{\nu\mu}(q,p,r)$ is approximated by \1eq{eq:H_truncated}, the corresponding form factors
will be obtained through 
\be
\label{eq:Ai_proj}
A_i(q,p,r)=\frac{\mathcal{T}^{\mu\nu}_{\!i}(q,r) \left[ g_{\nu\mu} +  (d_1)_{\nu\mu} + (d_2)_{\nu\mu} \right]}{2h^2(q,r)}\,.
\ee

The implementation of the above projections may be 
carried out using an algebraic manipulation program, such as the Mathematica \mbox{Package-X}~\cite{Patel:2015tea,Patel:2016fam};
the rather lengthy expressions produced from these projections are presented in Appendix~\ref{app:projections}.

\section{\label{inputs}Inputs and approximations}

For the evaluation of \1eq{eq:d1d2} we need the
following ingredients: ({\it i}) the gluon propagator $\Delta(q)$ and its ``kinetic'' term $J(q)$,
({\it ii}) the ghost dressing function $F(q)$, ({\it iii}) the three-gluon vertex, entering in $(d_2)_{\nu\mu}$, ({\it iv})
the ghost-gluon vertex, entering in both $(d_1)_{\nu\mu}$ and $(d_2)_{\nu\mu}$, and 
({\it v}) the value of the strong coupling $\alpha_s\equiv g^2/4\pi$ at the renormalization scale $\mu$.
The corresponding input quantities will be denoted by
$\Delta_{\inpt}(q)$, $J_{\inpt}(q)$, $F_{\inpt}(q)$, $\Gamma^{\inpt}_{\mu\alpha\beta}$, and $B_1^{\inpt}(Q)$, respectively. 
It is important to comment already at this point on a characteristic  feature shared by 
inputs ({\it i})-({\it iv}), which is implemented in order for the resulting $A_i$ to satisfy 
\1eq{eq:constraint1} as accurately as possible.
In particular, in the deep ultraviolet all aforementioned quantities will be
forced to tend to their tree-level values, \ie their one-loop perturbative corrections (logarithms and/or constants) will be suppressed. This, in turn, will
guarantee that, for large values of the momenta,
the emerging $A_i$ will correctly capture their one-loop perturbative
behavior [see also discussion in section~\ref{nun_const}].
In what follows we briefly review how the above input quantities  are obtained.

({\it i}) and ({\it ii}):  As was done in a series of previous works~\cite{Aguilar:2010cn,Aguilar:2011ux,Aguilar:2016lbe},
for $\Delta_{\inpt}(q)$ and $F_{\inpt}(q)$ we employ
fits to the numerical solutions of the corresponding SDEs,
which are in excellent agreement with the {\it quenched} $\rm SU(3)$ lattice data of~\cite{Bogolubsky:2007ud}, subject
to the particular ultraviolet adjustments mentioned above.
Below we consider the individual cases  ({\it i}) and ({\it ii})
separately.

\begin{figure}[t]
\begin{minipage}[b]{0.45\linewidth}
\centering
\includegraphics[scale=0.32]{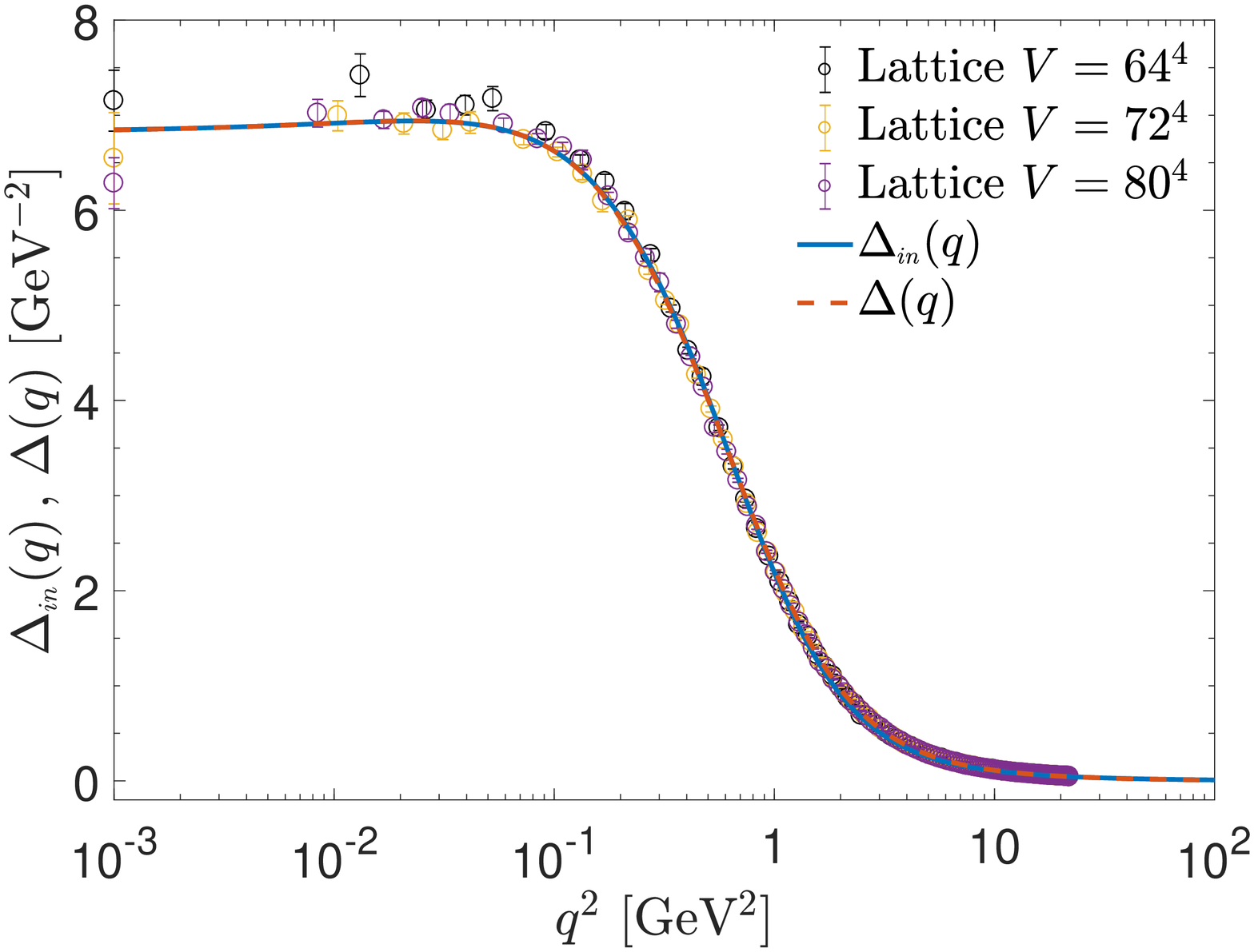}
\end{minipage}
\hspace{0.25cm}
\begin{minipage}[b]{0.45\linewidth}
\includegraphics[scale=0.32]{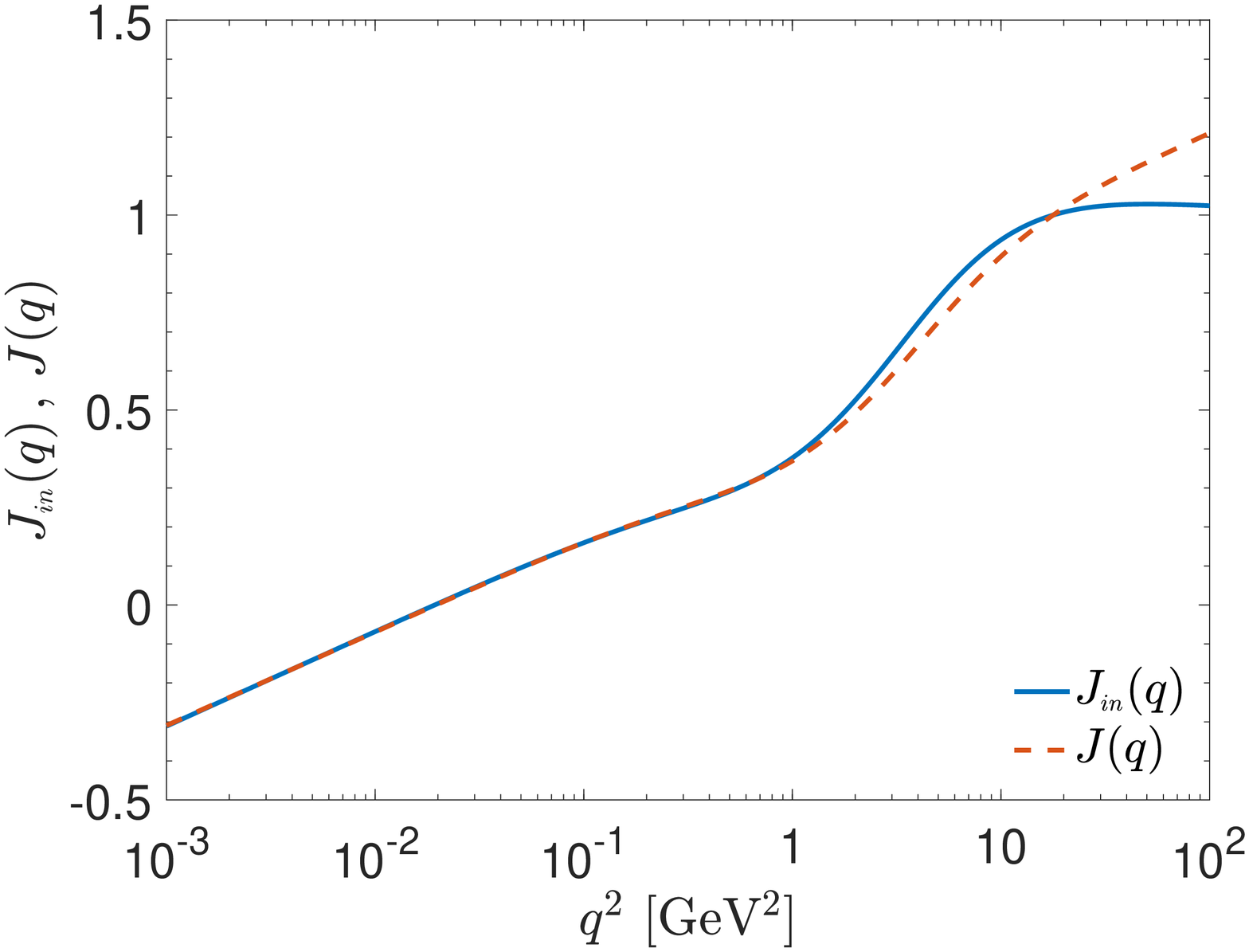}
\end{minipage}
\caption{The fits for $\Delta_{\inpt}(q)$ (left panel)
and  $J_{\inpt}(q)$ (right panel) given by Eqs.~\eqref{eq:gluon_m_J_euc}  and
~\eqref{eq:J_tuv}, respectively (blue continuous curves). The fits
for $\Delta(q)$ and $J(q)$  follow the 
same functional dependence but with $\tau_1/(q^2 + \tau_2)\to 1 $. The lattice data is from Ref.~\cite{Bogolubsky:2007ud}.}
\label{fig:props}
\end{figure}

({\it i}): The fit for $\Delta_{\inpt}(q)$ (in Euclidean space) is given by~\cite{Aguilar:2015bud}  
\be\label{eq:gluon_m_J_euc}
\Delta^{-1}_{\inpt}(q) = q^2J_{\inpt}(q) + m^2(q) \,,
\ee
where the kinetic term has the form 
\be\label{eq:J_tuv}
J_{\inpt}(q) = 1 + \frac{C_\mathrm{A}\alpha_s}{4\pi}\left( \frac{\tau_1}{q^2 + \tau_2} \right)\left[ 2\ln\left(  \frac{q^2 + \rho_l m^2(q)}{\mu^2} \right) + \frac{1}{6}\ln\left( \frac{q^2}{\mu^2} \right) \right] \,,
\ee
while the effective gluon mass $m^2(q)$ obeys a power-law running\footnote{The solutions for $m^2(q)$ found in~\cite{Aguilar:2017dco}
deviate slightly from the exact power law running, in compliance with the 
operator product expansion (see also~\cite{Aguilar:2007ie}). In particular, $m^2(q)=m_0^2/[1+(q^2/\rho_m^2)^{1+\gamma}]$, with $\gamma$ ranging between $0.1$ and $0.3$,
depending on a number of subtle assumptions and approximations. Here we use for simplicity the case $\gamma=0$; the dependence of our
results on variations of $\gamma$ (within the aforementioned range of values) is negligible.}~\cite{Aguilar:2017dco}, 
\be\label{eq:dyn_mass}
m^2(q) = \frac{m_0^2}{1 +  q^2/\rho_m^2} \,,
\ee
with the adjustable parameters given by \mbox{$\tau_1 = 12.68$}, \mbox{$\tau_2 = 1.05 \text{ GeV}^2$}, \mbox{$m_0^2 = 0.15 \text{ GeV}^2$}, \mbox{$\rho_m^2 = 1.18 \text{ GeV}^2$} and \mbox{$\rho_l = 102.3$}. 
On the left panel of Fig.~\ref{fig:props} we
show the lattice data for $\Delta(q)$ (circles)~\cite{Bogolubsky:2007ud}, together
with the corresponding fit (blue continuous curve) given by the combination of \3eqs{eq:gluon_m_J_euc}{eq:J_tuv}{eq:dyn_mass}.

On the right panel of Fig.~\ref{fig:props} we
present the $J_{\inpt}(q)$ of \1eq{eq:J_tuv}; the reason for displaying it in isolation is that it 
constitutes the main ingredient in the approximation implemented for the three-gluon vertex in item ({\it iii}), 
see \2eqs{eq:three_gluon_s}{Xmin}. 
Notice that the $J_{\inpt}(q)$ contains both massive and massless logarithms, which are crucial
for triggering three characteristic features, namely its suppression with respect to its tree-level value ($J^{(0)}(q) = 1$)
for a wide range of physically relevant momenta,   
the reversal of its sign (\emph{zero-crossing}), and its logarithmic divergence at the origin~\cite{Aguilar:2013vaa, Athenodorou:2016oyh}. These features, in turn, will be inherited by the components of the three-gluon vertex
constructed in ({\it iii}). Even though $J_{\inpt}(q)$ contains these logarithms, for large $q^2$ it tends to 1, in compliance with
the requirement discussed above, due to the inclusion of the function $\tau_1/(q^2 + \tau_2)$; note that 
this function becomes $1$ in the  
``{\it bona-fide}''  fit for $J(q)$, which is also displayed in Fig.~\ref{fig:props}, for direct comparison.

\begin{figure}[!ht]
\begin{minipage}[b]{0.45\linewidth}
\centering
\includegraphics[scale=0.32]{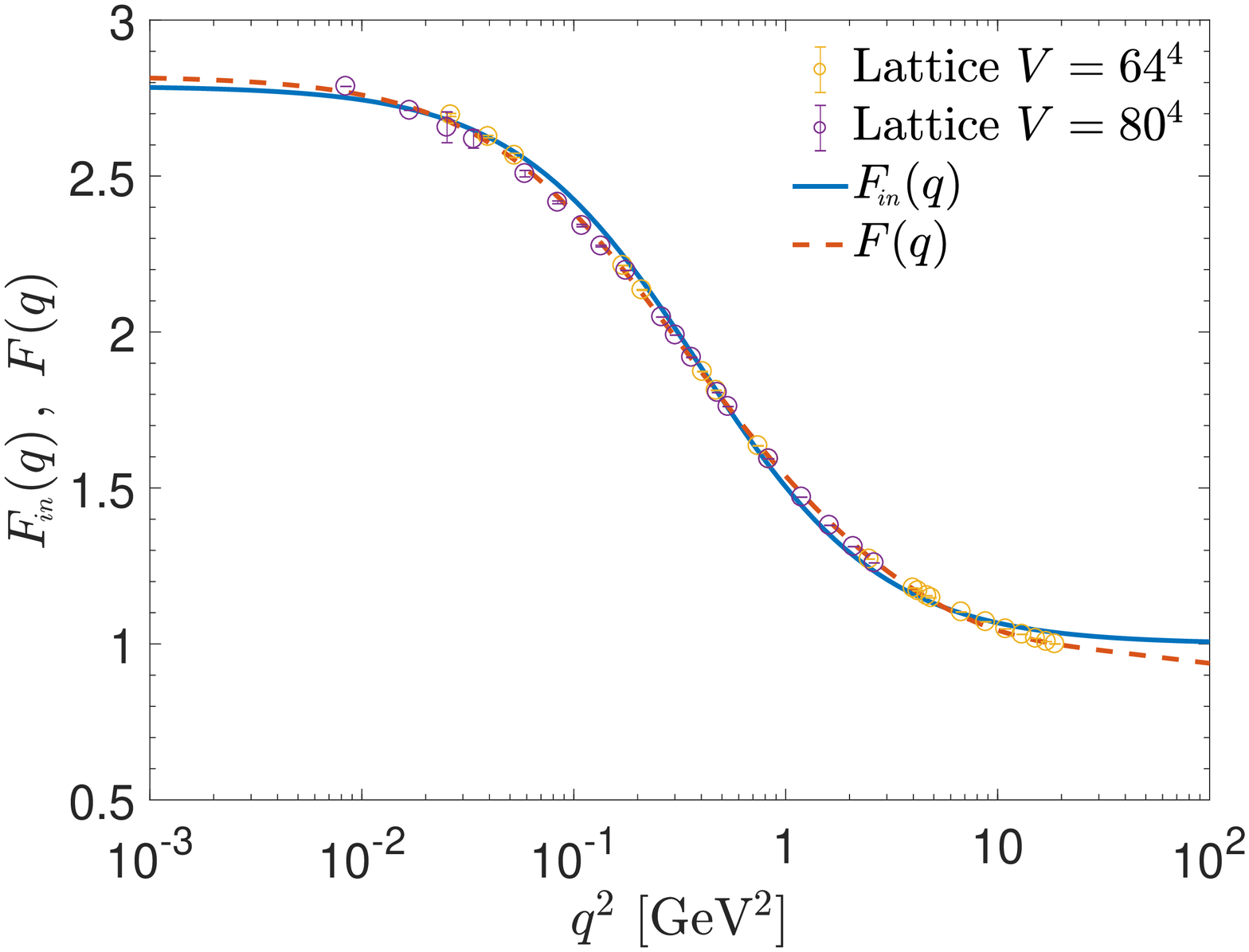}
\end{minipage}
\hspace{0.25cm}
\begin{minipage}[b]{0.45\linewidth}
\includegraphics[scale=0.32]{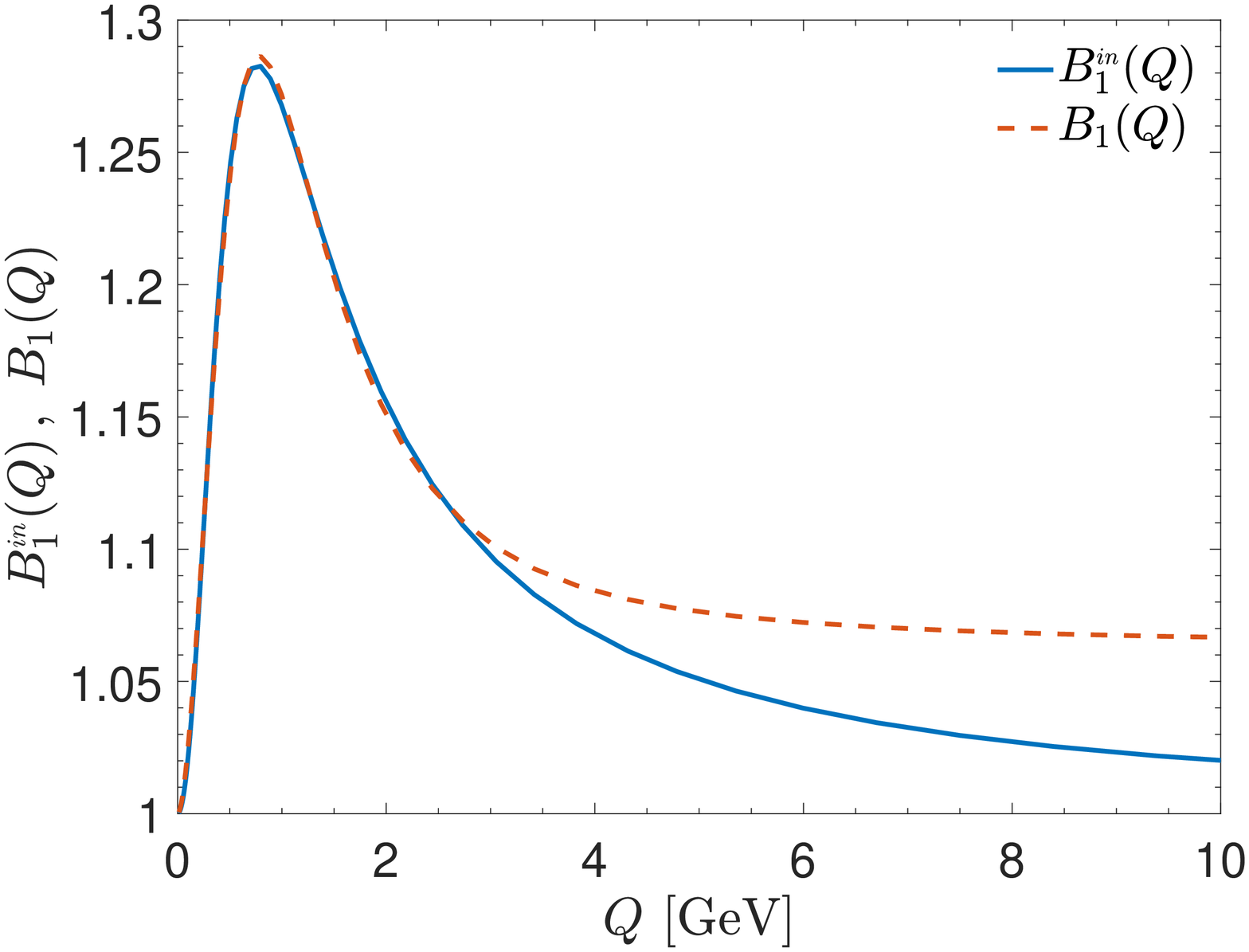}
\end{minipage}
\caption{Left panel: Fits for $F(q)$ without ultraviolet logarithms (blue continuous), corresponding to Eq.~\eqref{eq:ghost_TUV}, and with ultraviolet logarithms (red dashed), given by Eq.~\eqref{eq:F_UVlogs}, compared to the lattice data from~\cite{Bogolubsky:2007ud}. Right panel: The form factor $B^{\inpt}_1(Q)$ given by Eq.~\eqref{eq:b1_fit} (blue continuous), and its counterpart  $B_1(Q)$  with the one-loop correction (red dashed).}
 	\label{fig:b1_sym_fit}
\end{figure}

({\it ii}): The fit for $F_{\inpt}(q)$ is shown on the left panel of Fig.~\ref{fig:b1_sym_fit} (blue continuous line), together with the 
corresponding lattice data; its functional form is given by 
\be
\label{eq:ghost_TUV}
F_{\inpt}(q) = 1 + \frac{\sigma_1}{q^2 + \sigma_2} \,,
\ee
with $\sigma_1 = 0.70 \text{ GeV}^2$ and $\sigma_2 = 0.39 \text{ GeV}^2$.  Again, in the limit 
of large $q^2$, the above expression recovers 
the tree-level result, {\it i.e.} $F_{\inpt}(q)=1$. On that same plot, the red dashed line corresponds to the fit of 
$F(q)$ introduced in \1eq{eq:F_UVlogs}, which corresponds to the typical solution of the SDE
for $F(q)$~\cite{Aguilar:2013xqa}, and, as such,
contains the appropriate perturbative logarithms. 
Evidently, the difference between the two
fits becomes relevant in the deep ultraviolet, where the $F(q)$ of \1eq{eq:F_UVlogs} deviates gradually from unity, 
approaching eventually zero at a logarithmic rate. 

({\it iii}) and ({\it iv}):
The fully dressed vertex $\Gamma_{\mu\alpha\beta}$ and $\Gamma_{\mu}$
enter in \1eq{eq:d1d2} that controls $H_{\nu\mu}$, but, at the same time, 
the determination of their longitudinal parts from \1eq{stig} and \1eq{eq:Bi_from_Ai}
requires the knowledge of $H_{\nu\mu}$, converting the problem into an extended system of coupled equations\footnote{It should be clear that a fully self-consistent treatment
would also require information on the transverse parts of the vertices, which are {\it a-prior}i inaccessible to a gauge-technique based approach.}.
However, given the complexity of such an endeavor, we will employ instead a set of approximations for these two vertices.
We next analyze ({\it iii}) and ({\it iv}) separately.

({\it iii}): Let us first consider the three-gluon vertex, entering in  $(d_2)_{\nu\mu}$, and set \mbox{$t=-(\ell+r)$}.
Our way of approximating $\Gamma_{\mu\alpha\beta}(r,t,\ell)$ is the following.
First, we consider the STIs exemplified by \1eq{stig}, and ``abelianize'' them by turning off the ghost sector, 
\ie setting $F(r)=1$ and $H_{\nu\mu}=g_{\nu\mu}$. The resulting identities may then be ``solved'' following the BC procedure~\cite{Ball:1980ax},
thus furnishing the abelianized longitudinal form factors, \mbox{$X_i$ $(i=1...10)$}, which contain combinations of the function $J(q)$ only. 
Then, the ``input'' $\Gamma_{\mu\alpha\beta}(r,t,\ell)$, to be denoted by $\Gamma^{\inpt}_{\mu\alpha\beta}(r,t,\ell)$,
is chosen to contain only the three tensorial structures that comprise the tree-level vertex, multiplied by the
corresponding form factors, which are related to each other by the Bose symmetry. In particular,  
\begin{align}
\Gamma^{\inpt}_{\mu\alpha\beta}(r,t,\ell)=  (r-t)_\beta g_{\mu\alpha} {X}^{\inpt}_1(r,t,\ell)
+(t-\ell)_\mu g_{\alpha\beta}{X}^{\inpt}_1(t,\ell,r) 
+ (\ell - r)_\alpha g_{\beta\mu}{X}^{\inpt}_1(\ell,r,t) \,,
\label{eq:three_gluon_s}
\end{align}
with 
\be
{X}^{\inpt}_1(r,t,\ell) = \frac{1}{2}[ J_{\inpt}(r) + J_{\inpt}(t) ]\,.
\label{Xmin}
\ee
Notice that at  tree-level ${X}^{\inpt}_1=1$, and Eq.~\eqref{eq:three_gluon_s} reduces indeed to
\begin{align}
  \Gamma^{(0)}_{\mu\alpha\beta}(r,t,\ell)=  (r-t)_\beta g_{\mu\alpha}
+(t-\ell)_\mu g_{\alpha\beta} 
+ (\ell - r)_\alpha g_{\beta\mu} \,.
\label{eq:3gluon}
\end{align}
Thus, the $\Gamma^{\inpt}_{\mu\alpha\beta}$ of \1eq{eq:three_gluon_s} will be 
used as a ``seed'' for obtaining the one-loop dressed approximation for $H_{\nu\mu}$. Note that, in addition to the remaining  seven longitudinal form factors  
that have not been included into $\Gamma^{\inpt}_{\mu\alpha\beta}$ for simplicity,
the uncertainties associated with the omission of all transverse structures must be also kept in mind.

({\it iv}): Turning to the ghost-gluon vertex, as mentioned right after Eq.~\eqref{dqd}, two out of the three vertices have been naturally
replaced by their $B_1$ components, and only the $\Gamma_\mu(q - \ell, \ell + p, r)$ in $(d_1)_{\nu\mu}$ contains both $B_1$ and $B_2$.
In what follows we will set (by hand) $B_2=0$  for this vertex, and retain only $B_1$; thus, at this point,
all ghost-gluon vertices appearing in the problem have been replaced by their $B_1$ form factor.

The approximation used for $B_1(q,p,r)$ is obtained as follows. We start by carrying out 
the first iteration of \1eq{eq:d1d2}, using for $B_1$ its tree-level value.
This furnishes the first approximation for the $A_i(q,p,r)$, which, by means of the first relation in \1eq{eq:Bi_from_Ai},
yields the next approximation for $B_1(q,p,r)$. At this point we isolate from $B_1(q,p,r)$ the ``slice'' that
corresponds to the ``totally symmetric'' configuration
\be
q^2 = p^2= r^2 = Q^2 \,,\,\,\,\,
q\cdot p = q\cdot r = p\cdot r = -\frac{1}{2}Q^2 \,,
\label{symconf}
\ee
shown on the right panel of Fig.~\ref{fig:b1_sym_fit} (red dashed line).
Then, to get $B^{\inpt}_1(Q)$ we adjust the ``tail'' of the curve, such that it reaches the tree-level value 1 for large $Q$; 
the resulting functional form may be fitted by 
\be\label{eq:b1_fit}
B^{\inpt}_1(Q) = 1 + \frac{\tau_1 Q^2}{(1 + \tau_2 Q^2)^\lambda} \,,
\ee
where the  parameters $\tau_1 = 2.21\text{ GeV}^{-2}$, \mbox{$\tau_2 = 2.50 \text{ GeV}^{-2}$} and $\lambda = 1.68$.
Past this point, the iterative procedure described
above is discontinued, and the $B^{\inpt}_1(Q)$ of \1eq{eq:b1_fit} is fixed as the final input in \1eq{eq:d1d2}.

After the above simplification, \1eq{eq:sym_sub} becomes 
\begin{align}
{\mathcal V}_1(\ell,q,p,r) =& B^{\inpt}_1(\ell)\,, \nonumber\\
{\mathcal V}_2(\ell,q,p,r) =& \frac{1}{2}\left[ B^{\inpt}_1(\ell+r) + B^{\inpt}_1(\ell) \right]\,.
\label{eq:vgg}
\end{align}

({\it v}):
Finally, for most of the analysis, the strong charge 
will assume the value \mbox{$\alpha_s  = 0.22$} at the subtraction point \mbox{$\mu=4.3$ GeV},
where all Green's functions are renormalized. The determination of this
particular value is rather convoluted, involving the combination of  
4-loop perturbative results,
nonperturbative information included in the vacuum condensate of dimension two, and the extraction of $\Lambda_{\rm QCD}$ from lattice results of the ghost-gluon vertex in the Taylor kinematics~\cite{Boucaud:2008gn}.  
Given the theoretical uncertainties associated with some of the aforementioned ingredients, we consider the
value \mbox{$\alpha_s  = 0.22$} rather approximate; in fact, as we will see in section~\ref{sec:ggvertex},
the final analysis seems to favor slightly higher values of the charge, of the order of \mbox{$\alpha_s  = 0.25$}.
Note that the difference between using  \mbox{$\alpha_s  = 0.22$} or \mbox{$\alpha_s  = 0.25$}
is practically imperceptible at the level of the 3-D plots presented in the next section; 
however, it becomes visible when particular ``slices'' are isolated [see left panel of Fig.~\ref{fig:ghostSDE}].

\section{\label{Ai_res} Results for the form factors of the ghost-gluon kernel}

In this section we present the results for the five form factors $A_i$.
We will first present 3-D plots in general Euclidean kinematics, and then take a closer look at three
special kinematic limits.

\subsection{\label{3Dplots} 3-D plots}

First, we use the
standard conversion rules to pass  \1eq{eq:d1d2} and its ingredients from Minkowski to Euclidean space~\cite{Aguilar:2016lbe}.
In particular,  $(q^2, p^2, q\cdot p) \to -(q_{\s{\rm E}}^2, p_{\s{\rm E}}^2, q_{\s{\rm E}}\cdot p_{\s{\rm E}})$, and
\begin{align}
&\Delta(q^2) \xrightarrow{q^2\to-q^2_{\s{\rm E}}} - \Delta_{\s{\rm E}}(q^2_{\s{\rm E}})
\,; \quad D(q^2)  \xrightarrow{q^2\to-q^2_{\s{\rm E}}} - D_{\s{\rm E}}(q^2_{\s{\rm E}}) \,; \nonumber\\ 
&B_i(q,p,r) \rightarrow B_i(q_{\s{\rm E}},p_{\s{\rm E}},r_{\s{\rm E}}) \,; \quad A_j(q,p,r) \rightarrow A_j(q_{\s{\rm E}},p_{\s{\rm E}},r_{\s{\rm E}}) \,,
\label{eq:d_euc}
\end{align}
for $i = 1, 2$, $j = 1, \ldots, 5$.

In addition, the measure defined in Eq.~\eqref{dqd} becomes
\be
\int_{\ell} = i\int_{\ell_{\s{\rm E}}}\,,
\ee
which in spherical coordinates is given by
\begin{align}
  \int_{\ell_{\rm E}} =\frac{1}{32\pi^4}\int_{\Lambda_{\rm IR}^2}^{\Lambda_{\rm \s UV}^2}\!\! \mathrm{d}\ell_{\s{\rm E}}^2 \ell_{\s{\rm E}}^2
  \int_0^\pi\! \mathrm{d}\phi_1 \sin^2\phi_1 \int_0^\pi \mathrm{d}{\phi_2}\sin{\phi_2}\int_0^{2\pi} \mathrm{d}{\phi_3}\,.
\label{eucsph}
\end{align}
Note that, for numerical purposes, we have introduced in 
the radial integration a infrared and 
a ultraviolet cutoffs ${\Lambda_{\rm IR}^2}$ and ${\Lambda_{\rm UV}^2}$, respectively;
their numerical values will fix the overall size of our numerical grid, namely  
\mbox{$[ 5\times 10^{-5}\,\mbox{{GeV}}^2, 5\times 10^3\!,\mbox{{GeV}}^2]$} .

A standard choice for the orientation of the Euclidean four-momenta $q$ and  $p$ and the integration momentum $\ell$ is (from now on we suppress the subscript ``$\rm E$'')
\begin{align}
q &= |q|(1,0,0,0)\,; \nonumber \\ 
p &= |p|(\cos{\theta},\sin{\theta},0,0) \,;\nonumber \\ 
\ell &= |\ell|(\cos{\phi_1},\sin{\phi_1}\cos{\phi_2},\sin{\phi_1}\sin{\phi_2}\cos{\phi_3},\sin{\phi_1}\sin{\phi_2}\sin{\phi_3})\,.
\end{align}
Evidently, $q^2 =|q|^2$, $p^2 =|p|^2$, and $q\cdot p = |q||p|\cos{\theta}$.

%
\begin{figure}[!ht]
\vspace{-0.5cm}
\begin{minipage}[b]{0.45\linewidth}
\centering
\includegraphics[scale=0.39]{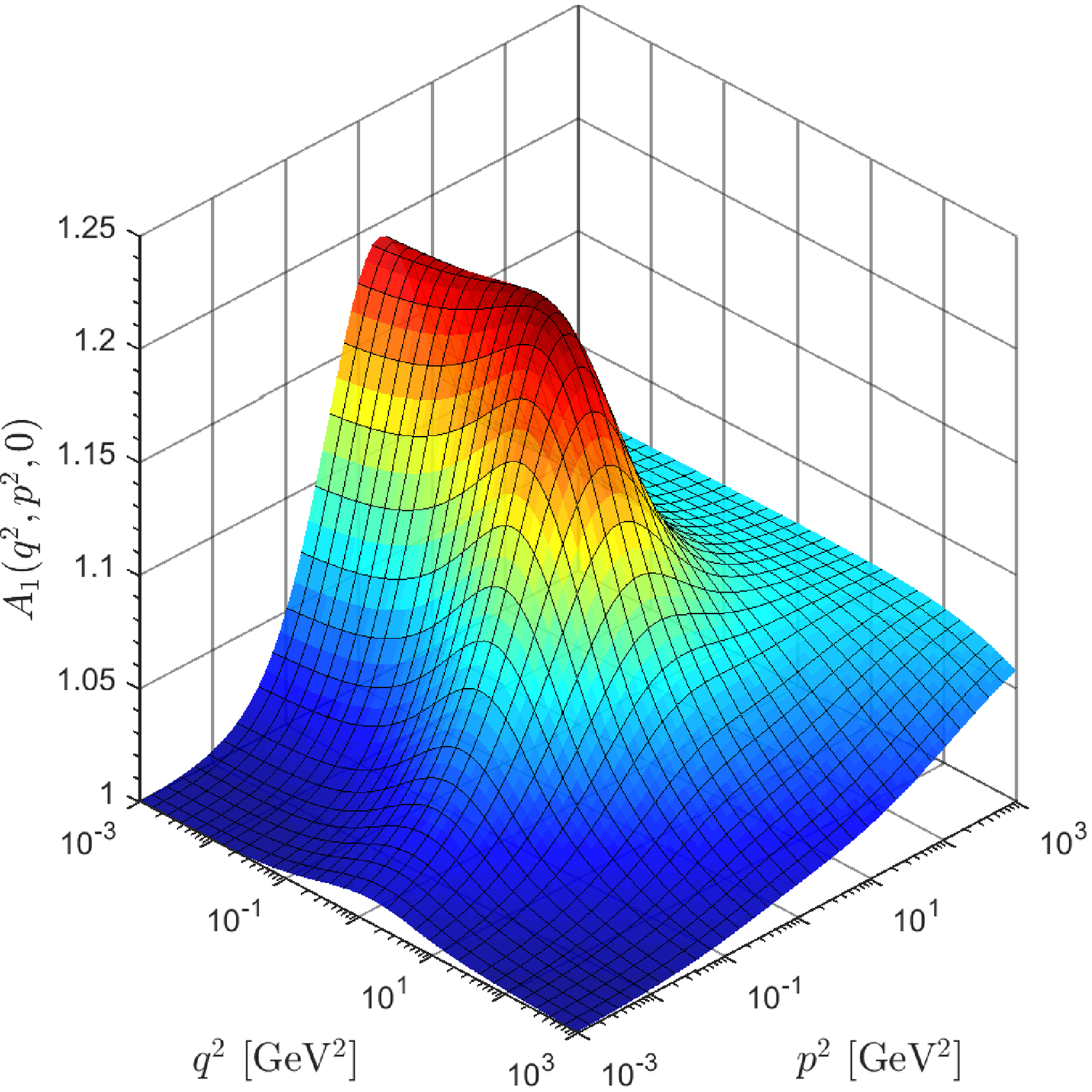}
\end{minipage}
\hspace{0.25cm}
\begin{minipage}[b]{0.45\linewidth}
\includegraphics[scale=0.39]{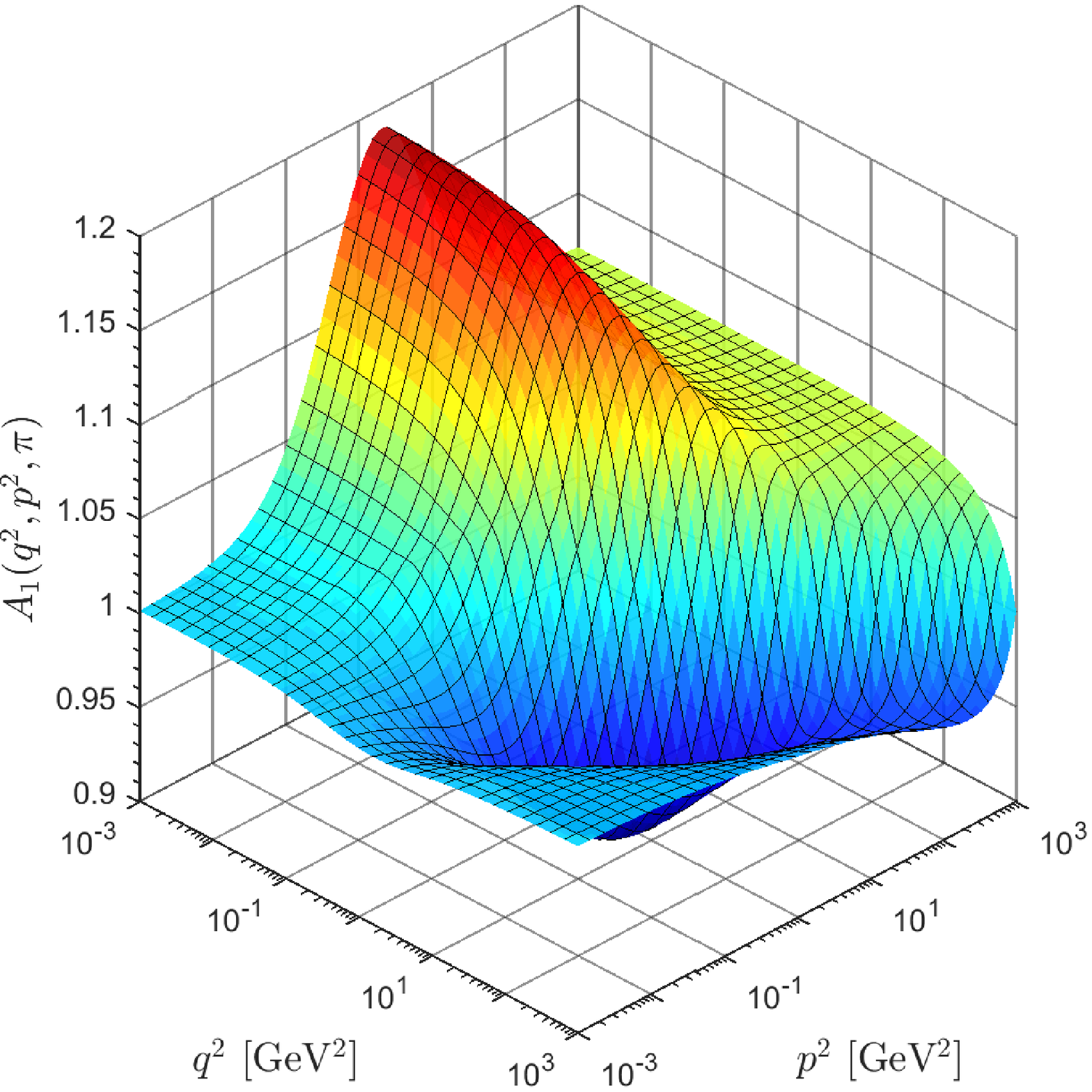}
\end{minipage}
\hspace{0.25cm}
\begin{minipage}[b]{0.45\linewidth}
\centering
\includegraphics[scale=0.39]{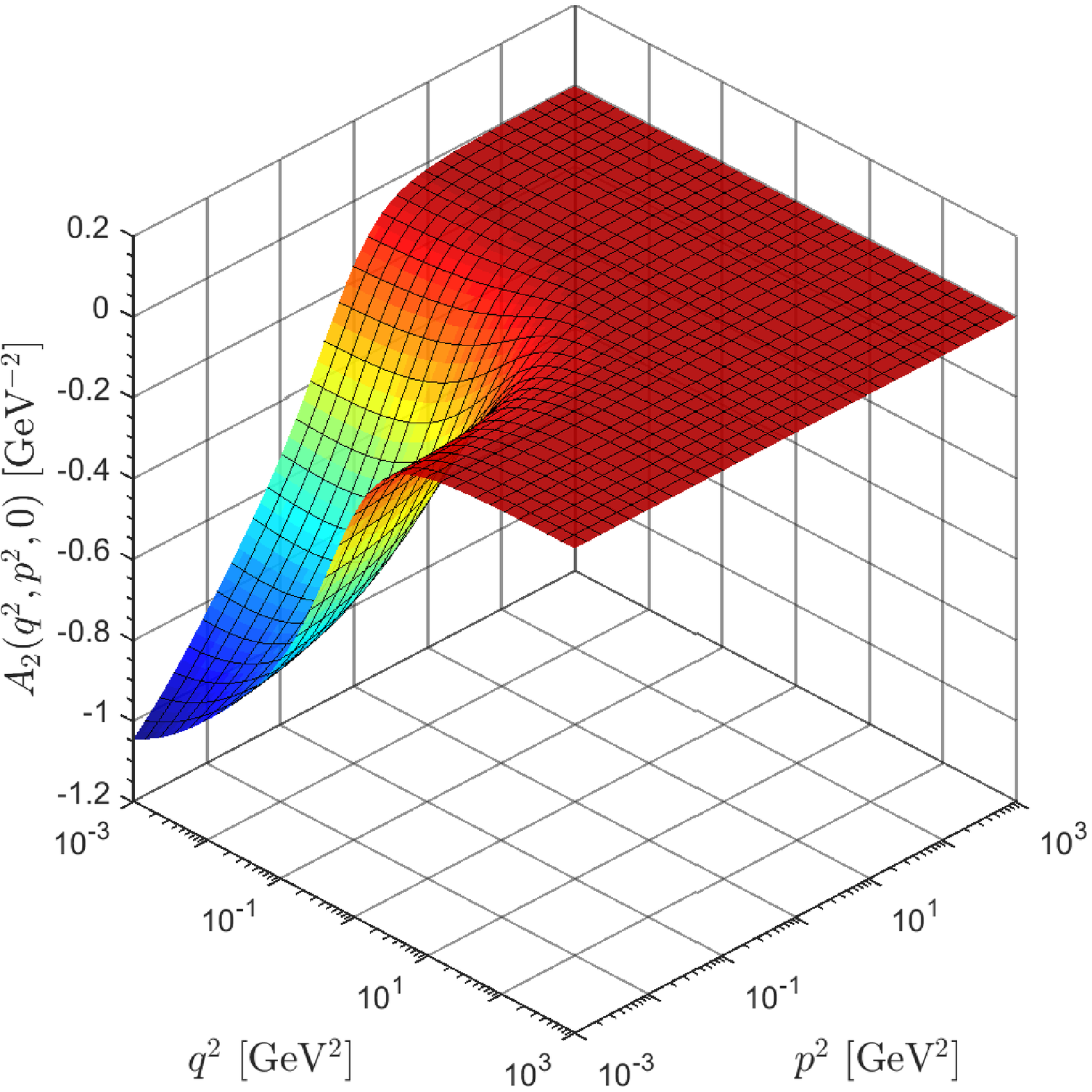}
\end{minipage}
\hspace{0.25cm}
\begin{minipage}[b]{0.45\linewidth}
\includegraphics[scale=0.39]{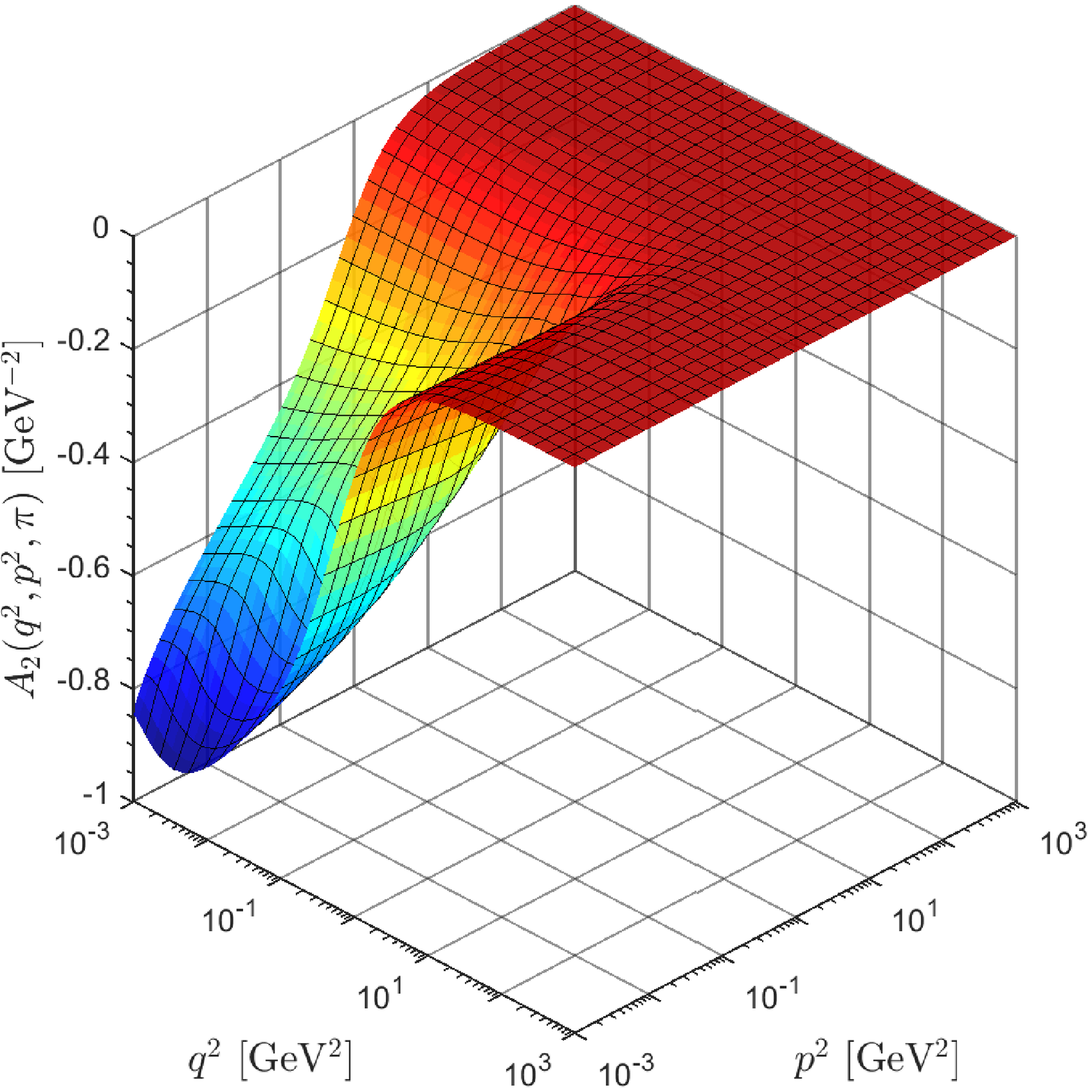}
\end{minipage}
\hspace{0.25cm}
\begin{minipage}[b]{0.45\linewidth}
\centering
\includegraphics[scale=0.39]{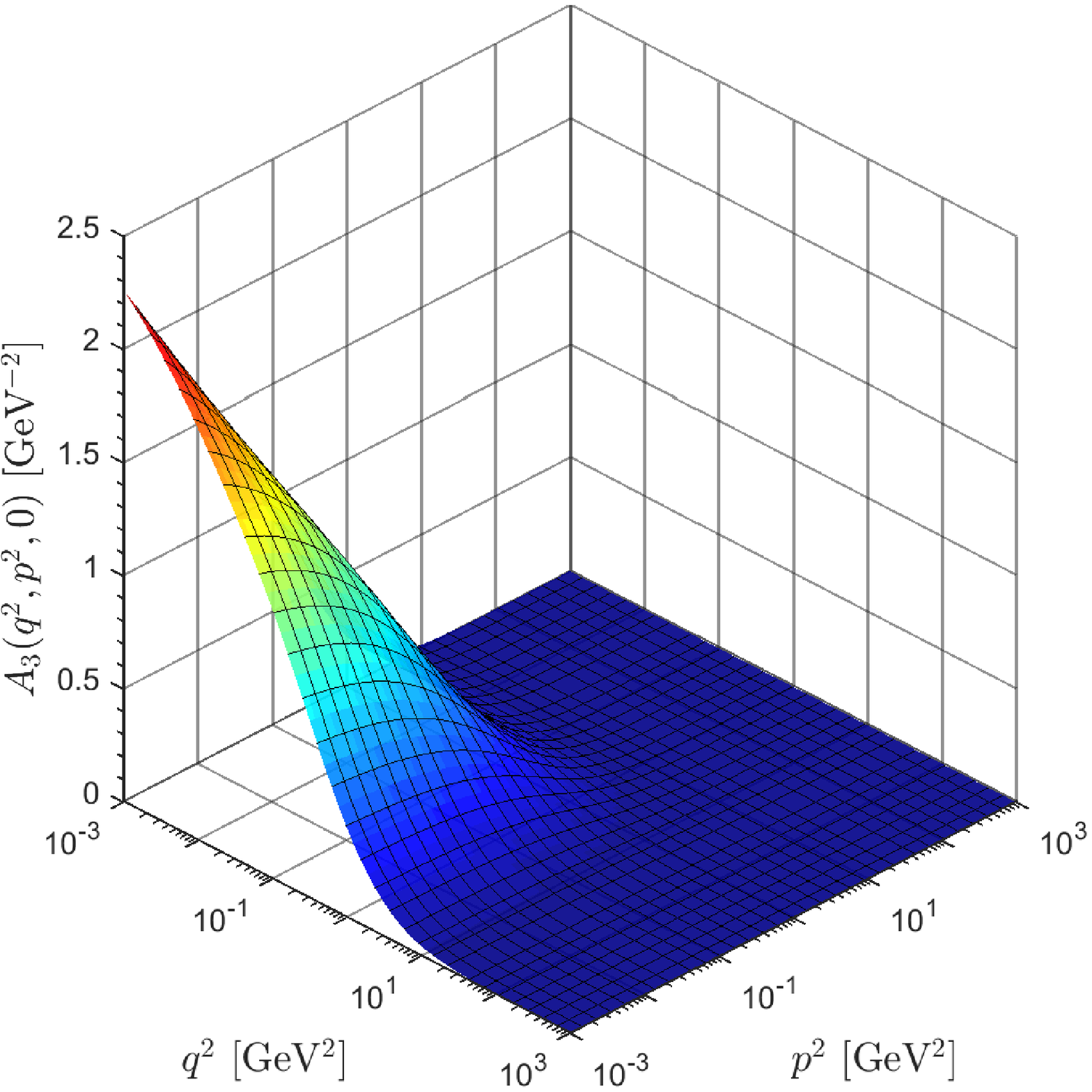}
\end{minipage}
\hspace{0.25cm}
\begin{minipage}[b]{0.45\linewidth}
\includegraphics[scale=0.39]{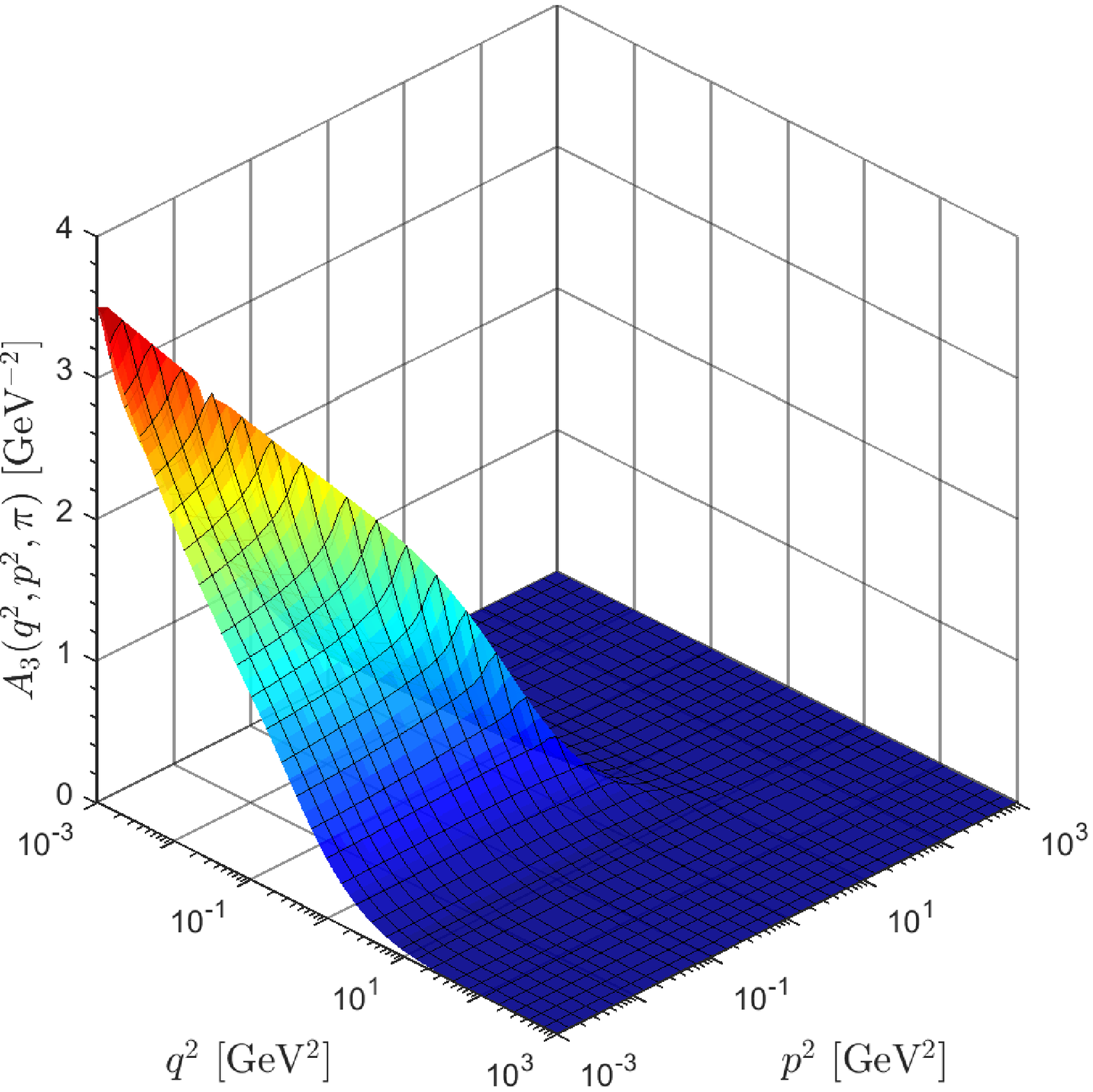}
\end{minipage}
\vspace{-0.5cm}
\caption{ The form factors of the ghost-gluon scattering kernel $A_1(q^2,p^2,\theta)$ (first row), $A_2(q^2,p^2,\theta)$ (second row), and  $A_3(q^2,p^2,\theta)$ (third row) for $\theta = 0$ and $\theta = \pi$
and $\alpha_s=0.22$.}\label{fig:A_fig1}
\end{figure}
%
\begin{figure}[!ht]
\begin{minipage}[b]{0.45\linewidth}
\centering
\includegraphics[scale=0.39]{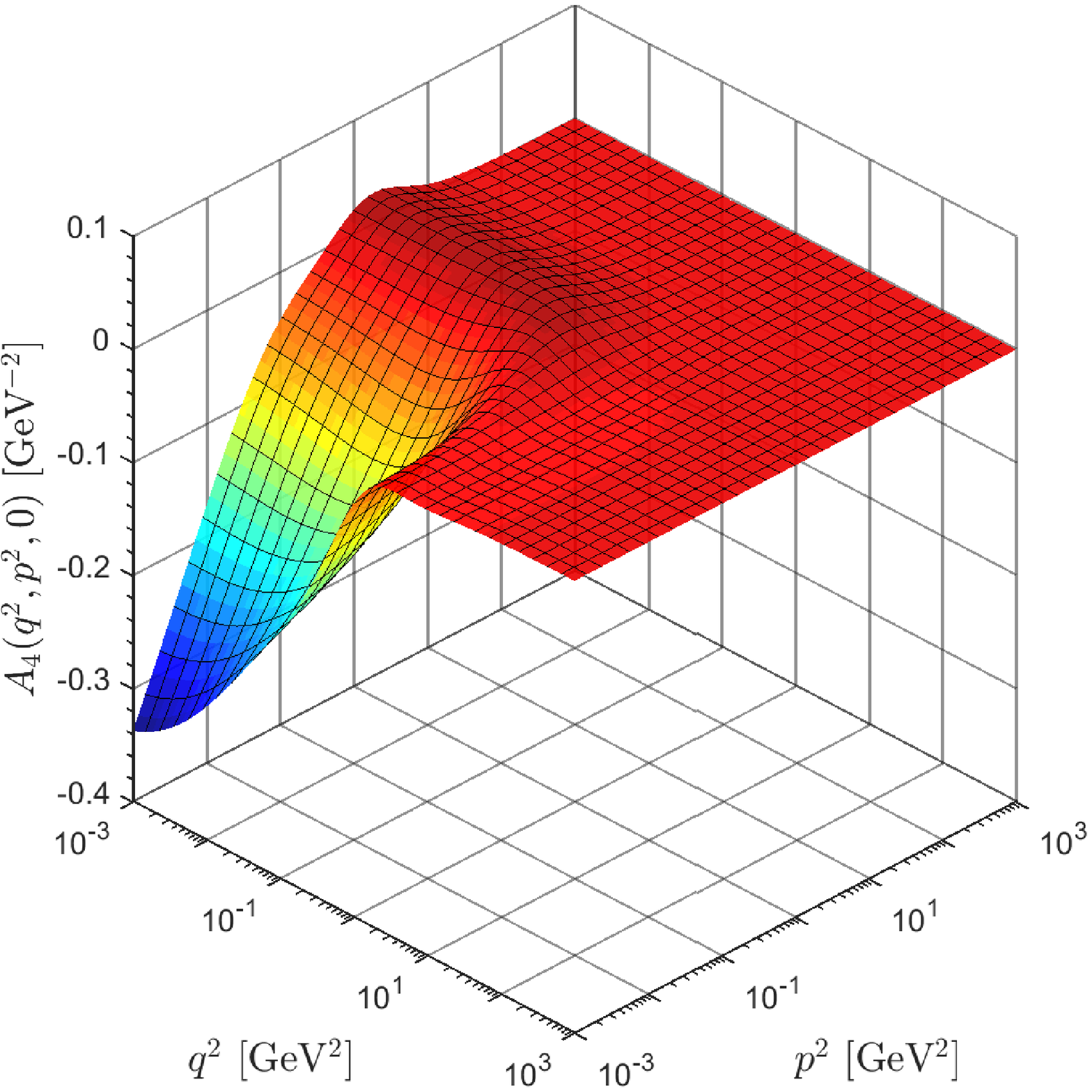}
\end{minipage}
\hspace{0.25cm}
\begin{minipage}[b]{0.45\linewidth}
\includegraphics[scale=0.39]{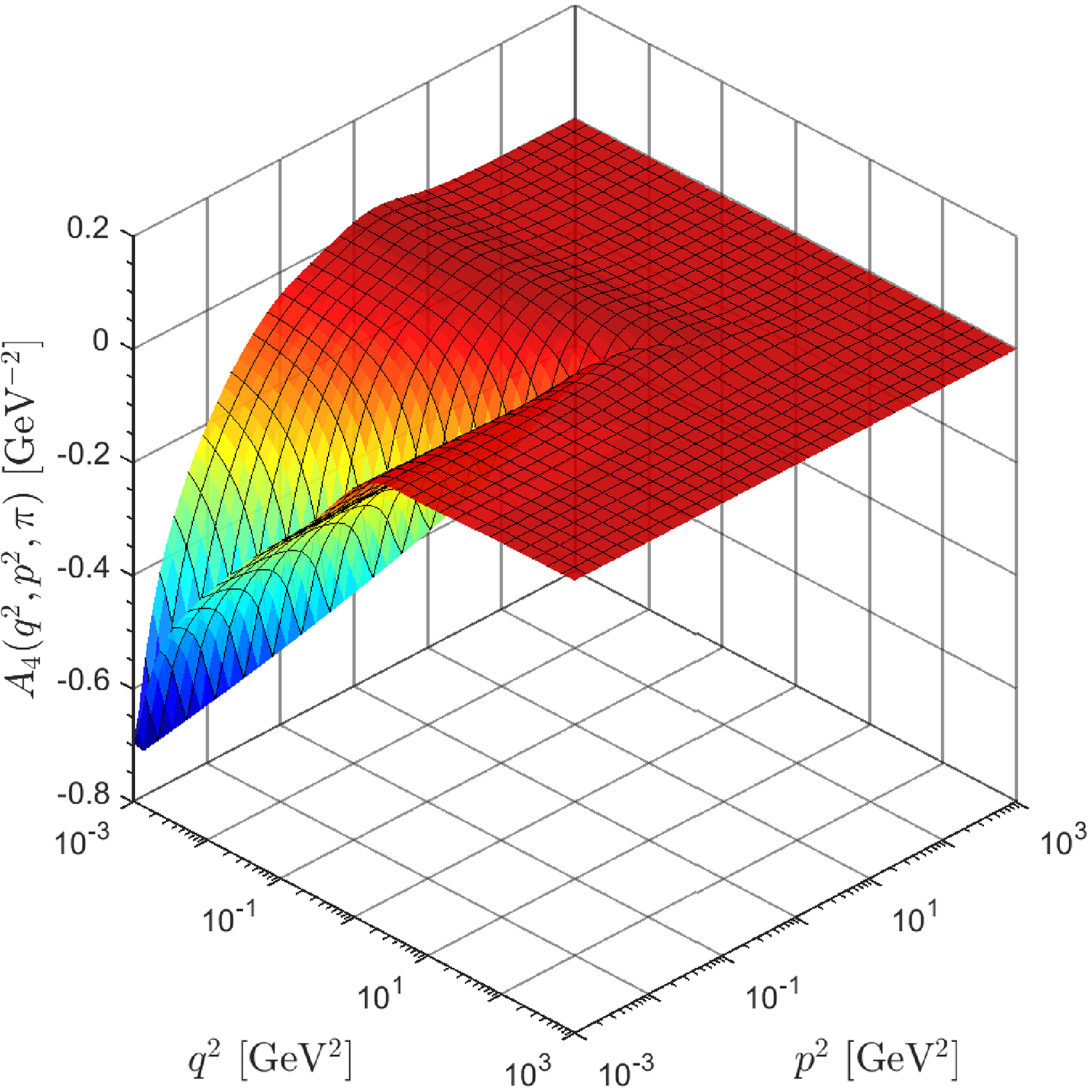}
\end{minipage}
\hspace{0.25cm}
\begin{minipage}[b]{0.45\linewidth}
\centering
\includegraphics[scale=0.4]{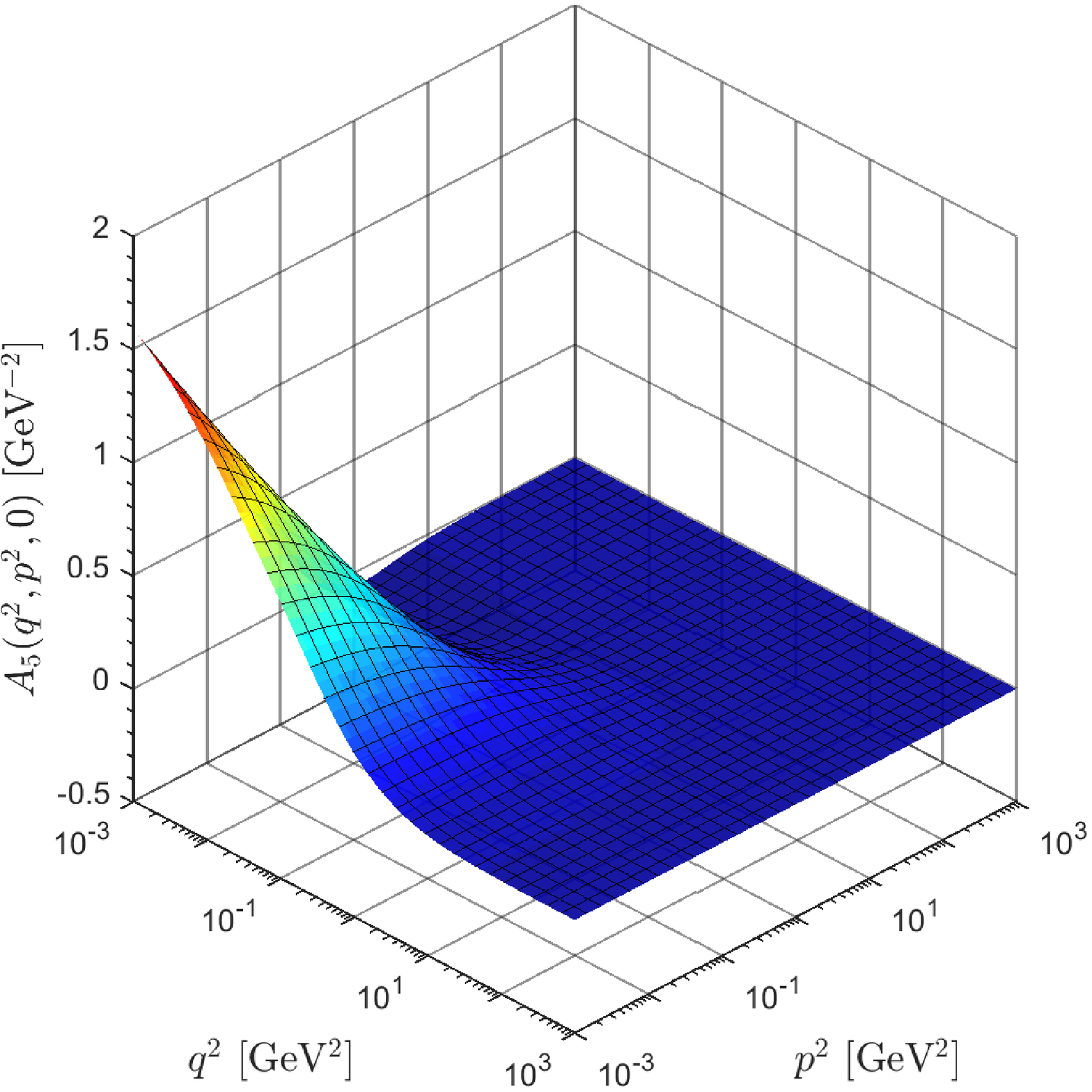}
\end{minipage}
\hspace{0.25cm}
\begin{minipage}[b]{0.45\linewidth}
\includegraphics[scale=0.39]{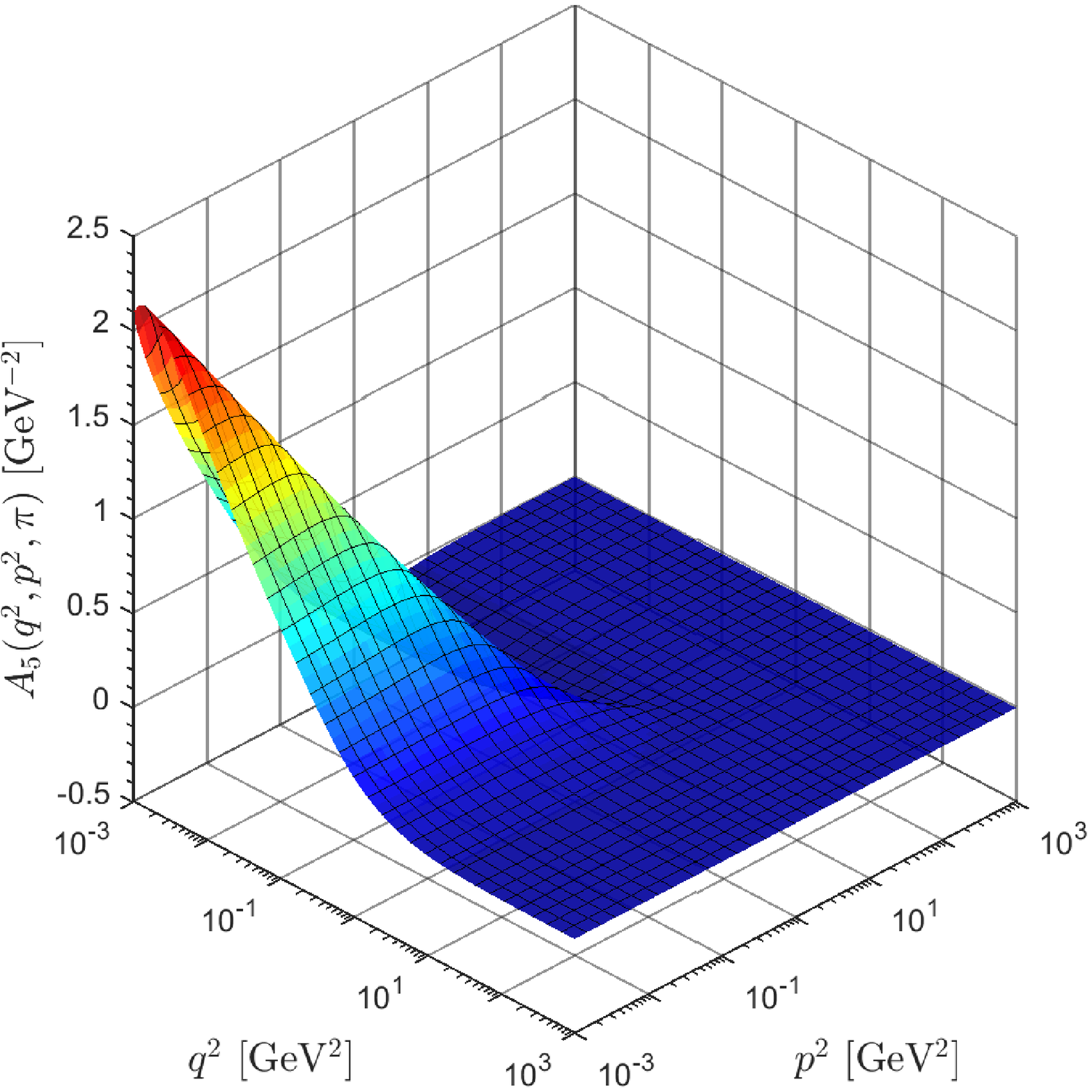}
\end{minipage}
\caption{ The form factors of the ghost-gluon scattering kernel $A_4(q^2,p^2,\theta)$ (top row) and $A_5(q^2,p^2,\theta)$ (bottom row) for $\theta = 0$ and $\theta = \pi$ and $\alpha_s=0.22$.}\label{fig:A_fig2}
\end{figure}

In what follows we will express all relevant form factors as functions of $q^2$,  $p^2$ and 
the angle $\theta$, namely $A_i(q,p,r)\to A_i(q^2,p^2,\theta)$.
Note also that since the quantities entering in the integrals do not depend on the angle $\phi_3$, the last integral
in~\eqref{eucsph} furnishes simply a factor of $2\pi$.

The evaluation of the ghost-gluon scattering kernel given by Eqs.~\eqref {eq:H_truncated} and~\eqref{eq:d1d2}  amounts to a three-dimensional integration for each combination of external momenta and angles,  namely $( q^2, p^2, \theta )$, and for each
 of the five $A_i$. These integrations were performed numerically with the adaptative algorithm of Ref.~\cite{Berntsen:1991:ADA:210232.210234}, employing an 11th degree polynomial rule. The results were computed with the external squared momenta distributed logarithmically on a grid with $80$ points, in the range \mbox{ $[ 5\times 10^{-5}\,\mbox{{GeV}}^2, 5\times 10^3\!,\mbox{{GeV}}^2]$ }, whereas for the angle $\theta$ the grid was composed of $19$ uniformly distributed points within $[0, \pi]$.

In the Figs.~\ref{fig:A_fig1} and \ref{fig:A_fig2}, we present a typical set of results for the form factors $A_i$,  for \mbox{$\theta = 0$} and 
\mbox{$\theta = \pi$}.  

It is important to notice that all form factors
exhibit the following common features:\\ ({\it i}) in the infrared,
they display considerable departures from their tree-level values, ({\it ii})  in the ultraviolet they approach 
the corresponding one-loop answers, given in Appendix~\ref{app:pert}\footnote{This particular property
  is expected, given that the input functions have been adjusted precisely to that purpose, as discussed in the previous section.
  Note, however, that possible deviations from this prescribed behavior may be produced,
 due to artifacts of the numerical treatment (see discussion in the third paragraph of the soft gluon limit in subsection~\ref{spec}).}; 
({\it iii}) in  general, they display a mild dependence on the angle $\theta$.

Moreover, we find that $A_1$ is {\it finite} in the infrared, whereas  
$A_2$, $A_3$, $A_4$, and $A_5$ {\it diverge logarithmically}.  
The origin of these divergences  
may be traced back to two different sources ({\it i}) the massless of the ghost propagators appearing $(d_1)_{\nu\mu}$ of the Fig.~\ref{fig:H_truncated}, or ({\it ii}) the ``unprotected'' logarithms 
contained in the $J_{\inpt}(q)$ that enter in the Ansatz of $\Gamma^{\inpt}_{\mu\alpha\beta}$ given in \1eq{eq:three_gluon_s}, thus altering the
behavior of the graph $(d_2)_{\nu\mu}$. 

 In the next subsection, we will carefully scrutinize the circumstances leading to 
 the aforementioned infrared logarithmic divergences, for each one of the four form factors.

\subsection{\label{spec}Special kinematics limits}

In this subsection we first extract from the general 3-D solutions for the $A_i$ reported above
three special kinematic configurations, corresponding to particular 2-D ``slices''.
Then, we compare them with 
({\it i}) the corresponding perturbative expressions computed at one loop; ({\it ii}) the one-loop ``massive'' results, obtained by using ``naive'' massive gluon propagators inside the one-loop diagrams (see Appendix~\ref{app:pert});
and  ({\it iii}) the results found when the three-gluon vertex appearing in $(d_2)_{\mu\nu}$ is kept at its tree-level value, \ie
setting ${X}^{\inpt}_1(r,t,\ell)=1$.
As we will see, the  comparisons ({\it ii}) and ({\it iii}) are
fundamental for identifying the origin of the infrared logarithmic divergences displayed by the four  $A_i$.
Specifically, by means of the one-loop massive calculation one can establish analytically
whether $(d_1)_{\mu\nu}$ and $(d_2)_{\mu\nu}$ are individually convergent or divergent,  
depending on the nature of the propagators comprising them.  
As for ({\it iii}), the use of  
$\Gamma^{(0)}$ instead of $\Gamma^{\inpt}$ helps us identify the dressing 
of the latter as the only reason for the infrared divergences encountered in $(d_2)_{\mu\nu}$.

Thus, through this entire subsection, we display four curves in all panels. The curves correspond to
the full case (2-D slices) [using $\Gamma^{\inpt}$] (blue continuous), 
the one-loop result (purple dotted), the one loop massive with \mbox{$m^2=0.15\, \mbox{GeV}^2$} (green dashed dotted), and the case where the
$\Gamma_{\mu\alpha\beta}^{(0)}$ of Eq.~\eqref{eq:3gluon} is used as input in $(d_2)_{\mu\nu}$ (red dashed).
We adopt the same color convention in all panels.

Before proceeding, let us emphasize that, in order to expedite the one-loop calculations, we have 
implemented the corresponding kinematic limits {\it directly} at the level of $H_{\nu\mu}$, \ie
{\it before} projecting out the corresponding form factors. As a result,
and depending on the details of the limit considered, 
certain tensorial structures, together with the accompanying form factors,
are completely eliminated from the decomposition of $H_{\nu\mu}$ given in \1eq{eq:H}.
Of course, the form factors that are eliminated are nonvanishing, as may be easily verified from the
appropriate ``slices'' of the corresponding 3-D plots.

\vspace{0.25cm}
{\it(i)} The  \emph{soft gluon limit},  which means that  
$r = 0$; then, the momenta $q$ and $p$ have the same magnitude, \mbox{$|p|=|q|$}, and are anti-parallel {\it i.e.}, $\theta=\pi$.
Our results are expressed in terms of the momentum $q$.

\begin{figure}[!t]
\begin{minipage}[b]{0.45\linewidth}
\centering
\includegraphics[scale=0.32]{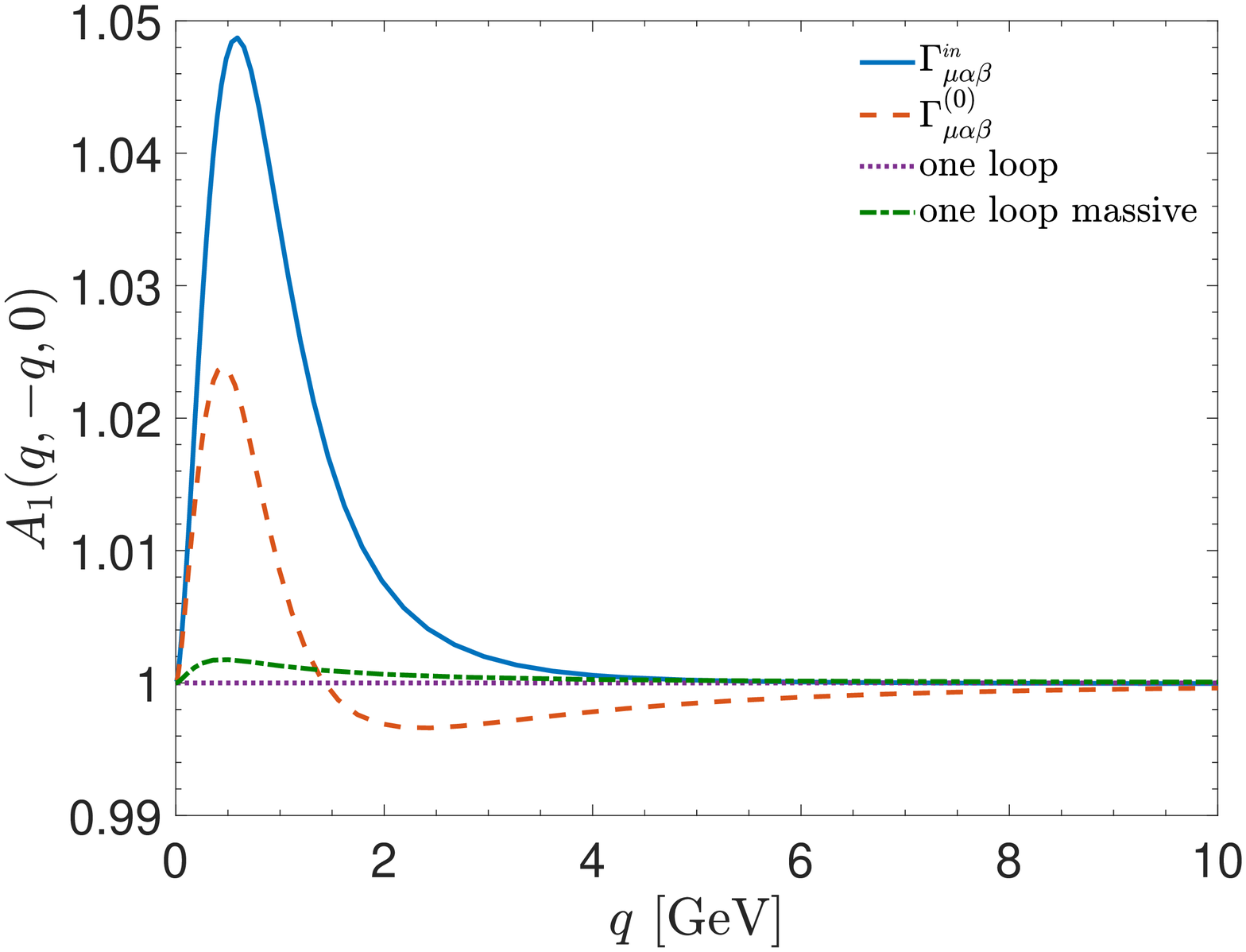}
\end{minipage}
\hspace{0.25cm}
\begin{minipage}[b]{0.45\linewidth}
\includegraphics[scale=0.32]{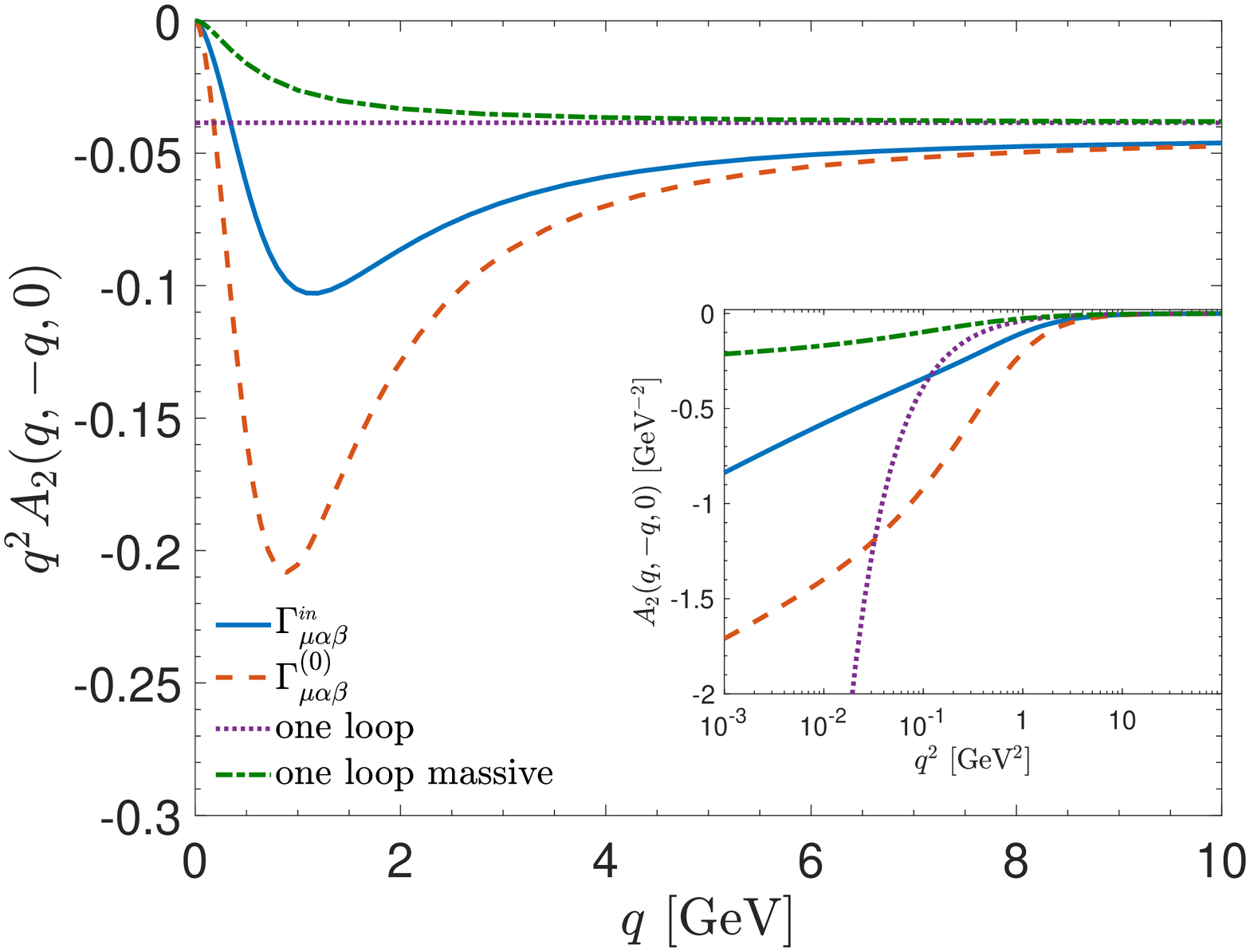}
\end{minipage}
\caption{(soft gluon kinematics) Left: Comparison between the  $A_1(q,-q,0)$ computed using $\Gamma^{\inpt}_{\mu\alpha\beta}$ 
(blue continuous) and the one obtained when ${\Gamma}^{(0)}_{\mu\alpha\beta}$ is used instead (red dashed). The massless (purple dotted) and the massive (green dashed dotted) one-loop perturbative results are given by Eqs.~\eqref{eq:Ai_sgl} and~\eqref{1lMsg}, respectively. Right: Same comparison for the dimensionless combination $q^2A_2(q,-q,0)$. In the inset we show 
  the corresponding logarithmic infrared divergence of $A_2(q,-q,0)$, using a logarithmic scale for $q^2$. Note that the purple dotted curve shows a much steeper (linear) divergence.}
\label{fig:pert_sgl}
\end{figure}
 
When this kinematic limit is implemented as described above, the 
only tensorial structures that survive are those associated with 
$A_1$ and $A_2$ [see Eq.\eqref{eq:H}]. These two form factors are shown
in Fig.~\ref{fig:pert_sgl}. 
$A_1(q,-q,0)$ (left panel)
displays only a mild  deviation  from its tree-level value in the
entire range of momenta. The maximum deviation is of the order of $5\%$, and is located around \mbox{$q\approx 1$ GeV}.  
It is interesting to observe that 
the one-loop massive and the nonperturbative calculation with $\Gamma_{\mu\alpha\beta}^{(0)}$ also display the peak around the same region of momenta, 
although there is a clear quantitative difference in their heights.  
Notice that $A_1$ is infrared finite, and for all curves we have  
$A_1(0,0,0)=1$. This particular value is recovered again for high values of $q$, as expected from the one-loop calculation of Eq.~\eqref{eq:Ai_sgl};
one may clearly observe how all curves approach each other and practically coincide around \mbox{$q\approx 10$ GeV}.  

  It is important to mention that, in the above analysis, the limit $\theta=\pi$ is rather subtle.
This happens because 
the projectors of the $A_i$ introduce a $\sin^4\theta$ in their
denominator [see Eq.~\eqref{eq:Ai_proj}], whose cancellation requires the proper Taylor expansion 
of the numerator around $\sin\theta\approx 0$. If instead of expanding one were to use
a configuration whose angle was slightly different from $\pi$, the resulting curve would fail to approach
the one-loop result, running instead ``parallel'' to it.

On the right panel of Fig.~\ref{fig:pert_sgl} we show the dimensionless combination $q^2A_2(q,-q,0)$,
which in the ultraviolet tends towards the constant value predicted by the one-loop result given by Eq.~\eqref{eq:Ai_sgl}.
Once again, the maximum deviation from its tree-level value is located around \mbox{$q\approx 1$ GeV}, and the nonperturbative calculation with $\Gamma_{\mu\alpha\beta}^{(0)}$ captures rather well the position of this minimum, although
its depth is bigger. In order to make apparent the infrared logarithmic divergence, in the inset we show the dimensionful $A_2(q,-q,0)$ alone,
using a logarithmic scale.
Notice that the one-loop massive analytical result [see Eq.~\eqref{sgmass0}] and the nonperturbative calculation with $\Gamma_{\mu\alpha\beta}^{(0)}$ also display the same type of divergence in the infrared. In addition, observe that the use of $\Gamma^{\inpt}_{\mu\alpha\beta}$ slows down the rate of the negative infrared divergence of $A_2$. 
It is interesting to mention that the infrared divergence of $A_2$ is due the the presence of the two massless ghost propagators in the diagrams $(d_1)$ of the Fig.~\ref{fig:H_truncated}.  In Table~\ref{tab:divergences} we summarize how each diagram behaves in the infrared  separately for the cases presented in the plot, except for the pure perturbative one-loop calculation.   

\vspace{0.25cm}

{\it(ii)}  The \emph{soft anti-ghost limit}, in which $q = 0$  and 
the momenta \mbox{$|p|=|r|$}; evidently, $|q||p|\cos{\theta} = 0$,
and any dependence on the angle $\theta$ is washed out .

\begin{figure}[!t]
\begin{minipage}[b]{0.45\linewidth}
\centering
\includegraphics[scale=0.32]{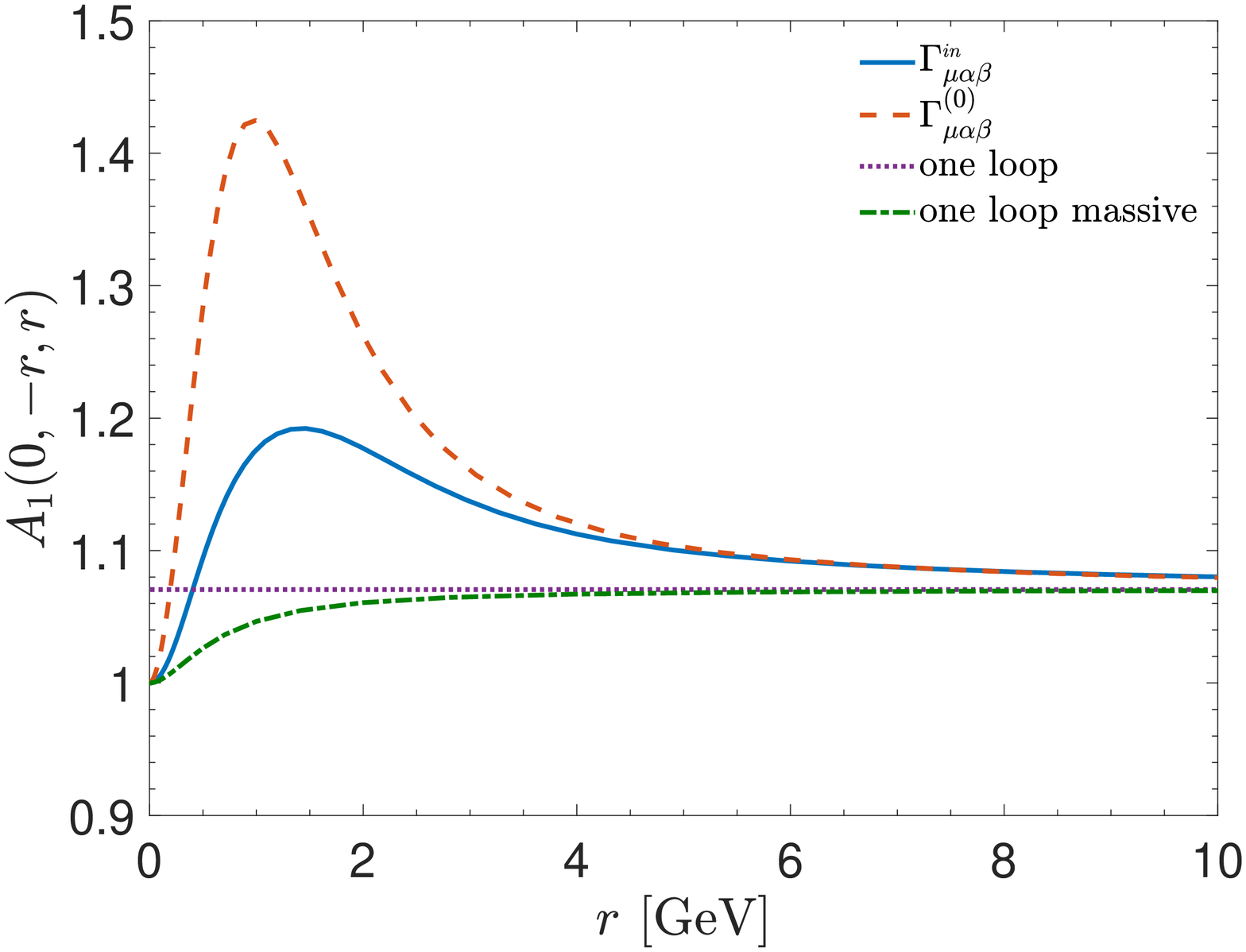}
\end{minipage}
\hspace{0.25cm}
\begin{minipage}[b]{0.45\linewidth}
\includegraphics[scale=0.32]{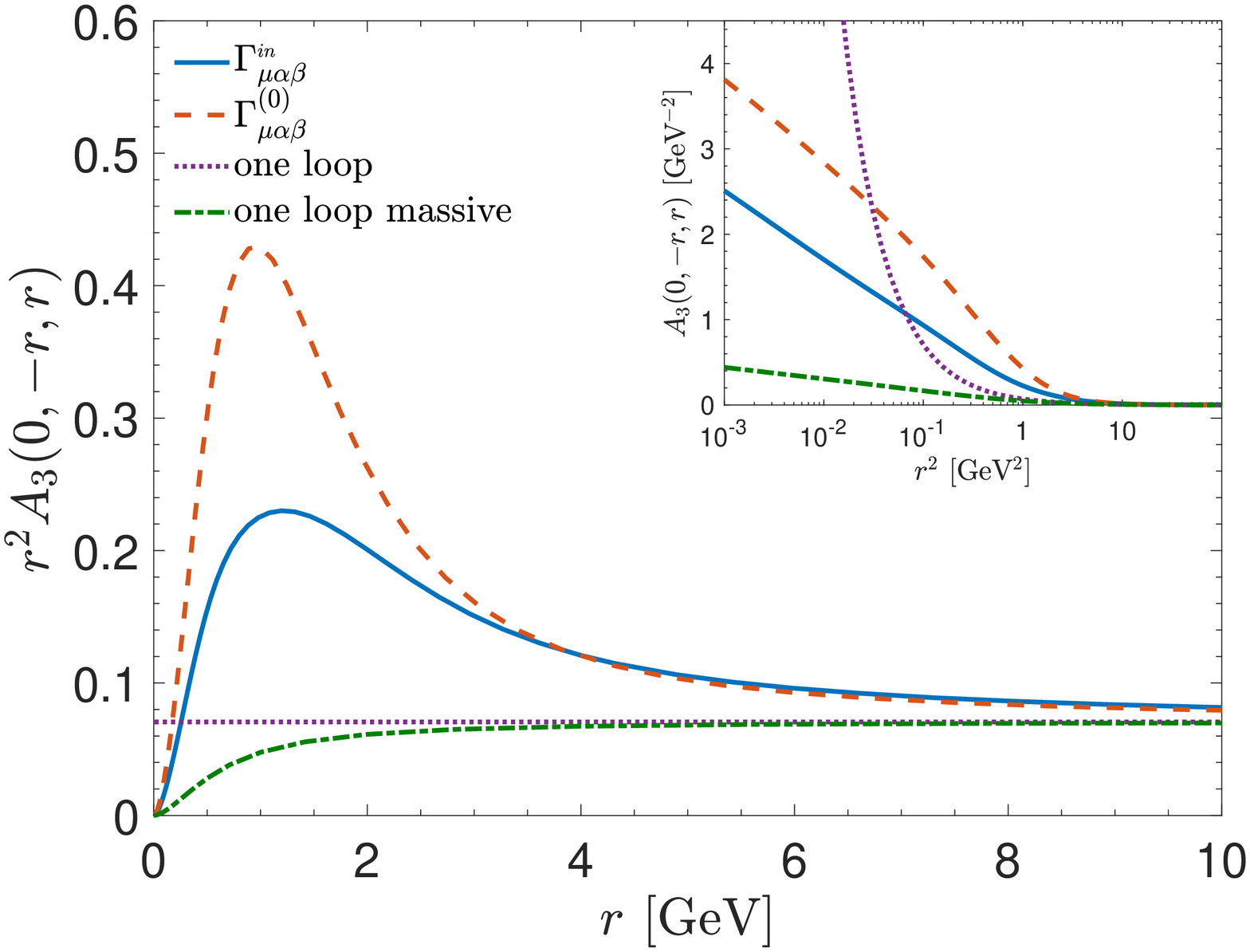}
\end{minipage}
\caption{(soft anti-ghost)
Left: Comparison of the  $A_1(0,-r,r)$  computed using $\Gamma^{\inpt}_{\mu\alpha\beta}$ (blue continuous) and ${\Gamma}^{(0)}_{\mu\alpha\beta}$  (red dashed) in the soft anti-ghost kinematics. The massless (purple dotted) and the massive (green dashed dotted) one-loop perturbative results are given by Eqs.~\eqref{eq:Ai_sagh} and~\eqref{1lmag}, respectively. Right: Same comparison for the dimensionless combination $r^2A_3(0,-r,r)$. In the inset we show  the corresponding logarithmic infrared divergence of the $A_3(0,-r,r)$, and the linear divergence of the massless one-loop case.}
\label{fig:pert_sagh}
\end{figure}

  In this limit, we may recover information only about $A_1$ and $A_3$, which 
depend on a unique momentum, namely $r$. 
In Fig.~\ref{fig:pert_sagh}, we can see that both form factors, $A_1(0,-r,r)$ and 
$A_3(0,-r,r)$ display  a sizable deviation
from their tree-level expressions around the region \mbox{$r\approx 1.0-1.5$ GeV}. Moreover, in the ultraviolet they are approaching the one-loop results of Eq.~\eqref{eq:Ai_sagh}.
$A_1$ is again infrared finite, while $A_3$ is logarithmically divergent, as shown in the inset.
 Note that the one-loop  massive result [see Eq.~\eqref{1mass_sag}] and the nonperturbative calculation with $\Gamma_{\mu\alpha\beta}^{(0)}$
 display the same qualitative behaviour; of course, the precise rates of each divergence are different.
 As can be seen in Table~\ref{tab:divergences}, 
the infrared divergence found in $A_3$ is due to both the massless ghost entering in the diagram $(d_1)$, and the unprotected logarithm present in the $\Gamma^{\inpt}_{\mu\alpha\beta}$ of graph $(d_2)$.

\vspace{0.25cm}

{\it(iii)} The \emph{totally symmetric limit}, defined in \1eq{symconf}.

\begin{figure}[!ht]
\begin{minipage}[b]{0.45\linewidth}
\centering
\includegraphics[scale=0.32]{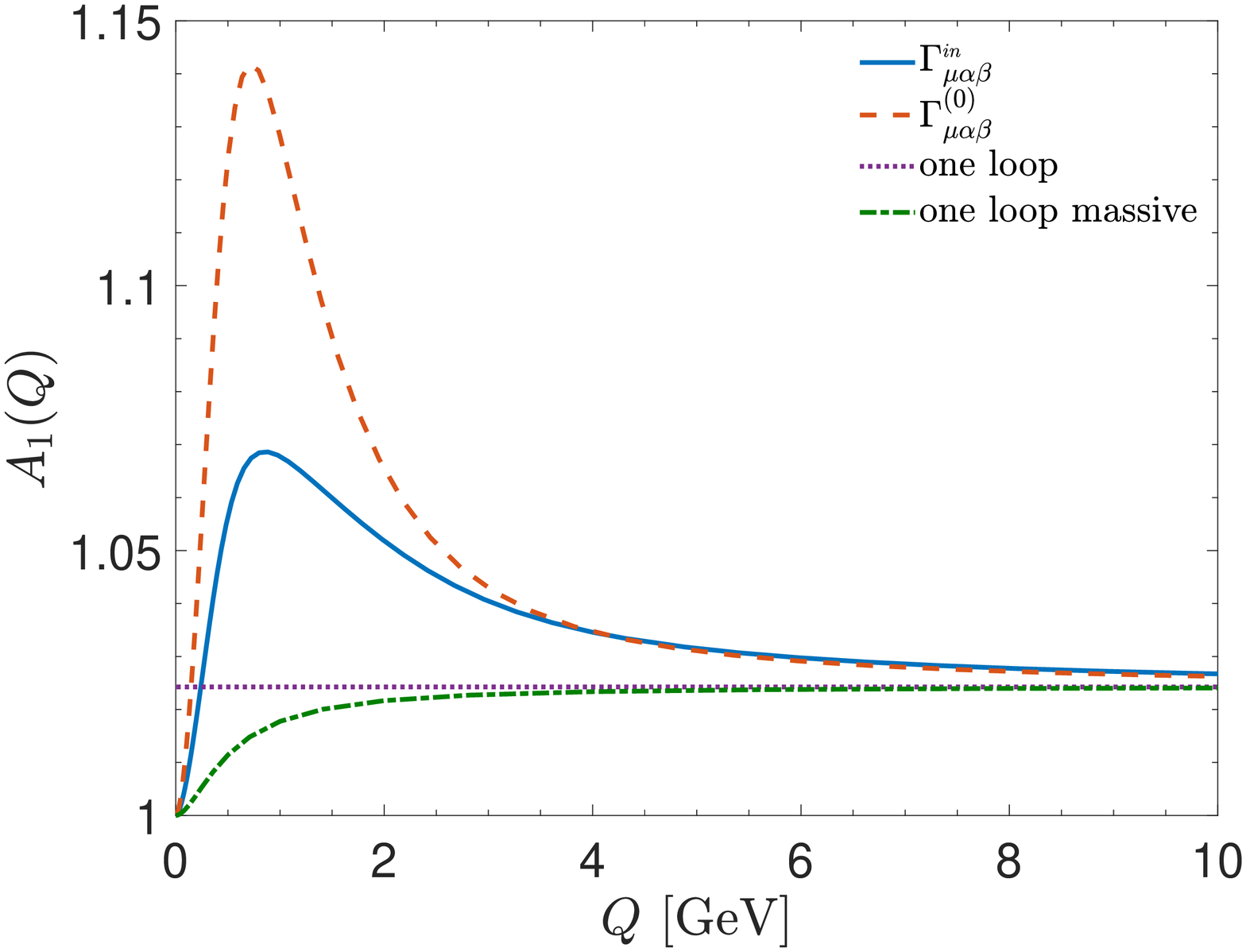}
\end{minipage}
\hspace{0.25cm}
\begin{minipage}[b]{0.45\linewidth}
\includegraphics[scale=0.32]{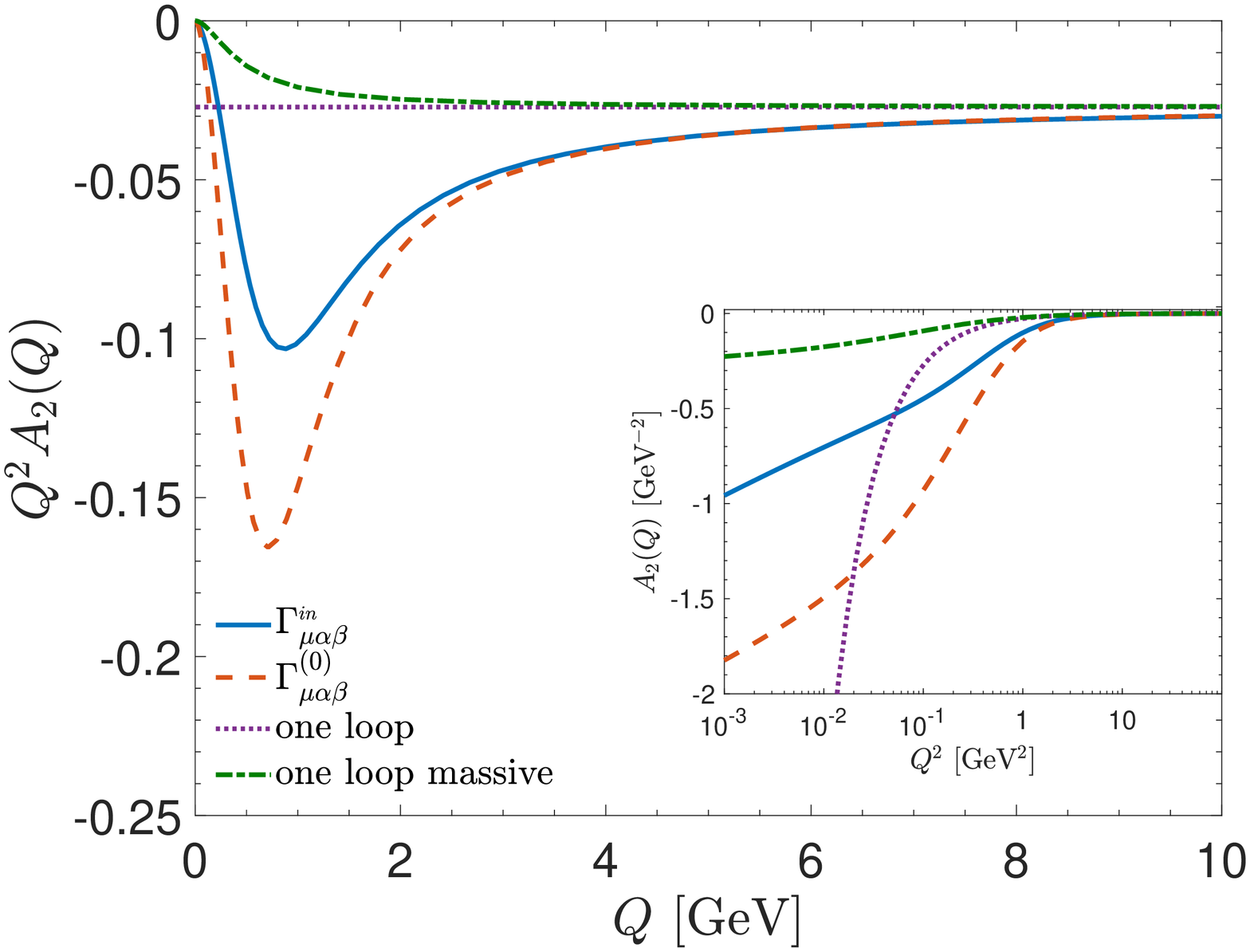}   
\end{minipage}
\begin{minipage}[b]{0.45\linewidth}
\centering
\includegraphics[scale=0.32]{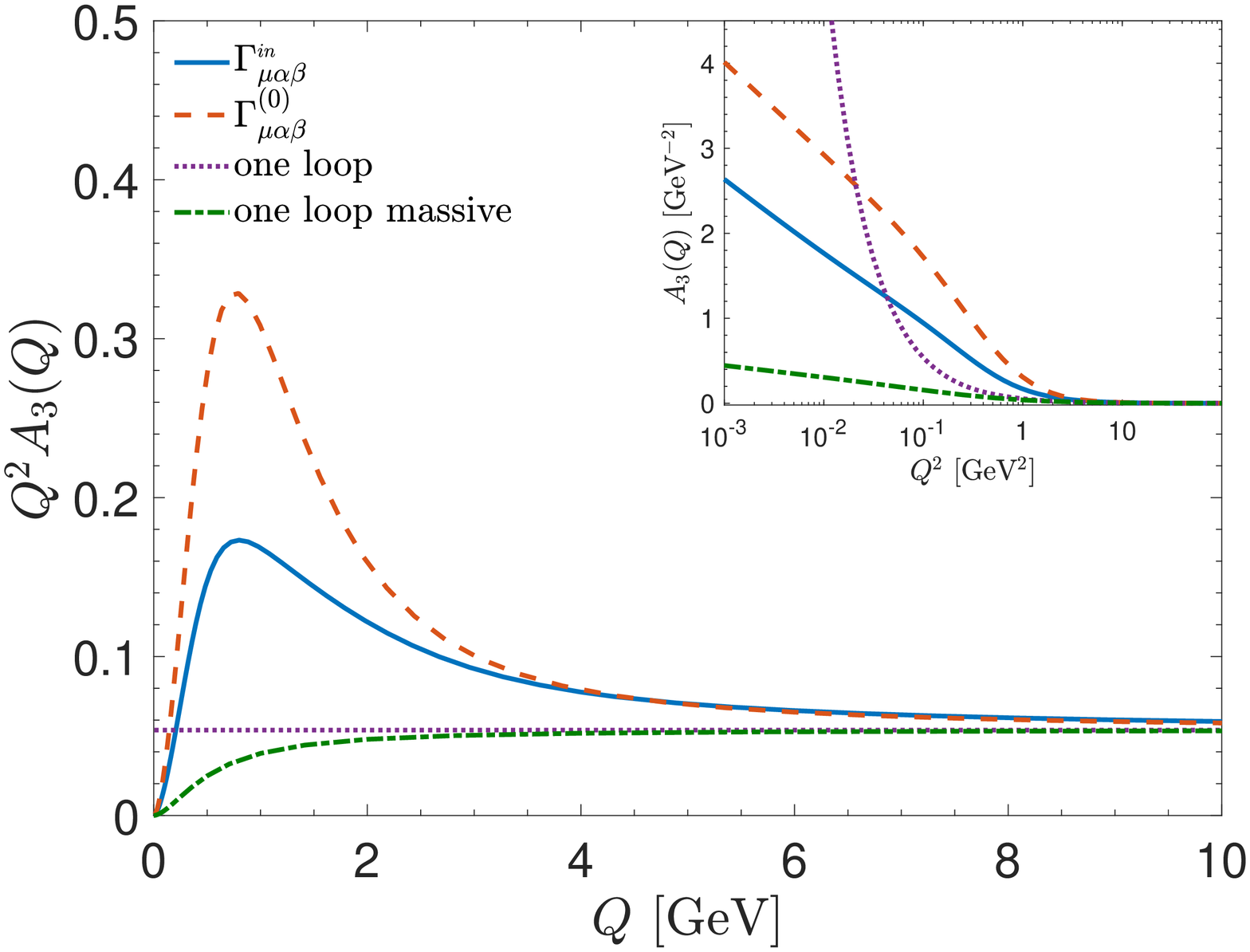}
\end{minipage}
\hspace{0.25cm}
\begin{minipage}[b]{0.45\linewidth}
\includegraphics[scale=0.32]{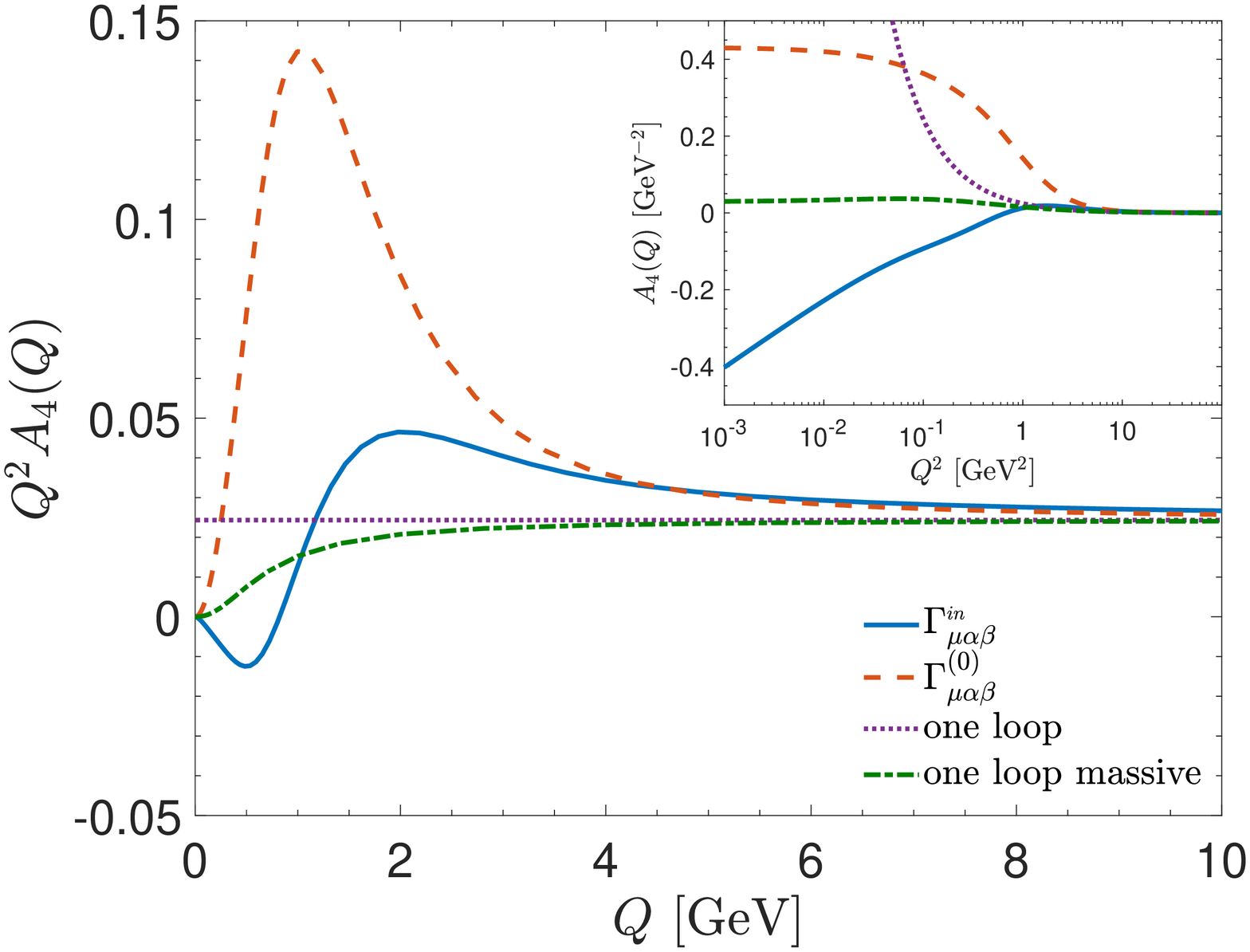}
\end{minipage}
\vspace{0.5cm}
\begin{centering}
\includegraphics[scale=0.32]{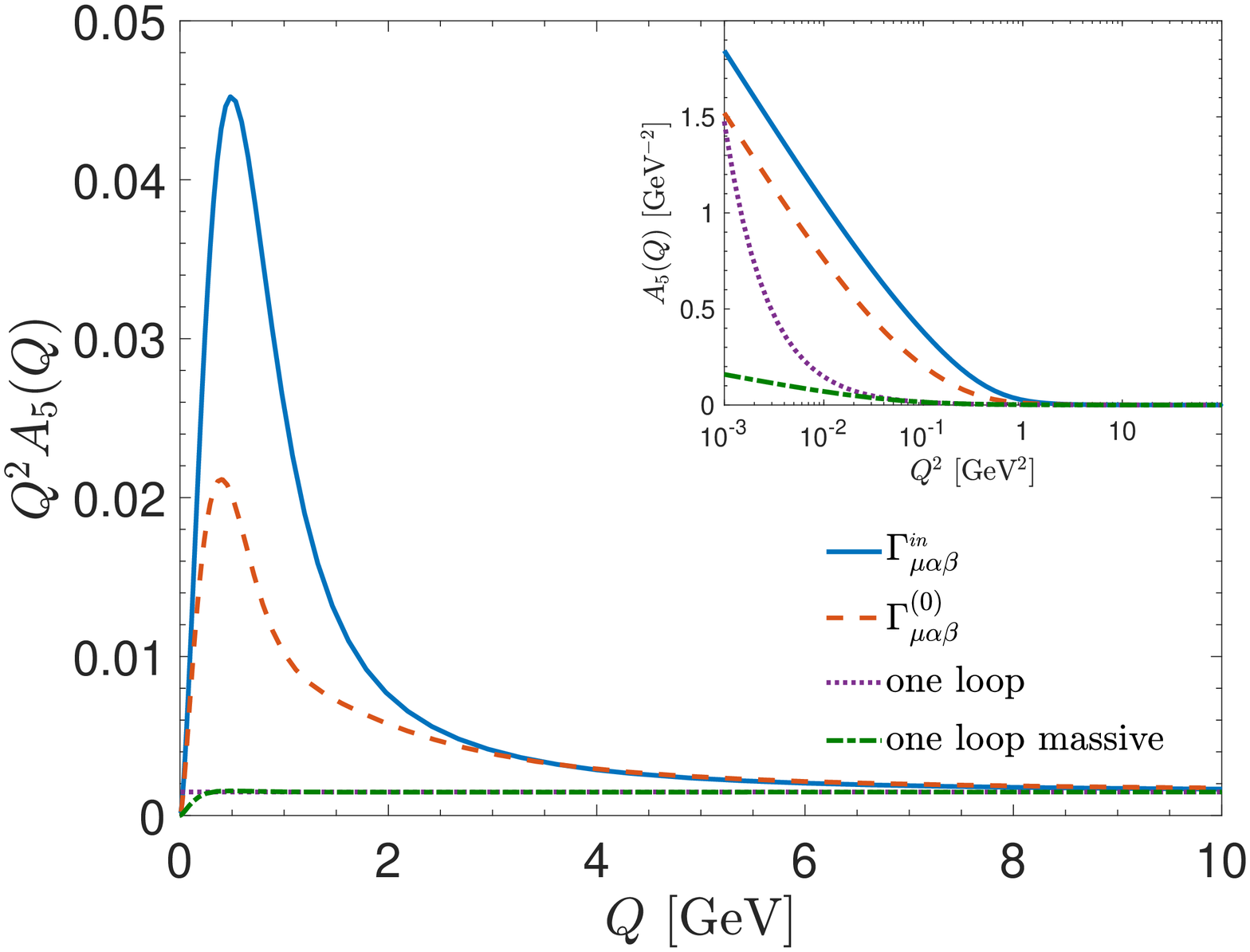}
\end{centering}
\vspace{-0.75cm}
\caption{(totally symmetric) 
The dimensionless combinations of the form factors $A_i(Q)$ in the totally symmetric configuration. The $A_i(Q)$ are computed using $\Gamma^{\inpt}_{\mu\alpha\beta}$ given  by Eq.~\eqref{eq:three_gluon_s}
(blue continuous) and ${\Gamma}^{(0)}_{\mu\alpha\beta}$ of the Eq.~\eqref{eq:3gluon} (red dashed). The one-loop results are given in Eqs.~\eqref{eq:Ai_sym} (purple dotted) while the infrared limits of the one-loop massive case 
are expressed by Eq.~\eqref{limitsym}. In the inset we show  the corresponding logarithmic divergences of the $A_i(Q)$, and the linear divergence of the
massless one-loop result.}
\label{fig:pert_sym}
\end{figure}

In Fig.~\ref{fig:pert_sym} we show the behaviour of the $A_i(Q)$ in this configuration; note
that in this configuration all form factors are accessible.

We clearly see that the $A_i$ obtained with either vertex 
display a sizable deviation from their tree-level value in the region of 
\mbox{$Q\approx 1-2$ GeV}, while for large values of $Q$ they recover the
ultraviolet behaviour expected from one-loop perturbation theory, given by Eqs.~\eqref{eq:Ai_sym}.
Interestingly enough, except for $Q^2A_5(Q)$, the use of $\Gamma^{\inpt}_{\mu\alpha\beta}$ yields $A_i$ that are more suppressed.

Moreover, one can notice that $A_4$ and $A_5$, whose forms were not presented for the 
previous configurations, also display a logarithmic divergence in the infrared (see the insets of Fig.~\ref{fig:pert_sym}).
In the case of $A_4$, the divergence is exclusively
associated with the unprotected logarithm present in the $\Gamma^{\inpt}_{\mu\alpha\beta}$ used in  $(d_2)$, while
the diagram responsible for the logarithm divergence of $A_5$ is $(d_1)$ [see Table~\ref{tab:divergences}].
Notice that, except for $A_4(Q)$, both the
 analytic one-loop massive results and the nonperturbative calculation with $\Gamma_{\mu\alpha\beta}^{(0)}$ reproduce the general pattern found when one uses $\Gamma^{\inpt}_{\mu\alpha\beta}$. More specifically, these cases capture whether the divergence is positive or negative and the finiteness of $A_1$.  In the case of $A_4$,  the impact of  $\Gamma^{\inpt}_{\mu\alpha\beta}$ is rather pronounced, and  it causes a negative logarithmic divergence.  
   
The  Table~\ref{tab:divergences} provides an overview of our main findings, specifying the different  origins of the infrared logarithmic divergences found in the form factors $A_2$, $A_3$, $A_4$, and $A_5$ in the three cases analyzed. 
%
\begin{table}[h]
\begin{center}
\begin{tabular}{|C{1.2cm}|C{1.6cm}|C{1.6cm}|C{1.6cm}|C{1.6cm}|C{1.6cm}|C{1.6cm}|C{1.6cm}|}
\hline
Form & \multicolumn{2}{c|}{one-loop~massive} & \multicolumn{2}{c|}{$\Gamma_{\mu\alpha\beta}^{(0)}$} & \multicolumn{3}{c|}{$\Gamma^{\inpt}_{\mu\alpha\beta}$} \\
\cline{2-8}
factors & $(d_1)$ & $(d_2)$ & $(d_1)$ & $(d_2)$ & $(d_1)$ & $(d_2)$ & Total \\ 
\hline
$A_1$ & F & F & F & F & F & F & F \\
\hline
$A_2$ & LD & F & LD & F & LD & F & LD \\
\hline
$A_3$ & LD & F & LD & F & LD & LD & LD \\
\hline
$A_4$ & F & F & F & F & F & LD & LD \\
\hline
$A_5$ & LD & F & LD & F & LD & F & LD \\
\hline
\end{tabular}
\caption{\label{tab:divergences}
The summary of the infrared limits
of the individual contributions of the diagrams $(d_1)$ and  $(d_2)$ appearing in the Fig.~\ref{fig:H_truncated}. The limits are for ({\it i}) the one-loop massive results [see Eqs.~\eqref{sgmass0}, \eqref{1mass_sag}, and \eqref{limitsym}]; ({\it ii}) the nonperturbative result obtained when
$\Gamma_{\mu\alpha\beta}^{(0)}$ is used as input in the diagram $(d_2)$; and 
({\it iii}) the nonperturbative result obtained with $\Gamma^{\inpt}_{\mu\alpha\beta}$.  The letter ``F'' stands for ``finite'', and the acronym ``LD'' for
``logarithmically divergent''.}
\end{center}
\end{table}

\section{\label{nun_const} The constraint from the STI}

The next item of our analysis is dedicated to the STI-derived 
constraint of Eq.~\eqref{eq:constraint1}.
The way this particular constraint becomes relevant for our considerations is two-fold.
First, a considerable degree of hindsight gained from this equation
has already been used in section~\ref{inputs}, in order to optimize the ultraviolet features of the 
input functions.
Second, as we will see below, the amount by which
the calculated value for $\mathcal{R}$ deviates from unity favors the use of dressed
rather than bare vertices in the graphs $(d_1)_{\nu\mu}$ and $(d_2)_{\nu\mu}$.

With respect to the first point, note that 
the relation of \1eq{eq:constraint1}, being a direct consequence of the BRST symmetry,
is satisfied exactly at any {\it fixed order} calculation in perturbation theory.
However, in general, our truncation procedure
does {\it not} reduce itself to a fixed order perturbative result, for any limit of the kinematic parameters.
This happens because 
certain of the (higher order) terms, generated after the integration of all ingredients,
ought to 
cancel/combine with contributions stemming from two- and higher-loop dressed diagrams of $H_{\nu\mu}$,
which, evidently, have been omitted from the outset.
The resulting mismatches, in turn, affect unequally the 
different kinematic configurations entering in $\mathcal{R}$, thus distorting the  
subtle balance that enforces Eq.~\eqref{eq:constraint1}.

A concrete manifestation of the underlying imbalances occurs when one uses input 
propagators and vertices containing perturbative information (\eg are of the general form  
$1+ c \alpha_s \log q^2/\mu^2)$. Since one may not intervene in the actual numerical evaluation
and discard ``by hand'' terms of ${\cal O}(\alpha^2_s)$ and higher, 
the final answer contains a certain amount of unbalanced contributions.
The clearest manifestation of this effect occurs when evaluating $\mathcal{R}$
for asymptotically large momenta: contrary to what one might expect, the ``tails'' of ${\mathcal R}$
deviate markedly from unity; in fact, the deviation increases as the momenta grow.


The use of input functions that tend to their tree-level values ameliorates the situation substantially,
because, in this way, the $A_i$ computed display at least their correct one-loop behavior.
This improvement, in turn, must be combined with a judicious choice for  
the $F(p)$ and $F(r)$
appearing explicitly in ${\mathcal R}$ [see \1eq{eq:constraint_ratio}]; in particular,
the function used must display asymptotically the logarithmic behaviour dictated
by one-loop perturbation theory. Specifically, we use the standard fit~\cite{Aguilar:2018epe} 
\be
F^{-1}(q) = 1 + \frac{9C_\mathrm{A}\alpha_s}{48\pi}\left[ 1 + D\exp\left( - \rho_4 q^2\right) \right ]\ln\left( \frac{q^2 + \rho_3 M^2(q)}{\mu^2} \right ) \,, \label{eq:F_UVlogs}
\ee
where
\be
M^2(q) = \frac{m^2_1}{1 + q^2/\rho_2^2} \,,
\ee
with 
\mbox{$m^2_1=0.16\,\mbox{GeV}^2$},  
\mbox{$\rho_2^2=0.69\,\mbox{GeV}^2$}, 
\mbox{$\rho_3=0.89$},
\mbox{$\rho_4=0.12\,\mbox{GeV}^{-2}$},
\mbox{$D=2.36$}, and  
\mbox{$\mu=4.3$\,GeV}.

Then, after these adjustments,
the ``tails'' of $\mathcal{R}$ display only a minuscule deviation from unity, which decreases slowly as the momenta increase.

We next turn to the second point, and consider what the STI constraint suggests regarding the vertices used in the calculation.

Clearly, for any kinematic
configuration where $|p|=|r|$, the numerator 
and the denominator of \1eq{eq:constraint_ratio}
become equal, and \1eq{eq:constraint1} is trivially satisfied. 
In particular, this is precisely what happens 
in the ``soft anti-ghost'' and ``totally symmetric'' limits, presented in the previous subsection.

\begin{figure}[!t]
\begin{minipage}[b]{0.45\linewidth}
\centering
\includegraphics[scale=0.32]{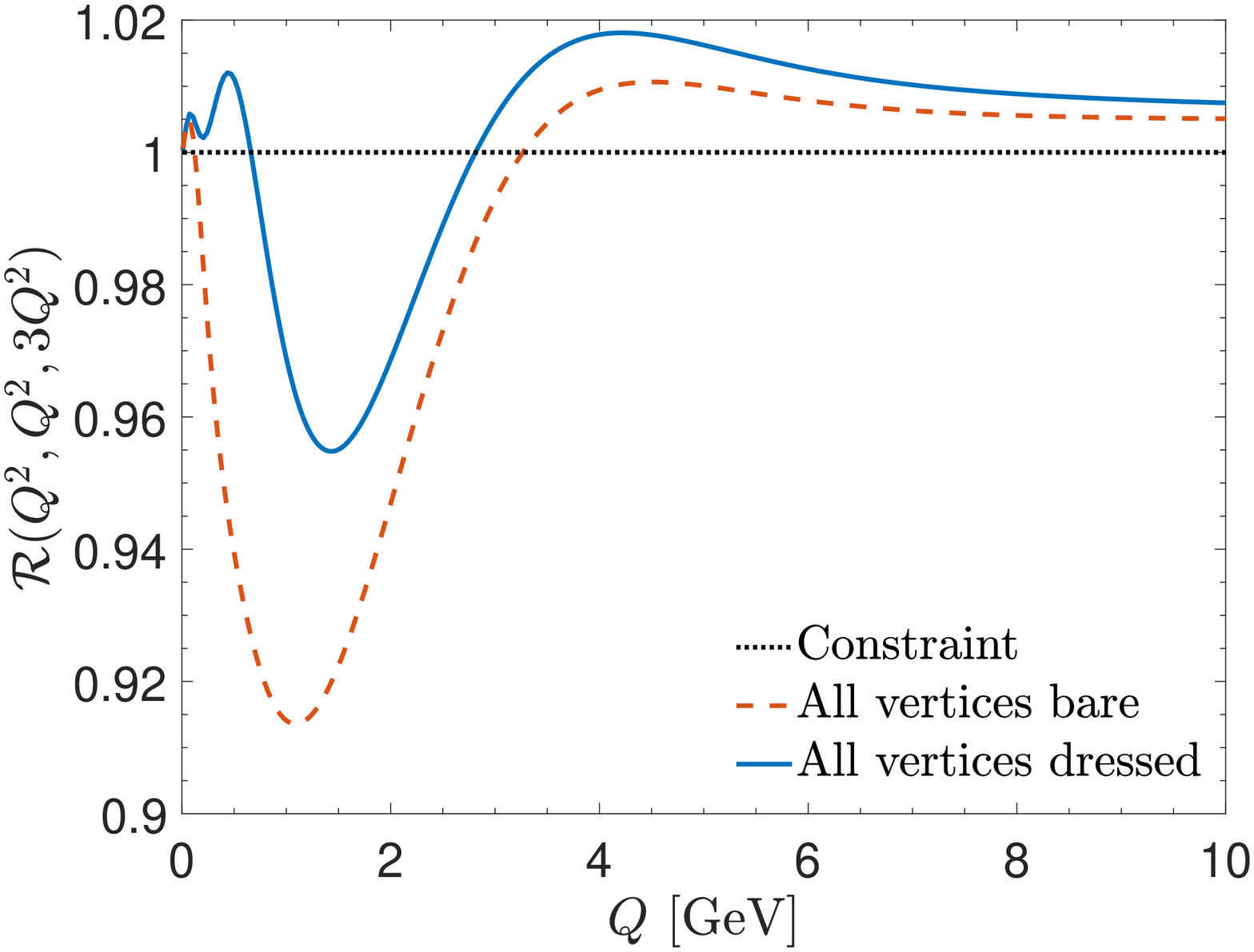}
\end{minipage}
\hspace{0.25cm}
\begin{minipage}[b]{0.45\linewidth}
\includegraphics[scale=0.32]{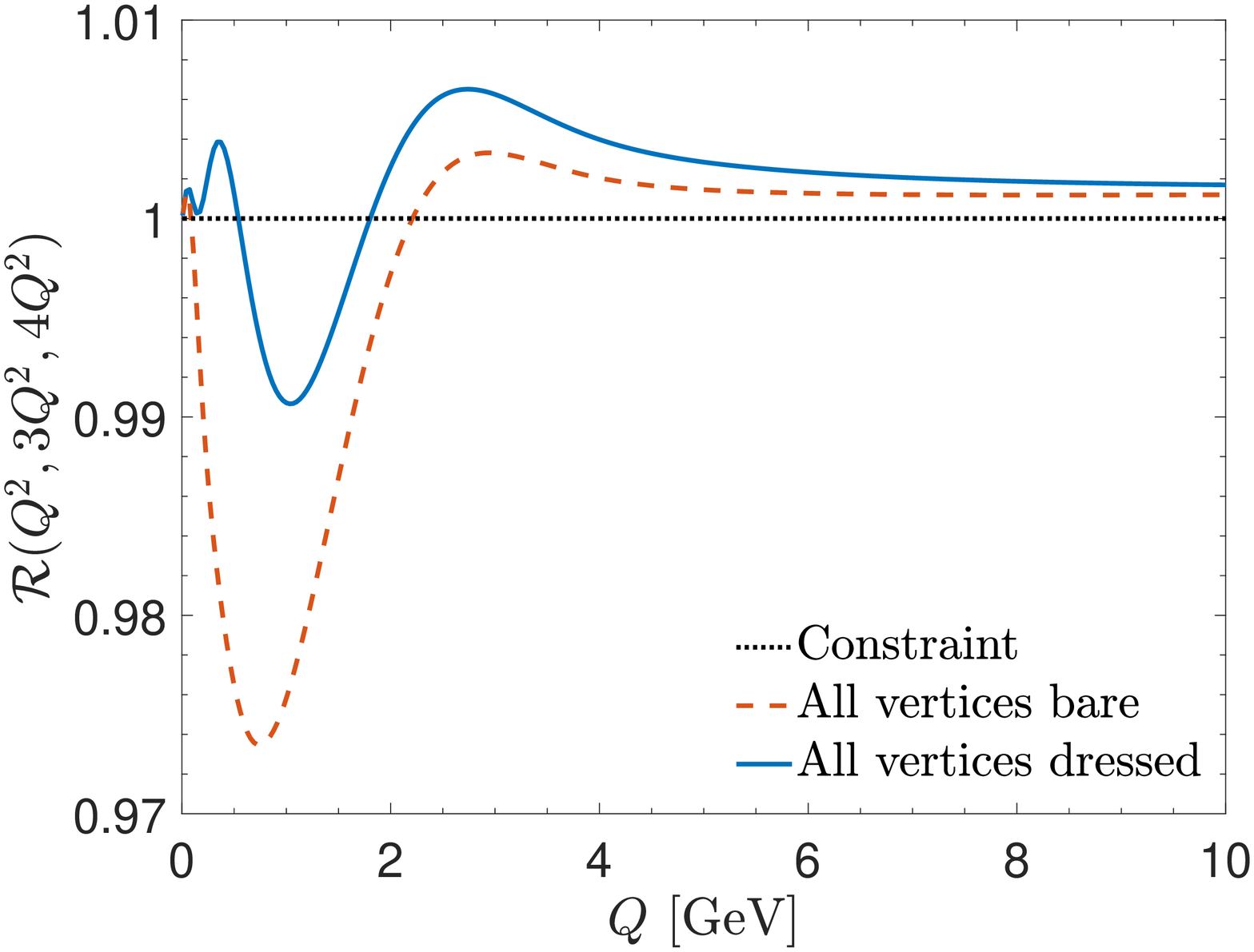}
\end{minipage}
\caption{ The ratio $\mathcal{R}(q^2,p^2,r^2)$, defined in Eq.~\eqref{eq:constraint_ratio} evaluated in two different kinematic
limits:  ({\it i})  $p^2 = q^2=Q^2$ and $r^2 = 3 Q^2$ (left panel) and 
({\it ii}) $q^2 =Q^2$, $p^2=3 Q^2$ and $r^2 = 4 Q^2$ (right panel). The blue continuous curve
represents the case where the $A_i$ are computed using all vertices
dressed, whereas the red dashed one is obtained when we employ bare
vertices. The black dotted line represents the exact value   
$\mathcal{R}(q^2,p^2,r^2)=1$, imposed by the STI.}
\label{fig:constraint_ab_13}
\end{figure}

Let us then consider two different kinematic limits, for which Eq.~\eqref{eq:constraint1} is 
not trivially fulfilled. Specifically, 
we compute $\mathcal{R}$ for two particular kinematic configurations, shown
in  Fig.~\ref{fig:constraint_ab_13}: ({\it i}) on the left panel we present
$\mathcal{R}$ when $p^2 = q^2 = Q^2$ and $r^2 = 3Q^2$, or, equivalently, $\theta=\pi/3$; we denote the
corresponding quantity by $\mathcal{R}(Q^2,Q^2,3Q^2)$ [alternatively, $\mathcal{R}(Q^2,Q^2,\pi/3)$)];
({\it ii}) on the right panel,  we present the case $q^2 = Q^2$, $p^2 = 3Q^2$, and $r^2 = 4Q^2$, which corresponds to 
$\theta=\pi/2$; we denote the result by $\mathcal{R}(Q^2,3Q^2,4Q^2)$.
As a reference, in Fig.~\ref{fig:constraint_ab_13} we plot the ideal result $\mathcal{R}(q^2,p^2,r^2)=1$ (black dotted),  
corresponding to the STI constraint of \1eq{eq:constraint1}.

Notice that in both cases we evaluate $\mathcal{R}(q^2,p^2,r^2)$ using two different approximations:  ({\it a}) the $A_i$
are computed using tree-level expressions for the full vertices appearing in the diagrammatic representation of $H^{\nu\mu}$
(red dashed curves), and 
({\it b}) the $A_i$ are computed with all vertices dressed [see \1eq{eq:d1d2}], using the
Ans\"atze discussed in section~\ref{sec:H_sde} (blue continuous curves).

On the left panel, one clearly observes that the maximum deviation from unity occurs for $q$ in the range \mbox{$1.0-1.5$ GeV},
being around $9\%$ when tree-level vertices are used, and dropping below $5\%$ when  
all vertices are dressed.  Then, in the perturbative region, for values of \mbox{$q \ge 5$\,GeV} the deviations
in both cases are smaller than $2\%$.

In the  kinematic configuration presented on the right panel, we notice that the deviations are milder.
Specifically, the maximum deviation appears in the momentum range  \mbox{$0.8$ - $1.1$ GeV}, and is less than $3\%$ when bare vertices are used, 
dropping to less than $1\%$ for dressed vertices. In the ultraviolet the deviation from unity is of the order of  $0.1\%$. 

The difference between the ideal and computed values of $\mathcal{R}$ 
may be quantified by means of a $\chi^2$ test. The test was implemented using the $80$ points of our logarithmic grid, defined in the entire range of momenta, {\ie} \mbox{ $[ 5\times 10^{-5}\,\mbox{{GeV}}^2, 5\times 10^3\!,\mbox{{GeV}}^2]$}. Note that the logarithmically spaced grid furnishes more weight to the  nonperturbative region because it has  a higher concentration of points in the infrared.

For the case of the bare vertices we obtain $\chi^2=0.057$ (left panel)
and $\chi^2=0.004$ (right panel), whereas for the dressed case one has 
$\chi^2=0.021$ (left panel) and $\chi^2=0.001$ (right panel); evidently, these results favor the  truncation scheme where all vertices are dressed. 

Alternatively, one may also use
as indicator of the similarity of the two curves, the integral over the 
absolute value of the difference of them. More specifically, we have evaluated
the following integral
\be
{\mathcal I}_{ab} = \int^{\Lambda_{\rm UV}}_{\Lambda_{\rm IR}} \left|{\mathcal R}(Q^2,aQ^2,bQ^2)-1\right|dQ\,,
\label{absint}
\ee
where the values of $a$ and $b$ are fixed by the choice of momenta in each configuration; for the two examples considered in
Fig.~\ref{fig:constraint_ab_13} we have ($a=1$, $b=3$) and ($a=3$, $b=4$), respectively. 
For the case of the bare vertices we find \mbox{${\mathcal I}_{13}=0.32$} and \mbox{${\mathcal I}_{34}=0.076$}, whereas for the dressed case one has \mbox{${\mathcal I}_{13}=0.29$} and \mbox{${\mathcal I}_{34}=0.068$}.
Evidently, this second indicator displays a slight preference for the truncation scheme where all vertices are dressed, but is considerably less
compelling compared to the  $\chi^2$ case. 
 
\section{\label{sec:ggvertex} Results for the ghost-gluon vertex}

As a direct application of the results obtained for the $A_i$ in the previous section, we  now turn our attention
to the determination of the form factors of the ghost-gluon vertex, for arbitrary Euclidean momenta.
To that end, we use the  exact expressions given by Eq.~\eqref{eq:Bi_from_Ai}, which was derived  
from Eq.~\eqref{eq:H_and_Gamma}~\cite{Aguilar:2009nf}.

\begin{figure}[!ht]
\begin{minipage}[b]{0.45\linewidth}
\centering
\includegraphics[scale=0.37]{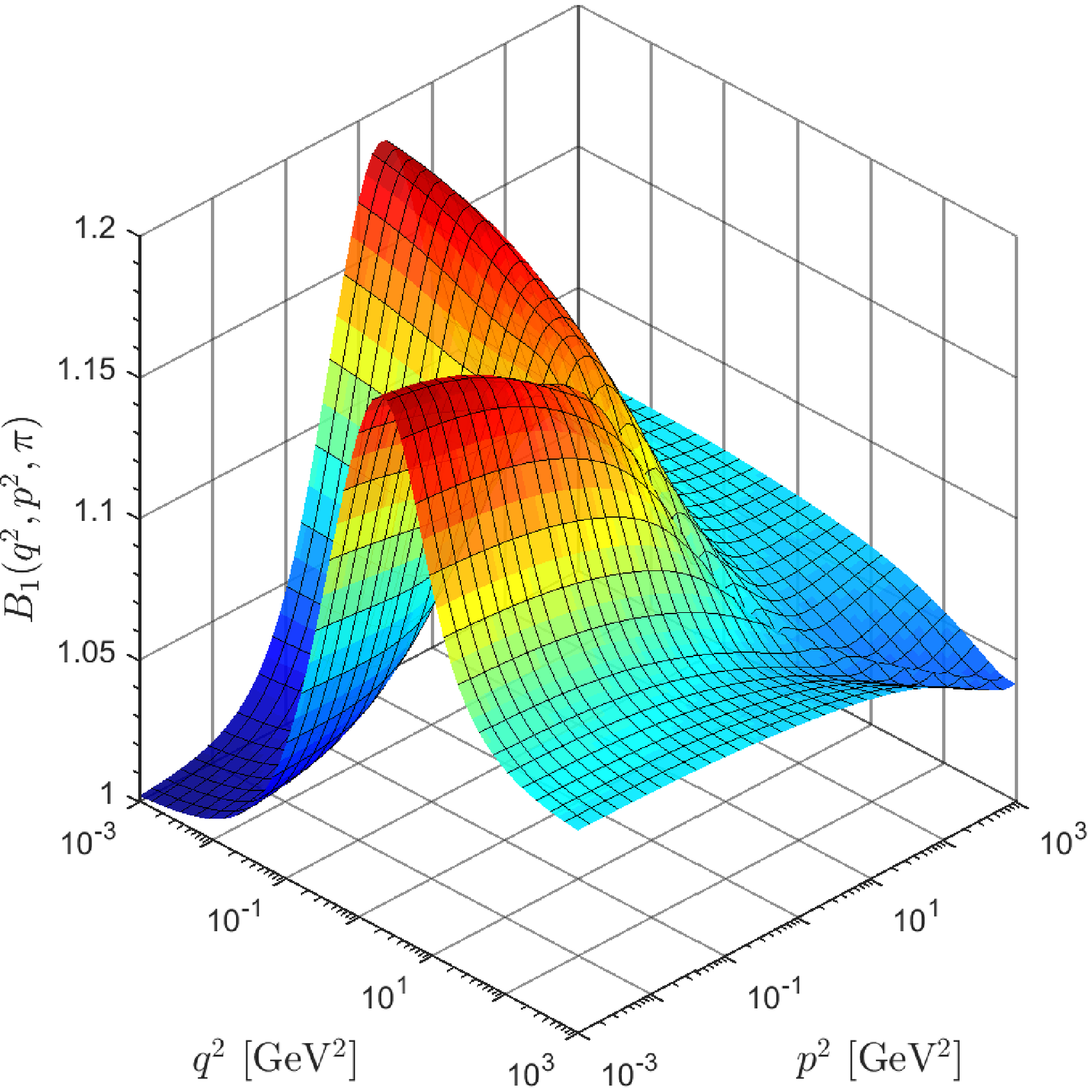}
\end{minipage}
\hspace{0.25cm}
\begin{minipage}[b]{0.45\linewidth}
\includegraphics[scale=0.39]{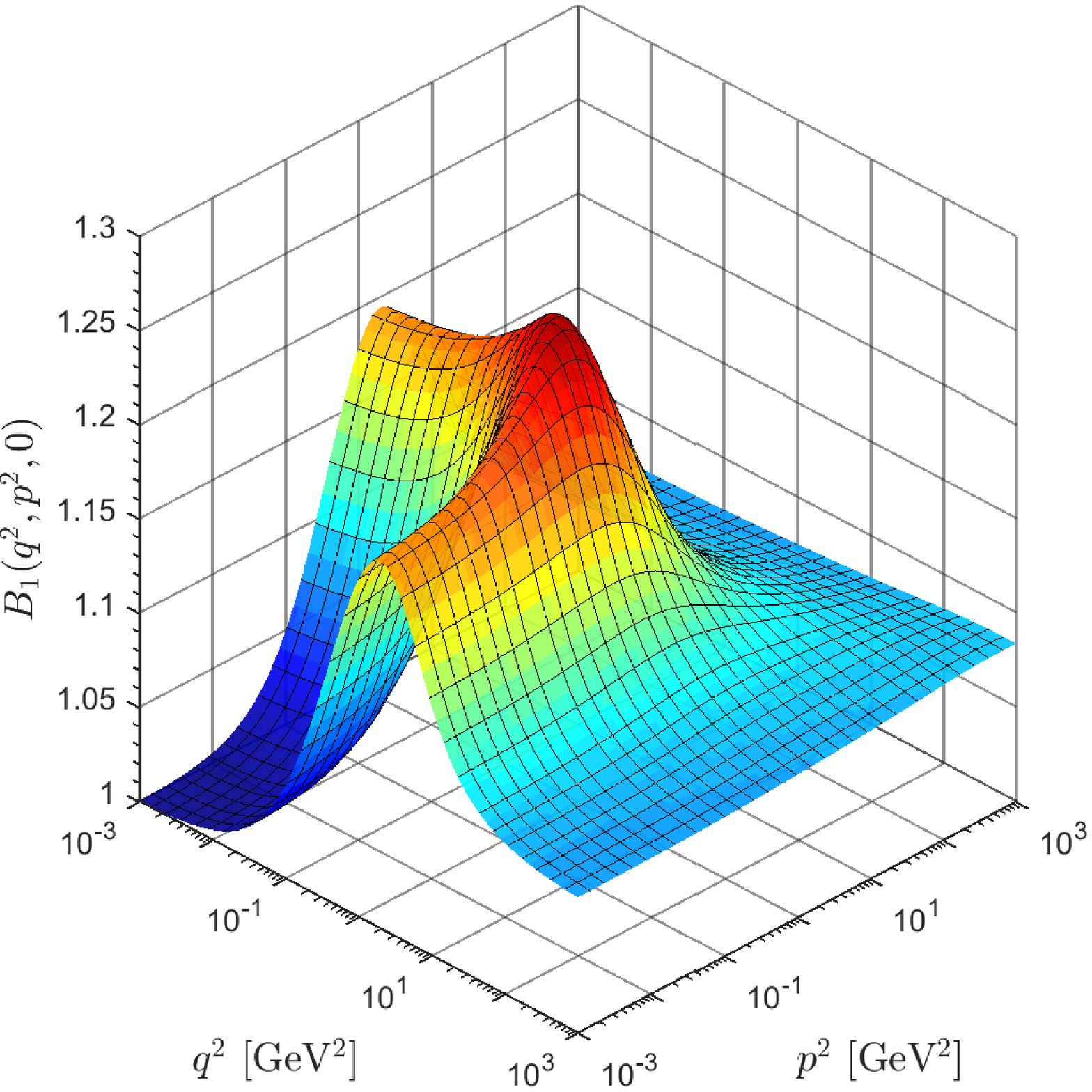}
\end{minipage}
\caption{Form factor $B_1(q^2,p^2,\theta)$ of the ghost-gluon vertex, for $\theta = 0$ (left panel) and $\theta = \pi$ (right panel).}\label{fig:B1_fig}
\end{figure}
\begin{figure}[!ht]
\begin{minipage}[b]{0.45\linewidth}
\centering
\includegraphics[scale=0.37]{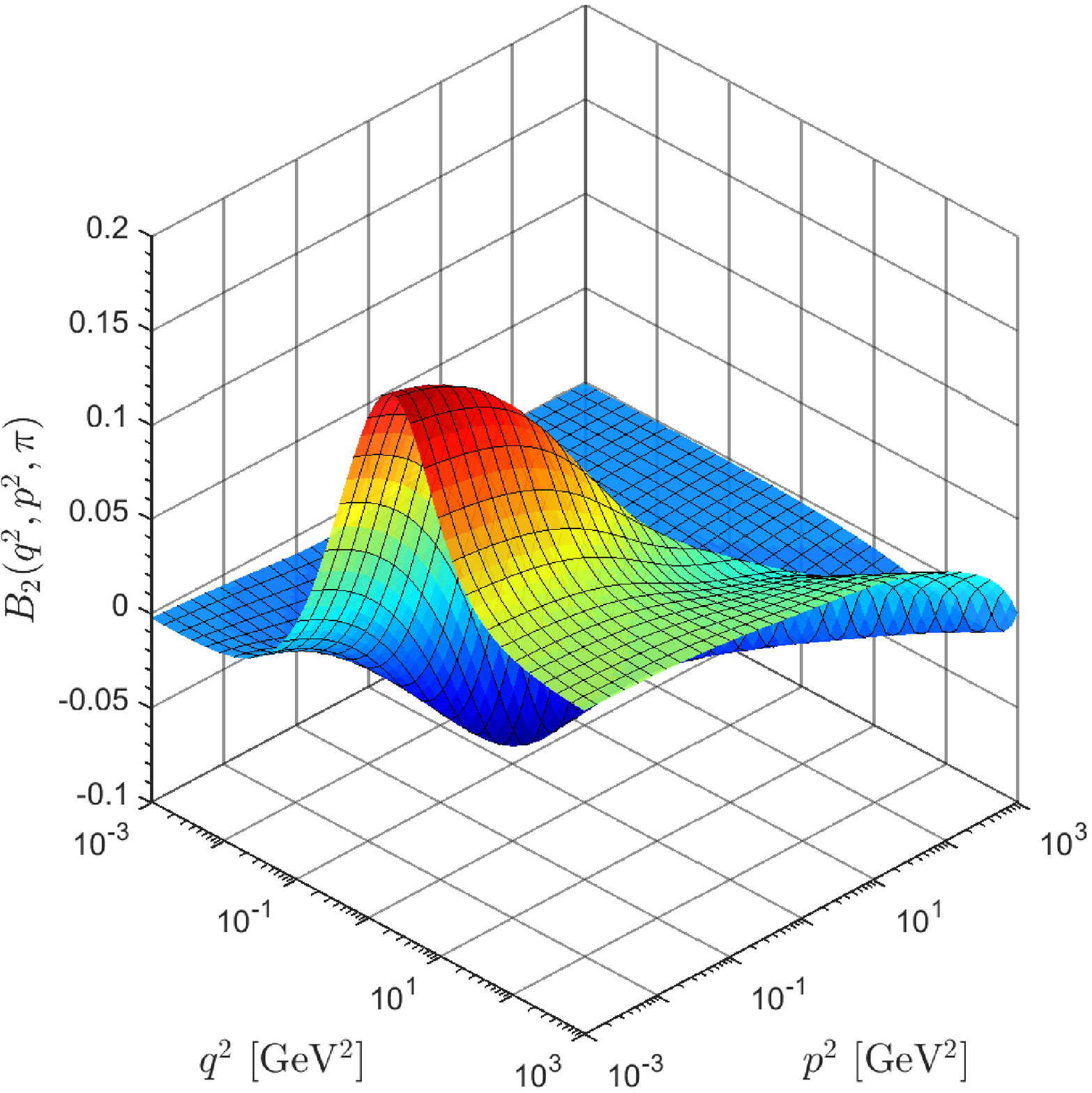}
\end{minipage}
\hspace{0.25cm}
\begin{minipage}[b]{0.45\linewidth}
\includegraphics[scale=0.39]{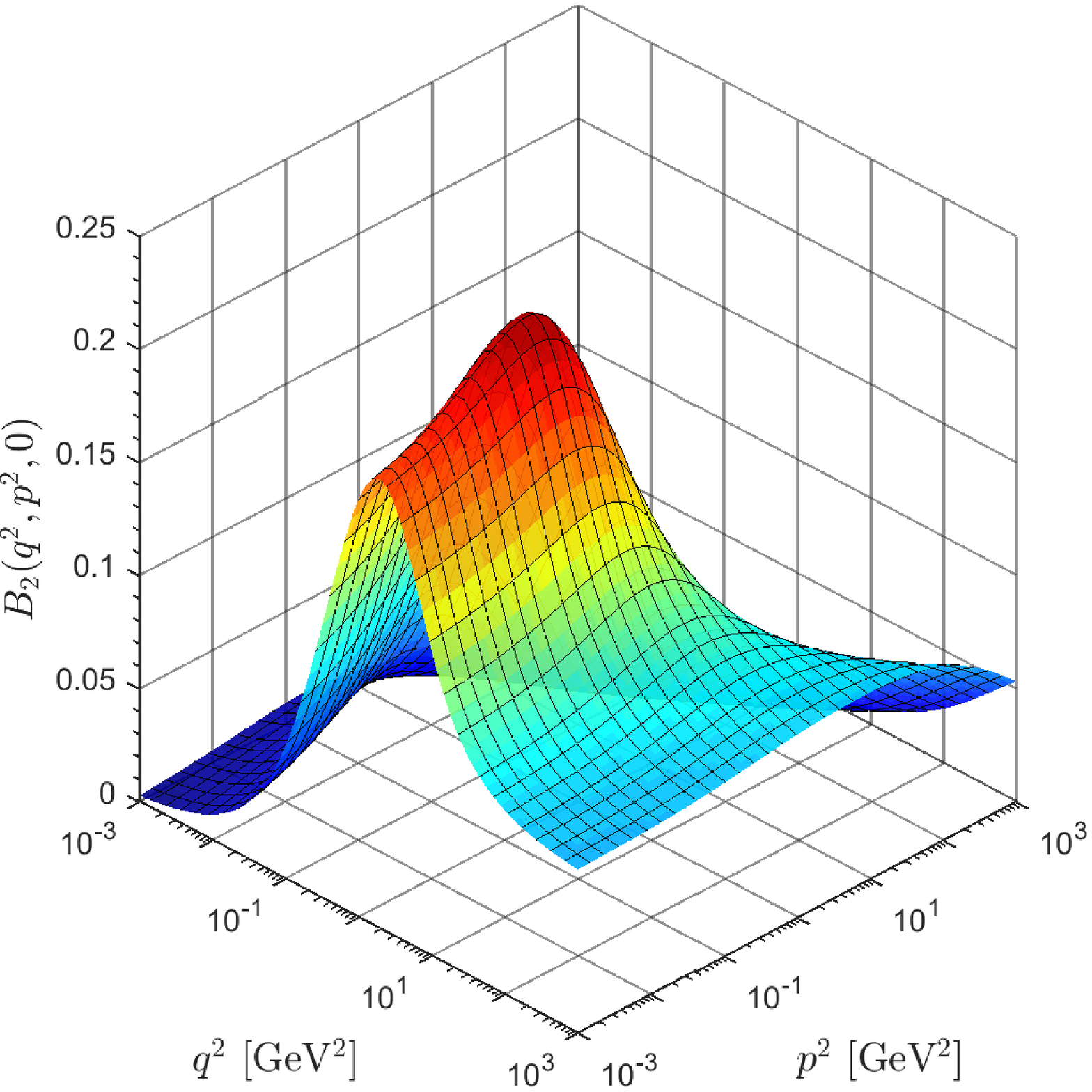}
\end{minipage}
\caption{Form factor $B_2(q^2,p^2,\theta)$ of the ghost-gluon vertex, for $\theta = 0$ (left panel) and $\theta = \pi$ (right panel).}\label{fig:B2_fig}
\end{figure}

\begin{figure}[!ht]
\begin{minipage}[b]{0.45\linewidth}
\centering
\includegraphics[scale=0.32]{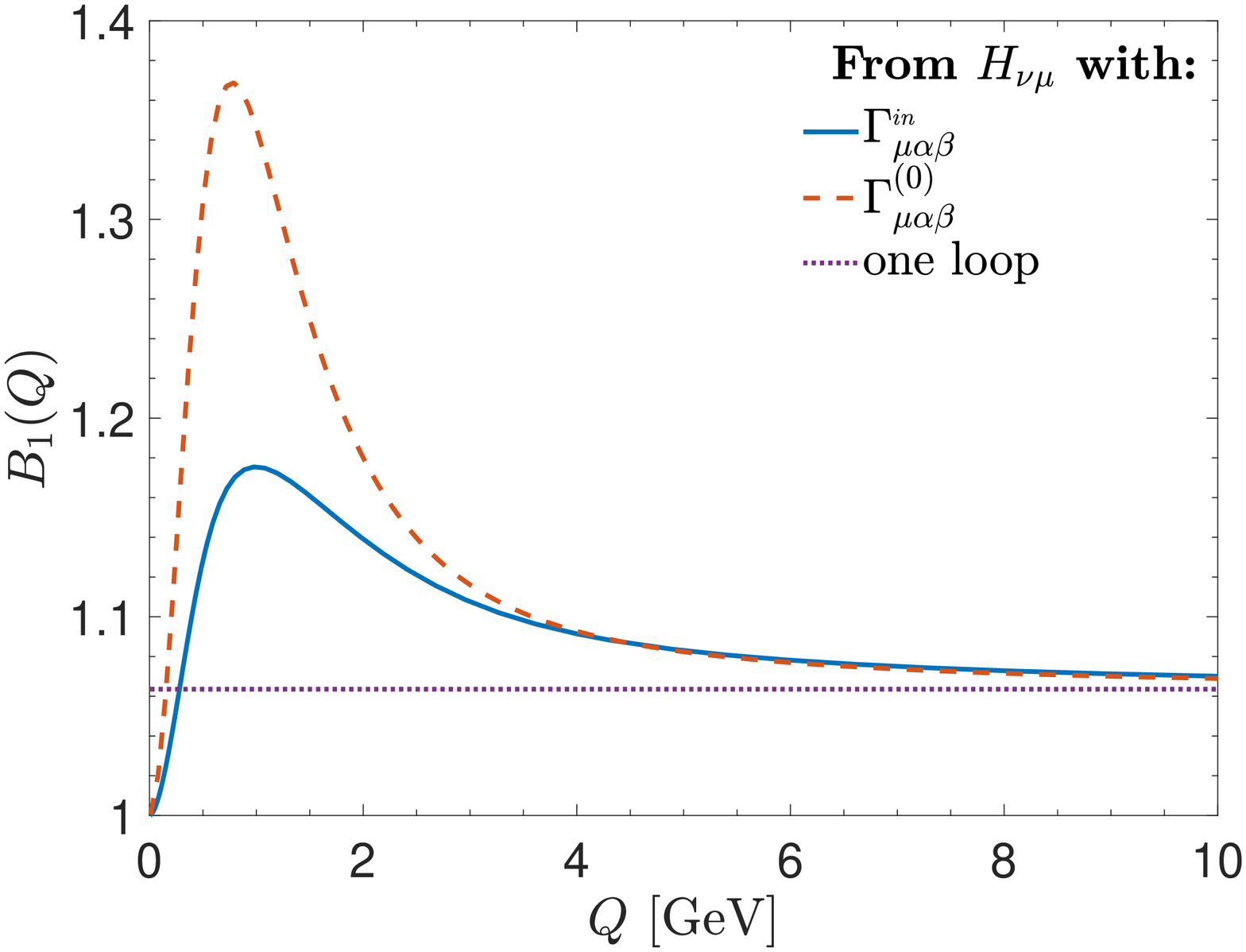}
\end{minipage}
\hspace{0.25cm}
\begin{minipage}[b]{0.45\linewidth}
\includegraphics[scale=0.32]{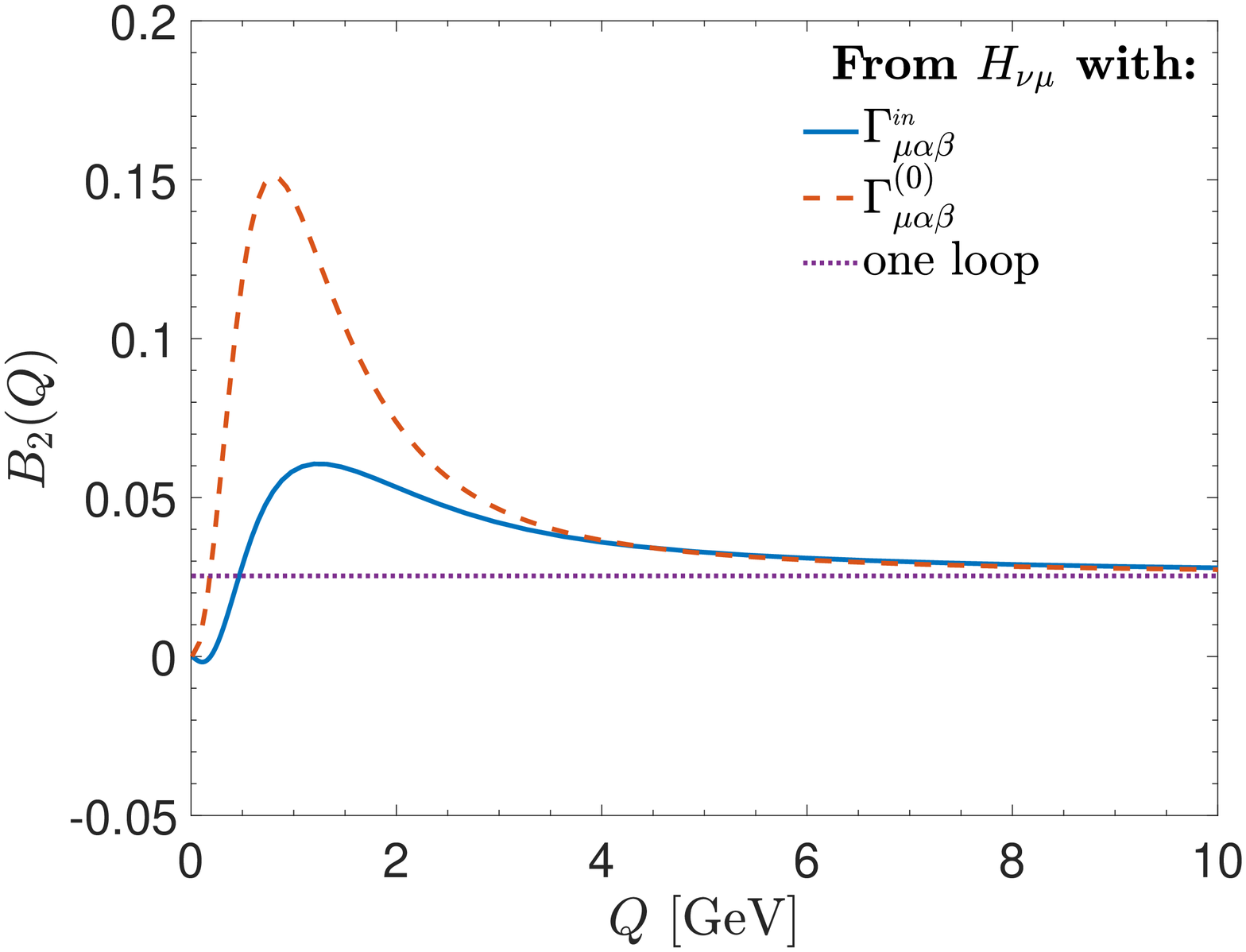}
\end{minipage}
\caption{ $B_1(Q)$ (left panel) and $B_2(Q)$ (right panel)  in the 
totally symmetric configuration  obtained when the $A_i$
entering in the ~Eq.~\eqref{eq:Bi_from_Ai} are computed using the three-gluon vertex dressed given by Eq.~\eqref{eq:three_gluon_s}
(blue continuous) or at tree-level given by Eq.~\eqref{eq:3gluon} (red dashed).     
The one-loop results for $B_1(Q)$ and $B_2(Q)$ (purple dotted) may be directly obtained combining Eqs.~\eqref{eq:Bi_from_Ai} and \eqref{eq:Ai_sym}.}
\label{fig:B1compv}
\end{figure}

In Fig.~\ref{fig:B1_fig} and Fig.~\ref{fig:B2_fig} we show, respectively, the form factors $B_1$ and $B_2$ as functions of $q^2$, $p^2$ and $\theta$.
In order to appreciate their  angular dependence, we present two representative cases:  $\theta = 0$ and $\theta = \pi$.  As we can see,
the angular dependence of $B_1$ is relatively weak, whereas  $B_2$ is clearly more sensitive to changes in $\theta$.
Note also that both form factors approach to their one-loop perturbative behaviour\footnote{Notice that the one-loop behavior for $B_1$ in the soft ghost, soft gluon, and totally symmetric configurations
deviates slightly from $1$, being $1.07$, $1.04$ and $1.06$, respectively. 
The corresponding relative errors between  our nonperturbative
computation and the expected one-loop behavior 
are smaller than $1\%$ for  momenta higher than \mbox{$8$\,GeV}, in the three
kinematic configurations mentioned - see for example Fig.~\ref{fig:B1compv}.}
whenever one of the ghost ($p$)  or anti-ghost ($q$) momenta becomes large. 

 In addition, for $p^2 = q^2 = 0$ they revert to their tree-level values, 
due to the fact that the one-loop dressed contributions to $H_{\nu\mu}$ vanish at the origin.
Moreover, we may visually verify that $B_1(q^2,p^2,\theta)$ is symmetric
under the exchange $q^2 \leftrightarrow p^2$, for any $\theta$, as required by the ghost-anti-ghost symmetry.

It is clear that $B_1$ and $B_2$ will depend through the $A_i$ on our choice for $\Gamma_{\mu\alpha\beta}$. 
In order to study this effect, 
we employ the results presented
in the section~\ref{spec},  where the $A_i$ were computed using as input
for $\Gamma_{\mu\alpha\beta}$  either the $\Gamma^{(0)}_{\mu\alpha\beta}$ of Eq.~\eqref{eq:3gluon} or
the $\Gamma^{\inpt}_{\mu\alpha\beta}$ of Eq.~\eqref{eq:three_gluon_s}.
In Fig.~\ref{fig:B1compv} we show  the results of this study for $B_1(Q)$ and $B_2(Q)$ in the totally symmetric configuration.
Clearly, when  the three-gluon vertex is dressed, 
the results for $B_1$ and $B_2$ are systematically suppressed. Notice that the relative difference is more pronounced 
in the  intermediate region of momenta, given that in the deep infrared we must have $B_1(0,0,0)=1$ and $B_2(0,0,0)=0$,
while in the ultraviolet $B_1$ and $B_2$ should recover the expected perturbative behaviour.  
In particular, around the region of 
\mbox{$0.9$ -$1.1$ GeV}, the deviations of $B_1$ and $B_2$ from their tree-level values are approximately $2-2.5$ times larger when
$\Gamma^{(0)}_{\mu\alpha\beta}$ is used.

\begin{figure}[!ht]
\includegraphics[scale=0.35]{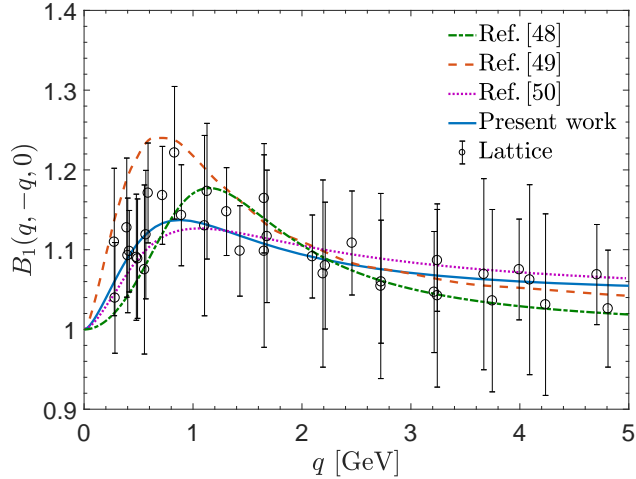}
\caption{ Our numerical result for $B_1(q,-q,0)$ (blue
continuous) compared with the results of~\cite{Boucaud:2011eh} (green
dashed-dotted),~\cite{Huber:2012kd} (red dashed), and~\cite{Mintz:2017qri} (magenta dotted). The lattice data  (circles) are from~\cite{Ilgenfritz:2006he,Sternbeck:2006rd}.}\label{fig:B1comp}
\end{figure}

Next, in Fig.~\ref{fig:B1comp}, we compare our results for $B_1$ in the soft gluon configuration
with those obtained in earlier works~\cite{Boucaud:2011eh,Huber:2012kd,Mintz:2017qri}; this 
configuration is the most widely explored in the literature, being the only one 
simulated on lattice for $\rm SU(3)$~\cite{Ilgenfritz:2006he,Sternbeck:2006rd}. The green dashed-dotted curve
represents the results for $B_1(q,-q,0)$, obtained from the approach developed in~\cite{Boucaud:2011eh}, based on the infrared completion of expressions derived using operator product expansion techniques. In the case of~\cite{Huber:2012kd}, $B_1$ was determined in general kinematics,
using a system of coupled SDEs, while in~\cite{Mintz:2017qri} the $B_1$ was determined exclusively in the soft gluon configuration.
It is interesting to notice that all analytical studies display the characteristic peak and converge to unity at the origin. Moreover, all of
them are in qualitative agreement with the lattice data (note, however, that the error bars are quite sizable).

\begin{figure}[!ht]
\begin{minipage}[b]{0.45\linewidth}
\centering
\includegraphics[scale=0.32]{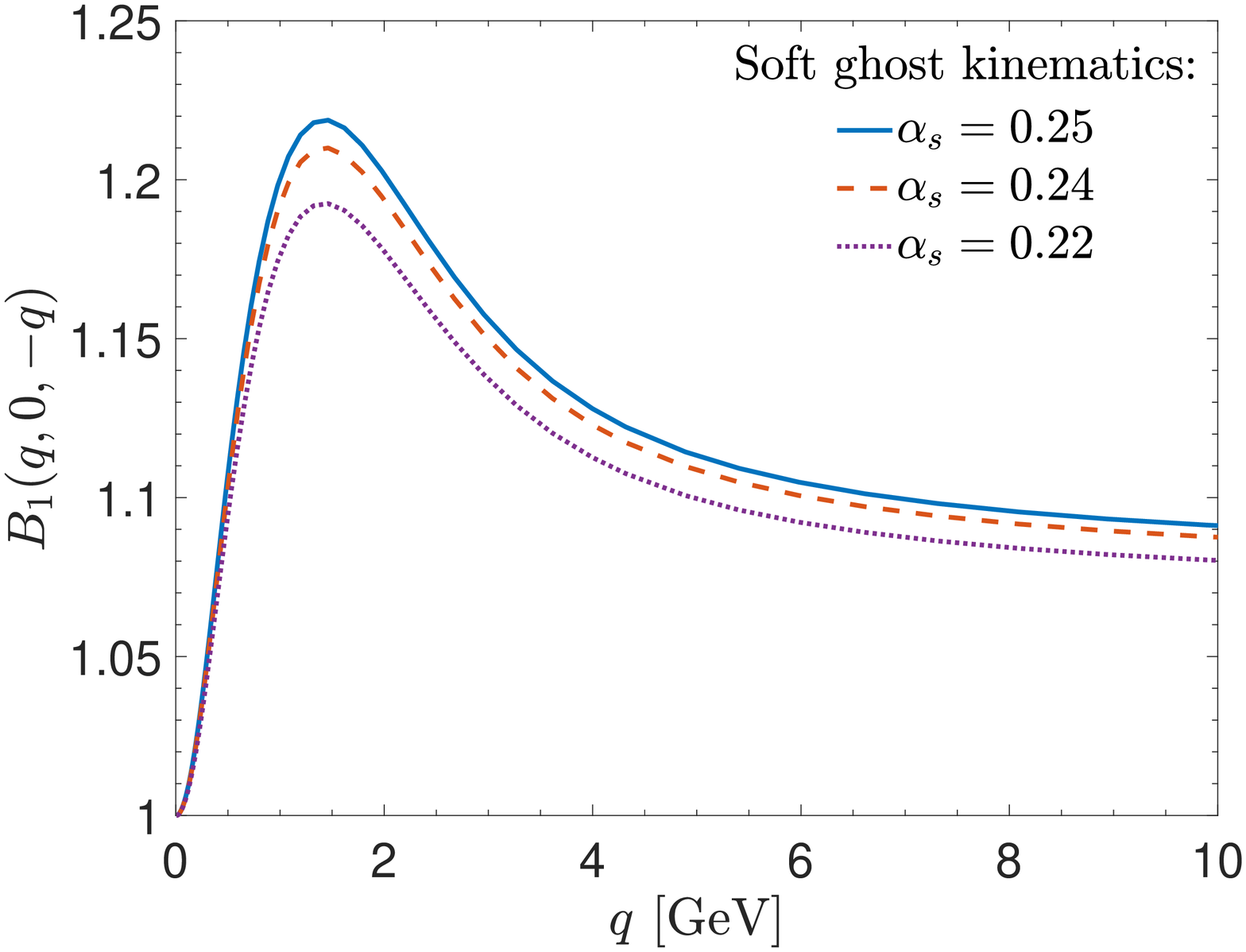}
\end{minipage}
\hspace{0.25cm}
\begin{minipage}[b]{0.45\linewidth}
\includegraphics[scale=0.32]{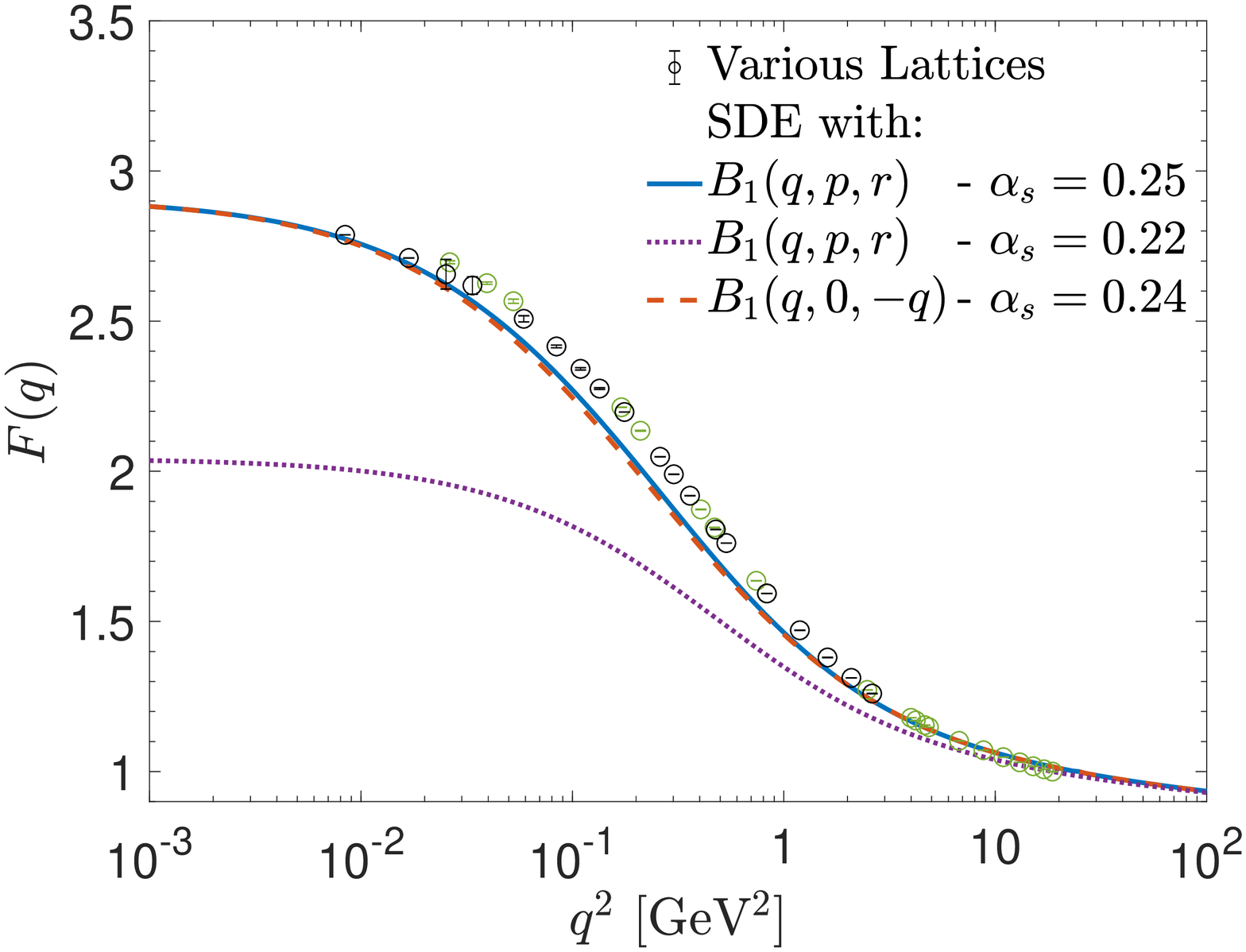}
\end{minipage}
\caption{The $B_1(q,0,-q)$ computed using three different values of
$\alpha_s$ (left panel). 
The $F(q)$ obtained by substituting into the ghost SDE: ({\it a})  
 the full $B_1(q,p,r)$ with $\alpha_s = 0.25$  (blue continuous), ({\it b})  the full $B_1(q,p,r)$ with $\alpha_s = 0.22$  (purple dotted), 
and ({\it c}) the soft gluon limit $B_1(q,0,-q)$ with  \mbox{$\alpha_s = 0.24$} (red dashed). 
 The lattice data are from~\cite{Bogolubsky:2007ud,Bogolubsky:2009dc} (right panel).
\label{fig:ghostSDE}}
\end{figure}

Finally, 
in Fig.~\ref{fig:ghostSDE}, we illustrate the impact 
that the full structure of $B_1(q^2,p^2,\theta)$ has on the SDE of the ghost dressing function. To that end, we
explore two scenarios:  ({\it i}) we  couple the entire momenta dependence of $B_1$ to the SDE for $F(q)$, 
carrying out the additional angular integration [see Eq.~(2.14) of~\cite{Aguilar:2013xqa}], and ({\it ii} ) we fix its momentum dependence to  
the soft ghost configuration~\cite{Dudal:2012zx,Aguilar:2013xqa}. 
We observe that with mild adjustments to the value of $\alpha_s$,
both scenarios  reproduce the standard lattice results of~\cite{Bogolubsky:2007ud} rather accurately;  in particular, 
while for the case ({\it i}) $\alpha_s=0.25$, for ({\it ii}) we obtain $\alpha_s=0.24$.

The reason for this small difference 
in the values of $\alpha_s$ can be easily understood. As mentioned in section~\ref{spec},
in the region of momenta of about two to three times the QCD
mass scale,  the soft ghost configuration maximizes the deviation from the tree-level value. 
Therefore, when we  approximate the entire momentum dependence of $\Gamma^{\mu}$ just by this configuration (instead of integrating over all
of them), we slightly overestimate the contribution of the ghost-gluon vertex to the ghost SDE.
 
It is also interesting to notice that, although the impact of changing the value of $\alpha_s$
is rather mild at the level of $B_1(q,0,-q)$,
it is rather pronounced when $F(q)$ is computed, as clearly seen in  Fig.~\ref{fig:ghostSDE}.
More specifically, the relative difference  between the $B_1(q,0,-q)$  computed with \mbox{$\alpha_s= 0.25$} and that
computed with \mbox{$\alpha_s= 0.22$} is less than $3\%$ around the region of the peaks. Instead, 
 in the case of $F(q)$, the relative difference between the corresponding curves increases to $30\%$ in the deep infrared; this, in turn, may be traced back to the known high sensitivity of the ghost SDE on the value of $\alpha_s$.

\section{\label{sec:conc} Conclusions}

We have presented a detailed nonperturbative study of the form factors, $A_i$, 
comprising the ghost-gluon kernel, $H_{\nu\mu}$,  using the ``one-loop dressed''
approximation of its dynamical equation, for general Euclidean momenta.
The results obtained have been presented in 3-D plots, and certain ``slices'',
corresponding to special kinematic limits, have been singled out and
inspected in detail. The $A_i$ obtained have been subsequently used for
the determination of the two form factors, $B_1$ and $B_2$, of the ghost-gluon vertex.

The ingredients entering in the calculations are the gluon and ghost propagators,
and the vertices $\Gamma_{\alpha\mu\nu}$ and $\Gamma_{\mu}$. Given that the
$H_{\nu\mu}$ itself is intimately connected to both these vertices,
a strictly self-consistent treatment would require to couple the dynamical equation governing $H_{\nu\mu}$
to the equations relating it to both $\Gamma_{\alpha\mu\nu}$ and $\Gamma_{\mu}$, and proceed to the
solution of the entire coupled system. Instead, we have treated the problem at hand by 
employing simplified versions of these vertices, whose use in recent studies~\cite{Binosi:2017rwj,Aguilar:2017dco}
yielded satisfactory results. Moreover, 
as has been explained in detail, there exists a subtle interplay between the truncation of the equations employed, the ultraviolet
behavior of the ingredients used for their evaluation, and the accuracy with which the resulting $A_i$ satisfy 
the STI constraint of \1eq{eq:constraint1}. Note in particular that while our input expressions for the two-point functions are in excellent
agreement with the lattice data of~\cite{Bogolubsky:2007ud} for infrared and intermediate momenta, their ultraviolet tails have been
adjusted to their tree-level values. 

We have paid particular attention to the impact that the structure of $\Gamma_{\alpha\mu\nu}$ may have on the results.
All our findings indicate that the use of a dressed $\Gamma_{\alpha\mu\nu}$, corresponding to the
so-called ``minimal BC solution'', ${\Gamma}^{\inpt}_{\mu\alpha\beta}$, induces an appreciable suppression with respect
to the results obtained by merely resorting to $\Gamma_{\alpha\mu\nu}^{(0)}$.
This happens because the form factor $X^{\inpt}_1$ 
is itself suppressed in the infrared, due to the form of the functions $J(q)$ that enter its definition [see \1eq{Xmin}]. 
This special feature of the three-gluon vertex, in turn,
appears to be favored by the STI-derived constraint, in the sense that the results obtained with ${\Gamma}^{\inpt}_{\mu\alpha\beta}$
are considerably closer to unity (see Fig.~\ref{fig:constraint_ab_13}).  

The information obtained on the structure of the ghost-gluon kernel 
opens the way towards the systematic nonperturbative construction of the
10 form factors comprising the ``longitudinal'' part of the three-gluon vertex, using 
the BC construction~\cite{Ball:1980ax} as a starting point.
The detailed knowledge of these form factors, in turn,
may have considerable impact on the study of the dynamical formation of gluon dominated bound states, such as 
glueballs and hybrids (see, \eg~\cite{Xu:2018cor}, and references therein).
We hope to be able to present results on this topic in the near future. 

\newpage

\appendix

\section{\label{app:pert} One-loop results for special kinematic configurations}

In this Appendix we present the one-loop results for the various $A_i$ in
the three special kinematic configurations considered in subsection~\ref{spec}~\cite{Celmaster:1979km, Davydychev:1996pb}. 
In addition, for two kinematic limits, we also show the corresponding results for the one-loop massive case, where the gluon propagator is endowed with a hard mass $m$. 
 
The one-loop calculations are performed {\it analytically}; the starting expressions may be obtained
from the $(d_1)_{\nu\mu}$ and $(d_2)_{\nu\mu}$ of \1eq{eq:d1d2}
by replacing the dressed quantities 
by their tree-level counterparts, {\it i.e.} \mbox{$D(q)=i/q^2$}, \mbox{${\mathcal V}_1={\mathcal V}_2=1$},
and \mbox{$J(q)=1$}.   In the case of the gluon propagator, the corresponding tree-level expressions used are either
\mbox{$\Delta(q)=1/q^2$} (for the ``conventional'' one-loop calculation) or \mbox{$\Delta^{-1}(q)=q^2-m^2$} (for the one-loop massive calculation). 

In addition, we employ  dimensional regularization, in which case 
the measure of \1eq{eq:d1d2} assumes the form
\be
\int_{\ell} \rightarrow \frac{\mu^{\epsilon}}{(2\pi)^{d}}\!\int\!\mathrm{d}^d \ell,
\label{dimreg}
\ee
where $d=4-\epsilon$, and $\mu$ is the 't Hooft mass scale.

Implementing the substitutions mentioned above, we obtain for the conventional one-loop case   
\begin{align}
\label{eq:1loop}
(d_1)_{\nu\mu}^{(1)}&=\frac{ig^2C_\mathrm{A}}{2}\int_{\ell} \frac{(\ell-q)_\mu}{\ell^2(\ell+p)^2(\ell-q)^2}\left[p_\nu -\ell_\nu\frac{(p\cdot \ell)}{\ell^2}\right]\,, \\
(d_2)_{\nu\mu}^{(1)}&=\frac{ig^2C_\mathrm{A}}{2}\int_{\ell}\frac{\Gamma_{\mu\sigma\alpha}^{(0)}}{\ell^2(\ell+r)^2(\ell-q)^2}\left[g^{\sigma}_\nu-\frac{\ell^\sigma \ell_\nu}{\ell^2}\right]\left[p^{\alpha}-\frac{p\cdot(\ell+r)(\ell+r)^\alpha}{(\ell+r)^2}\right]\,, \nonumber 
\end{align}
while for the one-loop massive case one has 
\begin{align}
\label{eq:1loopm}
(d_1)_{\nu\mu}^{\rm{(\Mass)}}&=\frac{ig^2C_\mathrm{A}}{2}\int_{\ell} \frac{(\ell-q)_\mu}{(\ell^2-m^2)(\ell+p)^2(\ell-q)^2}\left[p_\nu -\ell_\nu\frac{(p\cdot \ell)}{\ell^2}\right]\,, \\
(d_2)_{\nu\mu}^{(\Mass)}&=\frac{ig^2C_\mathrm{A}}{2}\int_{\ell}\frac{\Gamma_{\mu\sigma\alpha}^{(0)}}{(\ell^2-m^2)[(\ell+r)^2-m^2](\ell-q)^2}\left[g^{\sigma}_\nu-\frac{\ell^\sigma \ell_\nu}{\ell^2}\right]\left[p^{\alpha}-\frac{p\cdot(\ell+r)(\ell+r)^\alpha}{(\ell+r)^2}\right]\,, \nonumber 
\end{align}
where \mbox{$\Gamma_{\mu\sigma\alpha}^{(0)} = (2r+l)_\alpha g_{\mu\sigma}-(2l+r)_\mu g_{\alpha\sigma}+(l-r)_\sigma g_{\mu\alpha}$}. 

For numerical 
purposes, the mass  appearing in the one-loop massive calculation will be fixed \mbox{$m^2=0.15\, \mbox{GeV}^2$}. This value  coincides with the dynamical gluon mass at zero momentum, namely the value of $\Delta^{-1}(0)$, for \mbox{$\mu= 4.3$ GeV} [see Eqs.~\eqref{eq:gluon_m_J_euc} and~\eqref{eq:dyn_mass}].

Below we summarize the results (in Euclidean space) obtained
after introducing the Feynman parametrization and
using the \mbox{Package-X~\cite{Patel:2015tea,Patel:2016fam}}. 

\begin{enumerate}
\item {\emph{Soft gluon limit:}} To derive this configuration, we set $r\to 0$ directly into Eqs.~\eqref{eq:H} and~\eqref{eq:1loop}. It is straightforward to see that in this limit the tensorial structure of $H^{(1)}_{\nu\mu}(q,0)$ given by~\eqref{eq:H} reduces to 
\be
H^{(1)}_{\nu\mu}(q,-q,0) = A_1^{(1)}(q,-q,0)g_{\mu\nu} + A_2^{(1)}(q,-q,0)q_{\mu}q_{\nu}\,,
\label{eq:sgluonH}
\ee
and the form factors become  
\begin{enumerate}
\item {\emph{One loop:}}
\begin{align}
A_1^{(1)}(q,-q,0) = 1 \,; \qquad A_2^{(1)}(q,-q,0)=-\frac{3C_\mathrm{A}\alpha_s}{16\pi q^2} \,.
\label{eq:Ai_sgl}
\end{align}
Then, the one-loop result for $B_1^{(1)}(q,-q,0)$ may be directly obtained using Eq.~\eqref{eq:Bi_from_Ai}.

\item {\emph{One loop massive :}}
\begin{align}
A_1^{(\Mass)}(q,-q,0) =& 1 -\frac{ C_\mathrm{A} \alpha_s }{192 \pi  m^4 q^4} \left[ ( 10 m^8 + 8 m^6 q^2 ) \ln \left( \frac{m^2}{m^2 + q^2} \right) +10 m^6 q^2+ \right. \nonumber\\ 
&  \left.  3 m^4 q^4+2 m^2 q^6 
 + ( 4 m^2 q^6 - 2 q^8 ) \ln \left( \frac{q^2}{m^2 + q^2} \right)  \right] \,,  \nonumber\\
A_2^{(\Mass)}(q,-q,0) =& -\frac{ C_\mathrm{A} \alpha_s }{96 \pi m^4 q^6} \left[ 20 m^6 q^2 + 15 m^4 q^4 + q^6 \left( 4 q^2 - m^2 \right) \ln \left(\frac{q^2}{m^2}\right) \right. \nonumber\\
& \left. + \left(20 m^8+25 m^6 q^2-m^2 q^6+4 q^8\right) \ln \left(\frac{m^2}{m^2+q^2}\right)  + 4 m^2 q^6  \right] \,. 
\label{1lMsg}
\end{align}
 Evidently, in the  limit $m\to 0$, one recovers the one-loop results given by Eq.~\eqref{eq:Ai_sgl}.
 Moreover, in the infrared limit $q\to 0$, Eq.~\eqref{1lMsg} reduces to
\begin{align}
\lim_{q^2\to 0} A_1^{(\Mass)}(q,-q,0) =& 1  \,, \nonumber \\
\lim_{q^2\to 0}A_2^{(\Mass)}(q,-q,0) =& \frac{ C_\mathrm{A} \alpha_s }{ 576 \pi m^2 } \left[ 6 \ln \left( \frac{q^2}{m^2} \right) - 59 \right] \,.
\label{sgmass0}
\end{align}
Therefore, in the massive one-loop analysis, one finds that $A_1^{(\Mass)}$ is infrared finite, whereas  $A_2^{(\Mass)}$ displays a logarithmic divergence.

\end{enumerate}

\item{\emph{Soft anti-ghost limit:}} This limit is obtained by setting $q=0$. The one-loop expression for $H^{(1)}_{\nu\mu}$ becomes 
\be
H^{(1)}_{\nu\mu}(0,-r,r) = A_1^{(1)}(0,-r,r)g_{\mu\nu} + A_3^{(1)}(0,-r,r)r_{\mu}r_{\nu} \,,
\ee
with the two form factors given by: 
\begin{enumerate}
\item {\emph{One loop:}} 
\begin{align}
A_1^{(1)}(0,-r,r) = 1 +\frac{11C_\mathrm{A}\alpha_s}{32\pi} \,;  \qquad A_3^{(1)}(0,-r,r) = \frac{11C_\mathrm{A}\alpha_s}{32\pi r^2} \,.\label{eq:Ai_sagh}
\end{align}
\item {\emph{One loop massive:}}
\begin{align}
A_1^{(\Mass)}(0,-r,r) =& 1 -\frac{ C_\mathrm{A} \alpha_s }{192 \pi  m^6 r^4} \left[ 2 m^8 r^2-23 m^6 r^4 + r^6 \left( 2 m^4 - 6 m^2 r^2+r^4 \right) \ln \left( \frac{r^2}{m^2}\right) \right. \nonumber\\
& \hspace{-1.5cm}+ ( r^9 - 6 m^2 r^7 - 40 m^4 r^5 ) \sqrt{4 m^2+r^2} \ln \left[\frac{ \left(\sqrt{4 m^2r^2+r^4}+r^2\right)}{2 m^2}+1\right] \nonumber\\
& \hspace{-1.5cm}\left. + 2 \left(m^2+r^2\right)^2 \left(m^6-13 m^4 r^2-7 m^2 r^4+r^6\right) \ln
   \left(\frac{m^2}{m^2+r^2}\right) \right] \,, \nonumber \\
A_3^{(\Mass)}(0,-r,r) =& - \frac{ C_\mathrm{A} \alpha_s }{192 \pi  m^6 r^6} \left[ 8 m^8 r^2-20 m^6 r^4+6 m^4 r^6 + r^6 \left(8 m^4+r^4\right) \ln \left(\frac{r^2}{m^2}\right) \right. \nonumber\\
&  \hspace{-1.5cm} + ( r^9 - 6 m^2 r^7 - 40 m^4 r^5) \sqrt{4 m^2+r^2} \ln \left[\frac{ \left(\sqrt{4 m^2r^2+r^4}+r^2\right)}{2 m^2}+1\right] \nonumber\\
& \hspace{-1.5cm}\left. + 2 \left(m^2+r^2\right)^2 \left(4 m^6-16 m^4 r^2-4 m^2 r^4+r^6\right) \ln\left(\frac{m^2}{m^2+r^2}\right) \right] \,.
\label{1lmag}
\end{align}

  Note that, when we take the limit of $m\to 0$ in the above expressions, we recover the one-loop result given in Eq.~\eqref{eq:Ai_sagh}. Moreover, in the limit $q\to 0$,  the form factors of Eq.~\eqref{1lmag} reduce to 
\begin{align}
\lim_{r^2\to 0} A_1^{(\Mass)}(0,-r,r) =& 1  \,, \nonumber \\
\lim_{r^2\to 0}A_3^{(\Mass)}(0,-r,r) =& - \frac{ C_\mathrm{A} \alpha_s }{ 288 \pi m^2 } \left[ 12 \ln \left( \frac{r^2}{m^2} \right) - 31 \right] \,,
\label{1mass_sag}
\end{align}
where we confirm that $A_1^{(\Mass)}$ is infrared finite, while $A_3^{(\Mass)}$ is logarithmically  divergent.   

\end{enumerate}

\item{\emph{Symmetric configuration:}} This kinematic limit is defined in~\eqref{symconf}; in this case all form factors survive,
and are given by:

\begin{enumerate}
\item {\emph{One loop:}} 
\begin{align}
A_1^{(1)}(Q)=&1+\frac{C_\mathrm{A}\alpha_s}{96\pi}\left[9+\rm{I}\right] \,, \nonumber\\
A_2^{(1)}(Q)=&-\frac{C_\mathrm{A}\alpha_s}{48\pi Q^2}\left[4+\rm{I}\right] \,, \nonumber\\
A_3^{(1)}(Q)=&\frac{C_\mathrm{A}\alpha_s}{96\pi Q^2}\left[4+9\rm{I}\right] \,, \nonumber\\
A_4^{(1)}(Q)=&\frac{C_\mathrm{A}\alpha_s}{48\pi Q^2}\left[1+2\rm{I}\right] \,, \nonumber\\
A_5^{(1)}(Q)=&\frac{C_\mathrm{A}\alpha_s}{48\pi Q^2}\left[-2+\rm{I}\right] \,,\label{eq:Ai_sym}
\end{align}
where $\rm{I}$ is a constant~\cite{Celmaster:1979km} defined as
\be 
\rm{I} = \frac{1}{3}\left[ \psi_1 \left(\frac{1}{3}\right) - \psi_1 \left(\frac{2}{3}\right) \right] = 2.34391 \,,
\ee
with $\psi_1(z)$ being the ``trigamma function', expressed in terms of the standard $\Gamma(z)$ function  as
\be
\psi_1(z) = \frac{d^2}{dz^2}\ln [\Gamma(z)] \,,
\ee
and it has the following special values
\begin{align}
\psi_1 \left(\frac{1}{3}\right) = 10.0956 \,, \qquad  \psi_1\left(\frac{2}{3}\right) = 3.06388  \,.
\end{align}

\item {\emph{One loop massive:}}

The resulting expressions for the one-loop massive case are rather lengthy and will not be reported here. However, their infrared limits as $q\to 0$ are given by 
\begin{align}
\lim_{Q^2\to 0}A_1^{(\Mass)}(Q) =& 1 \,, \nonumber \\
\lim_{Q^2\to 0}A_2^{(\Mass)}(Q) =& \frac{ C_\mathrm{A} \alpha_s }{576 \pi  m^2} \left[ 6 \ln \left(\frac{Q^2}{m^2}\right) - 65 \right] \,, \nonumber \\
\lim_{Q^2\to 0}A_3^{(\Mass)}(Q) =& -\frac{ C_\mathrm{A} \alpha_s }{144 \pi  m^2} \left[ 6 \ln \left(\frac{Q^2}{m^2}\right) - 23 +  3 \mathrm{I} \right] \,, \nonumber \\
\lim_{Q^2\to 0}A_4^{(\Mass)}(Q) =& \frac{ C_\mathrm{A} \alpha_s }{48 \pi  m^2} \,, \nonumber \\
\lim_{Q^2\to 0} A_5^{(\Mass)}(Q) =& -\frac{ C_\mathrm{A} \alpha_s }{192 \pi 
   m^2} \left[ 6 \ln\left(\frac{Q^2}{m^2}\right) - 1 + 4 \mathrm{I} \right] \,.
\label{limitsym}
\end{align}

Therefore, for the one-loop massive case, one finds that 
$A_1^{(\Mass)}$ and $A_4^{(\Mass)}$ are infrared finite, while 
$A_2^{(\Mass)}$, $A_3^{(\Mass)}$ and $A_5^{(\Mass)}$ are logarithmically divergent.

\end{enumerate}

\end{enumerate}

\newpage

\section{\label{app:projections} Explicit expressions for the $A_i$}
We write the $A_i$ as the sum of their tree-level value 
and the contributions from $(d_1)_{\nu\mu}$ and $(d_2)_{\nu\mu}$, so that
$A_i=A_i^{(0)}+A_i^{(d_\s1)}+A_i^{(d_\s2)}$, 
where $A_1^{(0)}=1$ and $A_i^{(0)}=0$ for $i=2,3,4,5$.

We then introduce new kinematic variables \mbox{$\,s=q-\ell\,$}, \mbox{$\,t=-\ell-p\,,\,u=-p-q\,$}, and 
\mbox{$v=-\ell+p+q$}, the inner products \mbox{$\,a_1=\ell\cd p\,$}, \mbox{$\,a_2=\ell\cd q\,$}, 
and \mbox{$\,a_3=p\cd q\,$},
together with the combinations 
\begin{align}
T_1&=h_{pq}+3(p^2+a_3)^2\,,&\quad
T_2&=h_{pq}+3(q^2+a_3)^2\,,&\nonumber\\
T_3&=-p^2q^2+p^4-2 a_3(q^2+a_3)\,,&\quad
T_4&=-p^2q^2+q^4-2 a_3(p^2+a_3)\,,&\nonumber\\
T_5&= p^2 a_2^2 + q^2 a_1^2-2 a_1 a_2 a_3\,.
\end{align}
Moreover, as a short-hand expedient, we will denote the arguments of several functions as a super/subscript, \ie $f(x,y,z) = f_{xyz}$ or
$f(x,y,z) = f^{xyz}$.

Then, the action of the projectors~\eqref{project} on diagram $(d_1)_{\nu\mu}$ gives
\begin{align}
A_1^{(d_\s1)}&=\frac{ig^2C_A}{4}\int_\ell\K^{(d_1)}\left\{\frac{ a_1 \left[h_{pq}\, \ell^2-T_5\right]}{h_{pq}\, \ell^2 }\right\}\,,&\nonumber \\
A_2^{(d_\s1)}&=-\frac{ig^2C_A}{4}\int_\ell\frac{\K^{(d_1)}}{h_{pq}^2\, \ell^2}\left\{h_{pq} \ell^2 \Big[a_1 \left(4 a_3+p^2+3 q^2\right)-2 a_2 \left(a_3+p^2\right)+2 h_{pq}\Big]\right.&\nonumber\\
&-a_1\Big[a_2^2h_{pq}-2 a_2 \left(p^2 \left(3 a_1 a_3+2 a_1 q^2+h_{pq}\right)+a_3 \left(4a_1 a_3+3 a_1 q^2+h_{pq}\right)\right)&\nonumber\\
&\left.+a_1\left(q^2 \left(6 a_1 a_3+a_1 p^2+3a_1 q^2+2h_{pq}\right)+2 a_3(a_1 a_3+h_{pq})\right)+ 3a_2^2 \left(a_3+p^2\right)^2\Big]\right\}\,,&\nonumber \\
A_3^{(d_\s1)}&= \frac{i\,g^2C_A}{4}\int_\ell\frac{\K^{(d_1)}}{h_{pq}^2\, \ell^2} \left\{ 3 a_1^3 q^4+a_1 q^2 \left[a_2\left(a_2p^2-6a_1a_3\right)-3 h_{pq}
\ell^2\right]\right.&\nonumber\\
&\left.+2 a_2a_3\left(a_1a_2a_3+h_{pq} \ell^2\right) \right\}\,,&\nonumber\\
A_4^{(d_\s1)}&=-\frac{i\,g^2C_A}{4}\int_\ell\frac{\K^{(d_1)}}{h_{pq}^2\, \ell^2} \left\{h_{pq} \ell^2 \left[3 a_1\left(a_3+q^2\right)-2a_2 \left(a_3+p^2\right)+2h_{pq}\right]\right.&\nonumber\\
&+a_1 \left[-a_1 q^2 \left(3 a_1 a_3+3a_1 q^2-6a_2 a_3+2h_{pq}\right)+2 a_2 a_3 (2
a_1 a_3-a_2 a_3+h_{pq})\right.&\nonumber\\
&\left.\left.-a_2 p^2 \left(q^2 (a_2-2a_1)+3 a_2 a_3\right)\right]\right\}\,,&\nonumber\\
A_5^{(d_\s1)}&=\frac{i\,g^2C_A}{4}\int_\ell\frac{\K^{(d_1)}}{h_{pq}^2\, \ell^2} \Big\{a_1\Big[a_2(a_2-2a_1) \left(3 a_3^2+h_{pq}\right)+3 a_1 q^2 \left(a_1 a_3+a_1 q^2-2 a_2 a_3\right)&\nonumber\\
&+3 a_2^2 a_3 p^2\Big]
-h_{pq} \ell^2 \left[a_3(a_1-2 a_2)+3 a_1 q^2\right]\Big\}\,.&
\end{align}
where 
\be
\K^{(d_1)} =\frac{\Delta \left(\ell^2\right)F\left(t^2\right)F\left(s^2\right)B_1(s,-t,u)\mathcal{V}_1(\ell,q,p,r)}{s^2\, t^2}\,.
\ee


Turning to diagram $(d_2)_{\nu\mu}$, all $A_i^{(d_\s2)}$ may be cast in the common form 
\be
A_i^{(d_\s2)} =\frac{ig^2C_A}{2}\int_\ell\frac{\K_{\ell u v}S_i^{\ell u v}+\K_{u v\ell}S_i^{u v\ell}+\K_{v\ell u}S_i^{v\ell u}}{h_{pq}^2\, \ell^2 } \,,
\ee
where 
\be
\K_{x y z} =\frac{\Delta \left(\ell^2\right)\Delta\left(v^2\right)F\left(s^2\right)\mathcal{V}_2(\ell,q,p,r){X_1}^{\!\! x y z}}{s^2\, v^2}\,.
\ee

Then, the $S_i$ are given by
\begin{align}
S_1^{\ell uv}&=-h_{pq}\left\{a_{1} \left[\left(a_3+q^2\right)
   \left(T_5 + h_{pq}\, \ell^2\right)-a_2 \left(\ell^2 \left(2 a_{3}
   \left(a_3+p^2\right)+h_{pq}\right)+T_5\right)\right]\right.&\nonumber\\
&\left.+a_1^2 \left[a_3\, \ell^2 \left(a_3+q^2\right)-T_5\right]
+\left(a_{3}+p^2\right) \left[-a_2\, T_5 + a_2\, \ell^2 \left(a_2\,p^2-h_{pq}\right)+h_{pq}\, \ell^4\right]\right\}\,,&\nonumber\\
S_1^{uv\ell}&=-h_{pq}(a_1+a_2) \left(-a_1+a_3+p^2\right) \left[h_{pq}\, \ell^2-T_5\right]\,,&\nonumber\\
S_1^{v\ell u}&=-h_{pq}\left\{T_5 \left[a_1^2+a_1 \left(a_2-a_3-q^2\right)+a_2 \left(a_3+p^2\right)\right]+a_1h_{pq}\, \ell^4\right.&\nonumber\\
&\left.-\ell^2\left[a_1^2 (h_{pq}-2a_2 a_3)+a_1 q^2 \left(a_1^2-h_{pq}\right)+a_1 h_{pq} (a_2-a_3)+a_2\, p^2
   (a_1 a_2+h_{pq})+a_2 a_3 h_{pq}\right]\right\}\,,\nonumber\\
S_2^{\ell uv}&=-a_1^4\, T_2+a_1^3 \left[3 a_2 \left(2 a_3 \left(a_3+p^2\right)+p^2
   q^2-q^4\right)+\left(a_3+q^2\right) \left(3 \left(a_3+q^2\right)^2+h_{pq}\right)\right]&\nonumber\\
&-a_1^2 \Big[3 a_2^2\,T_3+a_2 \left(q^2 \left(20 a_3^2+17 a_3 p^2+p^4\right)+2 a_3^2 \left(5 a_3+4 p^2\right)+q^4
   \left(9a_3+7 p^2\right)\right)&\nonumber\\
&-\ell^2 \left(3 a_3 \left(3 q^2 \left(a_3+p^2\right)+a_3 p^2+q^4\right)+2h_{pq} \left(p^2+2 q^2\right)\right)\Big]
+\left(\ell^2-a_2\right) \left(a_3+p^2\right) \Big[a_2^2 T_1&\nonumber\\
&-h_{pq} \ell^2 \left(2a_3+p^2+q^2\right)\Big]
-a_1\Big[a_2^3\,T_1+h_{pq} \ell^2\left(a_3+q^2\right) \left(2 a_3+p^2+q^2\right)&\nonumber\\
&-a_2^2 \left(p^2 \left(20 a_3^2+17 a_3 q^2+q^4\right)+2 a_3^2
   \left(5a_3+4 q^2\right)+p^4 \left(9a_3+7 q^2\right)\right)&\nonumber\\
&+a_2\ell^2 \left(p^2 \left(15 a_3^2+8a_3 q^2-q^4\right)+a_3^2 \left(10 a_3+7 q^2\right)+3 p^4 \left(2 a_3+q^2\right)\right)\Big]\,,&\nonumber\\ %
S_2^{uv\ell}&=a_1^2 \Big[3a_2^2 T_3+a_2\left(6 a_3 p^4+p^2 \left(15a_3^2+9a_3 q^2-h_{pq}\right)+h_{pq}\left(5q^2-2a_3\right)-3a_3q^2\left(a_3+q^2\right)\right)&\nonumber\\
&-h_{pq} \left(\ell^2 \left(4 a_3+p^2+3
   q^2\right)+2 \left(a_3+q^2\right) \left(p+q\right)^2\right)\Big]
+a_1^3 \Big[q^4 \left(3a_2-3 a_3+p^2\right)&\nonumber\\
&-q^2 \left(3 p^2 (a_2+a_3)+10
a_3^2+p^4\right)-2 a_3\left(p^2 (3 a_2+a_3)+3 a_3 (a_2+a_3)\right)\Big]+a_1^4 T_2&\nonumber\\
&+a_1 \Big[a_2^3 T_1+a_2^2 \left(p^2 \left(14
a_3^2+5 a_3 q^2+4q^4\right)+2 a_3^2 \left(5 a_3+q^2\right)-p^4 \left(3 a_3+5 q^2\right)-3
   p^6\right)&\nonumber\\
&+a_2 h_{pq} \left(\ell^2 \left(-2 a_3+p^2-3 q^2\right)+2 \left(p^2-q^2\right) \left(p+q\right)^2\right)+h_{pq} \ell^2 \left(a_3+p^2\right) \left(p+q\right)^2\Big]&\nonumber\\
&+a_2 \left(a_3+p^2\right)
   \left[(p+q)^2\left(h_{pq}\ell^2+2a_2h_{pq}-3a_2^2p^2\right)+2a_2h_{pq}\left(\ell^2-a_2\right)\right]\,,
\end{align}%

\begin{align}
S_2^{v\ell u}&=a_1^4 T_2+a_1^3 \left[3 a_2T_4-\left(a_3+q^2\right) \left(3 \left(a_3+q^2\right)^2+h_{pq}\right)\right]
+a_1^2 a_2 \Big[3 a_2 T_3+2 a_3^2 \left(5 a_3+4 p^2\right)&\nonumber\\
&+q^4 \left(9 a_3+7 p^2\right)\Big]
+\ell^2 \Big[a_1\left(a_2 h_{pq} \left(-2 a_3+3 p^2-5 q^2\right)+3 h_{pq}\left(a_3+q^2\right) (p+q)^2-a_2^2 T_1\right)&\nonumber\\
&+a_1^2 \left(p^2 \left(6a_2 a_3+4a_2 q^2-h_{pq}\right)+q^2 (6 a_2 a_3-5h_{pq})+2 a_3(4a_2 a_3-3 h_{pq})\right)
-a_1^3T_2&\nonumber\\
&+a_2 h_{pq} \left(a_3+p^2\right) \left(4 a_2-3 (p+q)^2\right)\Big]
+a_1 a_2^2\left[a_2 T_1-2 a_3^2 \left(5 a_3+4 q^2\right)-p^4\left(9 a_3+7 q^2\right)\right]&\nonumber\\
&+a_1a_2\left[20a_3^2\left(a_1q^2-a_2p^2\right)+17a_3p^2q^2\left(a_1-a_2\right)+p^2q^2\left(p^2a_1-q^2a_2\right)\right]&\nonumber\\
&+h_{pq} \ell^4 \left[a_1 \left(4 a_3+p^2+3 q^2\right)-2 a_2 \left(a_3+p^2\right)\right]
+a_2^3T_1 \left(a_3+p^2\right)
\,,&\nonumber\\
S_3^{\ell uv}&=3 a_1^3 q^6-2 a_2^2 a_3^2 \left[a_1^2+a_1a_2-a_1a_3+a_2a_3+a_2 p^2-\ell^2 \left(a_3+p^2\right)\right]&\nonumber\\
&-a_1q^4 \left[3 a_1\left(a_1^2+a_1a_2-a_1a_3+3 a_2a_3\right)+\ell^2 \left(-3a_1a_3-3 a_1p^2+h_{pq}\right)-a_2^2 p^2\right]&\nonumber\\
&-q^2 \Big[-\ell^2
   \Big(h_{pq} \left(a_1^2+a_1(a_2-a_3)+a_2a_3\right)+a_2 p^2 \left(-6 a_1a_3+a_2a_3+a_2 p^2+h_{pq}\right)&\nonumber\\
&-6 a_1a_2a_3^2\Big)+h_{pq} \ell^4 \left(a_3+p^2\right)+a_2\Big(a_2p^2 \left(a_1^2+a_1a_2-7 a_1a_3+a_2a_3+a_2p^2\right)&\nonumber\\
&+a_1(a_3^2 (9 a_1-8 a_2)-6 a_1a_3 (a_1+a_2)+3 a_1h_{pq})\Big)\Big]\,,&\nonumber\\
S_3^{uv\ell}&=(a_1+a_2) \Big[2 a_2\left(a_3^2 (a_1 a_3+h_{pq})+q^2 \left(a_3 (a_1(a_3-3 a_1)+h_{pq})+2
a_1 p^2 \left(a_3+q^2\right)\right)\right)&\nonumber\\
&-a_1 q^2 \left(q^2 \left(a_1\left(-3 a_1+3 a_3+p^2\right)+2 h_{pq}\right)+2a_3(a_1 a_3+h_{pq})\right)&\nonumber\\
&+a_2^2 \left(\left(a_1-p^2\right) \left(3 a_3^2+h_{pq}\right)-3 a_3^3-5a_3h_{pq}\right)\Big]
+2 h_{pq}^2 \ell^4+\ell^2 \Big[q^2 \Big(a_1 a_3^2 (5 a_1+3a_2)&\nonumber\\
&+a_3 h_{pq} (3a_1+a_2)+2 h_{pq}^2\Big)
+2a_3 \Big(-a_2 a_3^2 (2a_1+a_2)+h_{pq}^2+a_3 h_{pq}(a_1+a_2)\Big)&\nonumber\\
&+p^2 \left(q^2 \left(2 a_2 a_3 (2 a_1+a_2)+h_{pq} (a_1-3 a_2)-a_1 q^2 (5
a_1+3 a_2)\right)-2 a_2 a_3 h_{pq}\right)\Big]
\,,&\nonumber\\
S_3^{v\ell u}&=-3 a_1^3 q^6+q^2 \Big[a_2 p^2 \left(a_1^2 (a_2+a_3)+a_1\left(a_2^2-6a_2 a_3-h_{pq}\right)+a_2^2 a_3+a_2^2 p^2\right)&\nonumber\\
&+a_1 a_3\left(a_1^2 (a_3-6 a_2)+a_1 \left(-6a_2^2+7 a_2 a_3+h_{pq}\right)-2 a_2(4 a_2 a_3+h_{pq})\right)\Big]&\nonumber\\
&+a_2 a_3\Big[a_3\left(a_1^2 (2 a_2-a_3)+a_1\left(2 a_2^2-3 a_2 a_3-h_{pq}\right)+a_2(2 a_2 a_3+h_{pq})\right)&\nonumber\\
&+a_2 p^2 (2 a_2 a_3+h_{pq})\Big]
+\ell^2 \Big[-a_3\left(a_1 a_3 (a_2 (2 a_2+a_3)+h_{pq})+(2 a_2 a_3+h_{pq})^2\right)&\nonumber\\
&+q^2 \Big(3 a_1 q^2 \left(h_{pq}-a_1^2\right)+p^2
   \left(a_1 (a_2 (a_3-a_2)+h_{pq})-a_1 q^2 (2 a_1+5 a_2)+a_2 (4 a_2 a_3+h_{pq})\right)&\nonumber\\
&+3a_3 h_{pq} (a_1-a_2)+a_1 a_3(2 a_1(3 a_2+a_3)+5 a_2 a_3)-h_{pq}^2\Big)\Big]+a_1 q^4 \Big[-p^2 (a_1-a_2)^2&\nonumber\\
&+a_1 (-3 a_3 (a_1-3a_2)+3 a_1 (a_1+a_2)+h_{pq})\Big]-h_{pq} \ell^4 \left(-3 a_1 q^2+2 a_2 a_3+h_{pq}\right)\,,&
\end{align}

\begin{align}
S_4^{\ell uv}&=-\left[a_1^2+a_1 \left(a_2-a_3-q^2\right)+a_2 \left(a_3+p^2\right)\right] \Big[a_2^2 \left(3a_3 \left(a_3+p^2\right)+h_{pq}\right)&\nonumber\\
&-2 a_1 a_2 \left(3 a_3 \left(a_3+q^2\right)+h_{pq}\right)+3 a_1^2
   q^2 \left(a_3+q^2\right)\Big]-h_{pq}\ell^4 \left(a_3+p^2\right) \left(a_3+q^2\right)&\nonumber\\
	&-\ell^2 \Big[a_1 \left(a_3+q^2\right) \left(q^2 \left(-3a_1a_3-4 a_1 p^2+h_{pq}\right)+a_3 (a_1
a_3+h_{pq})\right)&\nonumber\\
&+a_2 \Big(-p^2 \left(q^2 \left(-7 a_1 a_3+a_1 q^2+h_{pq}\right)+a_3(h_{pq}-4 a_1 a_3)\right)&\nonumber\\
&+a_3 \left(q^2 (7 a_1 a_3-h_{pq})+a_3 (5 a_1 a_3-h_{pq})\right)+2 a_1
p^4 q^2\Big)&\nonumber\\
&-a_2^2 \left(a_3+p^2\right) \left(3 a_3 \left(a_3+p^2\right)+h_{pq}\right)\Big]\,,&\nonumber\\
S_4^{uv\ell}&=2 h_{pq}^2 \ell^4+(a_1+a_2) \Big[a_1q^2 \Big(a_3\left(3 a_1^2-7 a_1a_3-4 h_{pq}\right)+p^2 \left(-3 a_1 a_3+a_1 q^2-2 h_{pq}\right)&\nonumber\\
&+q^2 (3 a_1 (a_1-a_3)-2h_{pq})\Big)
+a_2^2 \Big(2 a_1a_3^2+a_1 p^2 \left(3 a_3+q^2\right)-h_{pq} \left(2 a_3+p^2\right)&\nonumber\\
&-3a_3 p^2\left(2 a_3+p^2+q^2\right)\Big)
+2 a_2 \Big(a_3\left(2a_3\left(-a_1^2+2 a_1a_3+h_{pq}\right)+p^2 (2 a_1a_3+h_{pq})\right)&\nonumber\\
&+q^2 \left(a_1\left(\left(a_3+p^2\right)^2-a_1 \left(3 a_3+p^2\right)\right)+a_3h_{pq}\right)+2 a_1 p^2 q^4\Big)
\Big]&\nonumber\\
&+\ell^2 \Big[q^2 \left(-5a_1^2 h_{pq}+3 a_1 a_3 (a_2a_3+h_{pq})+h_{pq}(a_2a_3+2h_{pq})\right)-3 a_1a_2a_3^3&\nonumber\\
&+a_3h_{pq} \left(-3a_1^2+7 a_1a_3+a_2 (2 a_2+3 a_3)\right)-2 a_2 p^4 \left(h_{pq}-a_1
   q^2\right)+4 a_3h_{pq}^2&\nonumber\\
&-p^2 \Big(q^2 \left(-3 a_1a_2a_3+3 a_1a_2q^2+a_1h_{pq}+3a_2h_{pq}\right)+a_1a_3(2 a_2a_3-3 h_{pq})&\nonumber\\
&+h_{pq} (3 a_2a_3-2 h_{pq})\Big)\Big]\,,&\nonumber\\
S_4^{v\ell u}&=\left[a_1^2+a_1 \left(a_2-a_3-q^2\right)+a_2 \left(a_3+p^2\right)\right] \Big[3 a_1^2
   q^2 \left(a_3+q^2\right)-2 a_1a_2 \left(3 a_3 \left(a_3+q^2\right)+h_{pq}\right)&\nonumber\\
&+a_2^2 \left(3 a_3 \left(a_3+p^2\right)+h_{pq}\right)\Big]
-\ell^2 \Big[a_1^2 \left(q^2 \left(3 h_{pq}-2a_2
   \left(3 a_3+p^2\right)\right)
+a_3 (3 h_{pq}-4 a_2a_3)\right)&\nonumber\\
&+a_1 \left(a_2^2 \left(3 a_3\left(a_3+p^2\right)+h_{pq}\right)+a_2 \left(3 a_3-2 p^2+5 q^2\right) h_{pq}-3 h_{pq} \left(a_3+q^2\right)^2\right)&\nonumber\\
&+3 a_1^3 q^2 \left(a_3+q^2\right)+a_2 h_{pq} \left(a_3+p^2\right) \left(3 \left(a_3+q^2\right)-4 a_2\right)\Big]&\nonumber\\
&-h_{pq}\ell^4 \left[2 a_2 \left(a_3+p^2\right)-3 a_1 \left(a_3+q^2\right)\right]\,,&\nonumber\\
S_5^{\ell uv}&=-\left[a_1^2+a_1 \left(a_2-a_3-q^2\right)+a_2\left(a_3+p^2\right)\right] \Big[-2a_1 a_2\left(3a_3\left(a_3+q^2\right)+h_{pq}\right)&\nonumber\\
&+3a_1^2q^2 \left(a_3+q^2\right)+a_2^2 \left(3 a_3\left(a_3+p^2\right)+h_{pq}\right)\Big]
-h_{pq}\ell^4 \left(a_3+p^2\right) \left(a_3+q^2\right)&\nonumber\\
&-\ell^2 \Big[a_2\Big(-p^2 \left(q^2 \left(-7a_1 a_3+a_1 q^2+h_{pq}\right)+a_3
   (h_{pq}-4 a_1 a_3)\right)+2 a_1p^4 q^2&\nonumber\\
&+a_3 \left(q^2 (7 a_1 a_3-h_{pq})+a_3 (5 a_1 a_3-h_{pq})\right)\Big)
-a_2^2\left(a_3+p^2\right) \left(3 a_3\left(a_3+p^2\right)+h_{pq}\right)&\nonumber\\
&+a_1 \left(a_3+q^2\right) \Big(q^2 \left(-3a_1 a_3-4 a_1 p^2+h_{pq}\right)+a_3 (a_1 a_3+h_{pq})\Big)\Big]
\,,&
\end{align}

\begin{align}
S_5^{uv\ell}&=(a_1+a_2) \Big\{3a_1^3 q^2 \left(a_3+q^2\right)-a_1^2 \left(2a_2+a_3+q^2\right) \left[2a_3^2+q^2 \left(3a_3+p^2\right)\right]&\nonumber\\
&+a_1\Big[a_2^2 \left(3a_3 \left(a_3+p^2\right)+h_{pq}\right)+2 a_2
   \left(a_3+q^2\right) \left(3 a_3\left(a_3+p^2\right)+2 h_{pq}\right)&\nonumber\\
&+\left(a_3+q^2\right) \left(-3 \ell^2 h_{pq}-2 h_{pq}\left(a_3+q^2\right)\right)\Big]
-\left(a_3+p^2\right) \Big[\ell^2 \Big(a_3(2 a_2 a_3-h_{pq})&\nonumber\\
&-q^2 \left(2a_2 p^2+h_{pq}\right)\Big)+a_2\left(2h_{pq} \left(a_2-a_3-q^2\right)
+3 a_2 p^2\left(a_3+q^2\right)\right)\Big]\Big\}\,,&\nonumber\\
S_5^{v\ell u}&=-a_1^3 \left[q^2 \left(3 q^2 \left(-a_2+2 a_3+q^2\right)+3a_3 (a_2+a_3)+2a_2 p^2+h_{pq}\right)+a_3(4 a_2 a_3+h_{pq})\right]&\nonumber\\
&-a_1^2 \Big[-a_2\left(4 a_3^3+p^2 \left(5 a_3 q^2+h_{pq}+4 q^4\right)+a_3 q^2 \left(14 a_3+9 q^2\right)\right)-h_{pq} \left(a_3+q^2\right)^2&\nonumber\\
&+a_2^2 \left(3 a_3\left(a_3-p^2+2 q^2\right)+h_{pq}\right)\Big]
-\ell^2 \Big[a_3\Big(2a_2 a_3\left(-2a_1^2+a_1 (a_2+a_3)+2 a_2 a_3\right)&\nonumber\\
&+a_2 p^2 (3 a_1 a_2-a_1 a_3+3h_{pq})+a_3 h_{pq} (a_1+7a_2)\Big)+a_1q^4 \left(3 a_1^2+p^2 (4 a_1+5a_2)-3 h_{pq}\right)&\nonumber\\
&+h_{pq}^2(p+q)^2-q^2 \Big(p^2 \left(2 a_1^2a_2-a_1 a_2\left(a_2+p^2\right)+4 a_1 h_{pq}+a_2 h_{pq}\right)
+a_1 a_3^2 (4 a_1+5 a_2)&\nonumber\\
&+a_3 \left(-3 a_1^3+6 a_1^2 a_2+2a_2 p^2 (a_1+2a_2)+6 a_1 h_{pq}-3 a_2 h_{pq}\right)\Big)\Big]
+3a_1^4 q^2 \left(a_3+q^2\right)&\nonumber\\
&-a_1 a_2\Big[a_2p^2 \left(8 a_3^2+11a_3 q^2+q^4+p^2 q^2\right)+a_2a_3^2 \left(7 a_3+8 q^2\right)
-a_2^2\left(3 a_3\left(a_3+p^2\right)+h_{pq}\right)&\nonumber\\
&+2 h_{pq}\left(a_3+p^2\right) \left(a_3+q^2\right)\Big]
+h_{pq} \ell^4 \left[a_3(a_1-2 a_2+a_3)+q^2 \left(3
a_1-p^2\right)\right]&\nonumber\\
&+a_2^2 \left(a_3+p^2\right) \left[a_2 \left(3 a_3 \left(a_3+p^2\right)+h_{pq}\right)+h_{pq}
   \left(a_3+p^2\right)\right]\,.&
\end{align}

Finally, it is understood that, for the numerical evaluation of the above expressions,
all relevant quantities are to be replaced by their ``input'' expressions, namely
$\Delta_{\inpt}(q)$, $X_1^{\inpt}(r,t,\ell)$, $F_{\inpt}(q)$, and $B_1^{\inpt}(Q)$, introduced in section~\ref{inputs}. 

\acknowledgments 

The research of J.~P. is supported by the Spanish MEyC under grants FPA2017-84543-P and SEV-2014-0398, and Generalitat Valenciana  
under grant Prometeo~II/2014/066.
The work of  A.~C.~A and M.~N.~F. are supported by the Brazilian National Council for Scientific and Technological Development (CNPq) under the grants 305815/2015 and  142226/2016-5, respectively. A.~C.~A and C.~T.~F. also acknowledge the financial support
from  S\~{a}o Paulo Research Foundation (FAPESP) through the projects  2017/07595-0, 
2017/05685-2, 2016/11894-0, and 2018/09684-3.  This study was financed in part by the Coordena\c{c}\~{a}o de Aperfei\c{c}oamento de Pessoal de N\'{\i}vel Superior - Brasil (CAPES) - Finance Code 001 (M.~N.~F.). This research was performed using the Feynman Cluster of the
John David Rogers Computation Center (CCJDR) in the Institute of Physics ``Gleb
Wataghin", University of Campinas.


\end{document}